\documentclass[%
 aip,
 amsmath,amssymb,
reprint, onecolumn 
]{revtex4-1}

\usepackage[T1]{fontenc}
\usepackage{mathptmx}
\usepackage{etoolbox}

\usepackage{amssymb} 
\usepackage{bm}

\usepackage{graphicx}
\usepackage{dcolumn}
\usepackage{multirow}
\usepackage{array} 
\usepackage{tabularx}
\usepackage{caption}
\usepackage{subcaption}
\usepackage{threeparttable}
\usepackage{booktabs}
\usepackage{adjustbox}

\usepackage{tikz}
\usepackage{pgfplots}
\usetikzlibrary{external, arrows.meta}
\usepgfplotslibrary{fillbetween}
\pgfplotsset{compat=1.9}
\pgfplotsset{
    every axis/.append style={
        line width=0.6pt,
    },
    mygridstyle/.style={
        grid style=solid,
        line width=0.35pt,
        color=gray!50,
    },
}

\usepackage{CJKutf8}

\usepackage{hyperref}
\usepackage{xcolor}
\usepackage{cleveref}

\usepackage[normalem]{ulem}
\newcolumntype{L}{>{\raggedright\arraybackslash}X} 
\newcolumntype{C}{>{\centering\arraybackslash}X}   
\definecolor{hcorange}{RGB}{255,127,80}
\definecolor{hcblue}{RGB}{30,144,255}
\definecolor{hcgreen}{RGB}{229,242,229}
\definecolor{hcpurple}{RGB}{229,229,255}
\definecolor{hcpurple1}{RGB}{155,110,188}
\definecolor{hcred}{RGB}{130,118,213}
\definecolor{hc03}{RGB}{255,99,72}
\definecolor{hc15}{RGB}{47,53,66}
\definecolor{Purple9}{RGB}{112,48,160} 
\definecolor{Purple8}{RGB}{125,60,173}
\definecolor{Purple7}{RGB}{138,72,186}
\definecolor{Purple6}{RGB}{151,84,199}
\definecolor{Purple5}{RGB}{164,96,212} 
\definecolor{Purple4}{RGB}{177,108,225}
\definecolor{Purple3}{RGB}{190,120,238}
\definecolor{Purple2}{RGB}{203,132,251}
\definecolor{Purple1}{RGB}{216,144,255} 

\definecolor{Orange1}{RGB}{255,235,213} 
\definecolor{Orange2}{RGB}{255,215,183}
\definecolor{Orange3}{RGB}{255,195,153}
\definecolor{Orange4}{RGB}{255,175,123}
\definecolor{Orange5}{RGB}{255,155,93}  
\definecolor{Orange6}{RGB}{245,135,75}
\definecolor{Orange7}{RGB}{235,115,55}
\definecolor{Orange8}{RGB}{225,95,35}
\definecolor{Orange9}{RGB}{215,75,15}   

\crefname{equation}{Eq.}{Eqs.}
\Crefname{equation}{Eq.}{Eqs.}
\crefname{figure}{Fig.}{Figs.}
\Crefname{figure}{Fig.}{Figs.}
\crefname{algorithm}{Algorithm}{Algorithms}
\Crefname{algorithm}{Algorithm}{Algorithms}
\crefname{section}{Sec.}{Sections}
\Crefname{section}{Sec.}{Sections}
\crefname{table}{Tab.}{Tables.}
\Crefname{table}{Tab.}{Tables.}

\makeatletter
\def\@email#1#2{%
 \endgroup
 \patchcmd{\titleblock@produce}
  {\frontmatter@RRAPformat}
  {\frontmatter@RRAPformat{\produce@RRAP{*#1\href{mailto:#2}{#2}}}\frontmatter@RRAPformat}
  {}{}
}%
\makeatother

\begin{document}

\preprint{AIP/123-QED}

\title[Intelligent AFC for Ellipsoid]{Deep reinforcement learning-based active flow control of an elliptical cylinder: transitioning from an elliptical cylinder to a circular cylinder and a flat plate}

\author{Wang Jia (\begin{CJK*}{UTF8}{gbsn}贾旺\end{CJK*})}
\author{Hang Xu (\begin{CJK*}{UTF8}{gbsn}徐航\end{CJK*})}
\email{hangxu@sjtu.edu.cn}
\affiliation{School of Ocean and Civil Engineering, Shanghai Jiao Tong University, Shanghai, 200240, China}

\date{\today}

\begin{abstract}

We study the adaptability of deep reinforcement learning (DRL)-based active flow control (AFC) technology for bluff body flows with complex geometries.
It is extended from a cylinder with an aspect ratio $Ar = 1$ to a flat elliptical cylinder with $Ar=2$, slender elliptical cylinders with $Ar$ less than 1, and a flat plate with $Ar=0$.
We utilize the Proximal Policy Optimization (PPO) algorithm to precisely control the mass flow rates of synthetic jets located on the upper and lower surfaces of a cylinder to achieve reduction in drag, minimization of lift, and suppression of vortex shedding.
Our research findings indicate that, for elliptical cylinders with $Ar$ between 1.75 and 0.75, the reduction in drag coefficient ranges from 0.9\% to 15.7\%, and the reduction in lift coefficient ranges from 95.2\% to 99.7\%. 
The DRL-based control strategy not only significantly reduces lift and drag, but also completely suppresses vortex shedding while using less than 1\% of external excitation energy, demonstrating its efficiency and energy-saving capabilities. 
Additionally, for $Ar$ from 0.5 to 0, the reduction in drag coefficient ranges from 26.9\% to 43.6\%, and the reduction in lift coefficient from 50.2\% to 68.0\%. 
This reflects the control strategy's significant reduction in both drag and lift coefficients, while also alleviating vortex shedding.
The interaction and nonlinear development of vortices in the wake of elliptical cylinders lead to complex flow instability, and DRL-based AFC technology shows adaptability and potential in addressing flow control problems for this type of bluff body flow.

\end{abstract}

\maketitle

\section{INTRODUCTION}

The development of active flow control (AFC) technology progresses from theoretical exploration to practical application, gradually becoming an important branch within the domains of fluid mechanics and engineering.\cite{collis2004issues,jahanmiri2010active,en14113058} 
This technology actively intervenes the flow field to improve flow characteristics, such as reducing drag, controlling vortex shedding, enhancing mixing efficiency, or preventing flow separation, thereby enhancing system performance and efficiency.\cite{collis2004issues,Detached} 
In its early stages, AFC research primarily focused on theoretical analysis and simple experiments to explore the influence of external inputs, such as jets, vibrations, or magnetic fields, on fluid flow dynamics.\cite{jahanmiri2010active,Detached,en14113058} 
With the advancement of computational fluid dynamics (CFD) and increased computational power, researchers are able to delve deeper into complex flow responses through numerical simulations and develop more effective control strategies.\cite{Aram2018,jain2010computational,LI202214}
However, due to the high degree of nonlinearity and multiscale nature of fluid problems, which involve high-dimensional state spaces and complex system dynamics, many engineering problems are often intractable using existing models.\cite{pr8111379,reichstein2019deep,vinuesa2024perspectives}
Sophisticated control strategies that require consideration of a wide range of spatiotemporal scales are indispensable when addressing phenomena such as unstable flows\cite{vinuesa2024perspectives}, turbulence\cite{renApplying}, and multiphase flows.\cite{Subramaniam}

State-of-the-art machine learning technologies profoundly impact the field of active fluid control.\cite{annurevfluid,bruntonClosed,ren2020active}
Deep Reinforcement Learning (DRL) combines deep learning and reinforcement learning to optimize decision-making through interactions with the environment.\cite{arulkumaran2017deep,franccois2018introduction,9904958}
DRL excels at extracting complex features from large datasets and making optimal decisions based on these features.\cite{Janiesch2021,lecun2015deep}
By using deep neural networks, DRL handles high-dimensional state spaces and enables machines to tackle previously intractable decision-making challenges.\cite{arulkumaran2017deep,franccois2018introduction,9904958}
DRL's nonlinear modeling capabilities and adaptive learning mechanisms allow it to learn complex strategies directly from raw sensory data without manual feature engineering.\cite{dargazany2021drl}
It continuously improves its strategies through interactions with the environment, making it advantageous for dynamic decision-making problems.\cite{franccois2018introduction,9904958}
DRL finds applications in various fields including gaming, autonomous vehicles, robotics, natural language processing, and computer vision.
In particular, it achieves distinguished success, such as AlphaGo's victory over world champions in gaming, route planning and decision-making optimization in autonomous driving, and learning complex physical tasks in robotics.\cite{granter2017alphago,9308468,Sallab_2017}

With progressive advancements in DRL technology, there is a significant increase in the application of DRL in the field of fluid dynamics as well.\cite{annurevfluid, bruntonClosed, GARNIER2021104973, Popat}
DRL methods are highly suited to handle the high-dimensional and complex state spaces intrinsic to fluid dynamics.\cite{Aerospace2023}
Through the function approximation capabilities of deep neural networks, DRL is adept at processing fluid systems and discerning flow patterns and structures.\cite{ZHAO2024122836}
DRL models, through their interaction with the environment and trial-and-error processes, independently learn and refine strategies to meet diverse fluid conditions.\cite{Husen2023, collis2004issues, jahanmiri2010active}
For instance, DRL is applied to shape optimization of fluid systems, enabling the automatic discovery and application of effective shape adjustments to achieve optimization goals by intelligently exploring the design space and learning optimal strategies.\cite{BHOLA2023112018}
Moreover, the long-term reward mechanism intrinsic to DRL allows for the consideration of future impacts when formulating strategies.
This unique feature allows DRL to excel in many AFC problems, where DRL's focus on long-term rewards enables control models to balance short-term gains with future outcomes, thereby better addressing control and optimization challenges within fluid mechanics.\cite{vinuesa2022flow, Vignon2023, Viquerat, vinuesa2024perspectives}

In the literature, DRL attracts considerable interest for flow control around bluff bodies, as flow around these bodies tends to involve strong nonlinearities such as flow separation and recirculation. Traditional control methods often struggle to achieve control objectives, whereas DRL models show promising results. Pioneered by \citeauthor{rabault2019artificial} \cite{rabault2019artificial}, various researchers such as \citeauthor{tangRobustActiveFlow2020} \cite{tangRobustActiveFlow2020}, \citeauthor{heess2017emergence} \cite{heess2017emergence}, \citeauthor{renApplying} \cite{renApplying}, and \citeauthor{jia2024optimal} \cite{jia2024optimal} apply the Proximal Policy Optimization (PPO) algorithm to AFC in the flow around a circular cylinder. Consistently, they achieve an approximate 8\% reduction in drag in their validation experiments. 
Furthermore, \citeauthor{liReinforcementlearning}\cite{liReinforcementlearning}, \citeauthor{wangDRLinFluids}\cite{wangDRLinFluids}, \citeauthor{jia2024robust}\cite{jia2024robust}, \citeauthor{wang2022deep}\cite{wang2022deep}, and \citeauthor{he2023policy}\cite{he2023policy} respectively conduct flow control studies on confined circular cylinders, square cylinders, and airfoils using DRL. In addition, \citeauthor{Dixiapnas}\cite{Dixiapnas} initiate the application of DRL to AFC experiments, optimizing power efficiency by adjusting the rotational speed of auxiliary circular cylinders positioned behind another circular cylinder. \citeauthor{Bluffbody}\cite{Bluffbody} propose an AFC strategy to conceal the hydrodynamic characteristics of circular cylinders, such as strong shear forces and periodic shedding vortices.
Studies on flow around constrained circular cylinders under different blockage ratios also receive extensive attention in the field.\cite{rabault2019artificial, tangRobustActiveFlow2020, liReinforcementlearning, wangDRLinFluids, jia2024optimal} The application of DRL-based AFC technology in bluff body flows is summarized in \cref{tab:bluff}. DRL-based AFC control technology is successfully applied to square, circular, and airfoil flows. Additionally, multiple studies demonstrate that DRL-based flow control strategies exhibit robustness across different Reynolds numbers and achieve favorable control outcomes in fluid mechanics experiments. This indicates that DRL-based AFC technology provides effective control strategies in challenging high-dimensional and nonlinear scenarios.

\begin{table}[ht]
\centering
\caption{Application of AFC technology based on DRL in the flow around bluff body.}
\label{tab:bluff}
\vspace{-\baselineskip}
\begin{tabularx}{\textwidth}{
  >{\centering\arraybackslash}p{0.12\linewidth}
  >{\centering\arraybackslash}p{0.12\linewidth}
  >{\centering\arraybackslash}p{0.15\linewidth}
  >{\centering\arraybackslash}p{0.15\linewidth}
  >{\centering\arraybackslash}p{0.15\linewidth}
  >{\centering\arraybackslash}p{0.25\linewidth}
}
\toprule
\toprule
Re & Bluff Body & Control measures & DRL algorithm & Drag Reduction & Reference \\
\midrule
100     & Cylinder & Synthetic Jets & PPO & 8\%  & \citeauthor{rabault2019artificial}\cite{rabault2019artificial} \\
100     & Cylinder & Synthetic Jets & PPO & 8\%  & \citeauthor{jia2024optimal}\cite{jia2024optimal}  \\
100     & Cylinder & Synthetic Jets & PPO & 8\%  & \citeauthor{wangDRLinFluids}\cite{wangDRLinFluids} \\
120     & Cylinder & Synthetic Jets & S-PPO-CMA & 18.4\% & \citeauthor{paris}\cite{paris}  \\
100-300 & Cylinder & Synthetic Jets & PPO & 6\%-24\% & \citeauthor{he2023policy}\cite{he2023policy} \\
100-400 & Cylinder & Synthetic Jets & PPO & 6\%-39\% & \citeauthor{tangRobustActiveFlow2020}\cite{tangRobustActiveFlow2020} \\
1,000   & Cylinder & Synthetic Jets & PPO & 30\% & \citeauthor{renApplying}\cite{renApplying} \\
10,160  & Cylinder & Two fast-rotating smaller cylinders & TD3 & 30\% & \citeauthor{Dixiapnas}\cite{Dixiapnas} \\ 
100     & Square  & Synthetic Jets & SAC & 14\% & \citeauthor{wangDRLinFluids}\cite{wangDRLinFluids}  \\
100     & Square  & Synthetic Jets & SAC & 14\% & \citeauthor{chen2023deep}\cite{chen2023deep}  \\
100     & Square  & Synthetic Jets & SAC and TQC & 17\% & \citeauthor{Xia2024}\cite{Xia2024}  \\
100-400 & Square  & Synthetic Jets & SAC & 14\%-47\% & \citeauthor{jia2024robust}\cite{jia2024robust}  \\
100-500 & Square  & Synthetic Jets & SAC & 14\%-51\% & \citeauthor{jia2024effect}\cite{jia2024effect} \\
500-2,000 & Square & Synthetic Jets & SAC & 44\%-61\% & \citeauthor{yan2023stabilizing}\cite{yan2023stabilizing} \\
\bottomrule
\bottomrule
\end{tabularx}
\end{table}

The application of AFC strategies based on DRL is widely validated across various bluff body flow scenarios. Among these, research on confined circular cylinder flow under different blockage ratios is the most thorough. In contrast, studies related to elliptical cylinder flow are exceedingly rare in this domain. The aspect ratio of an elliptical cylinder significantly affects flow stability and vortex structure development, with the slender shape's structural effects directly increasing the difficulty of flow control.
While active flow control strategies for circular and square cylinders are crucial for managing aerodynamic and hydrodynamic performance in various engineering applications, extending these strategies to more complex, non-ideal geometries remains a major challenge.\cite{Vignon2023,rabault2020deep} Under these circumstances, the effectiveness of DRL strategies is largely unexplored, particularly regarding whether they can produce physically reasonable and adaptable control measures.
Current literature indicates a lack of comprehensive studies on the variability of DRL-based flow control strategies across different geometries, such as from streamlined bodies to various bluff bodies.\cite{GARNIER2021104973,Popat} Therefore, this study preliminarily investigates whether it is possible to effectively reduce drag and potentially suppress vortex shedding behind an elliptical cylinder at \(Re=100\). This lower Reynolds number allows for a detailed examination of the fundamental principles and mechanisms involved in the control strategy, enabling a rigorous evaluation without the additional complexities introduced by turbulence. The experiences and insights gained from this work are crucial for applying DRL to more practical engineering problems in fluid dynamics.

The structure of this paper is organized as follows: 
\cref{sec:Methodology} details the problem addressed in this study and describes the numerical simulation methods and deep reinforcement learning techniques used. Additionally, it presents the framework for solving the AFC problem using DRL, providing a thorough description of the fundamental elements involved.
\Cref{sec:results} presents the main results and analysis. 
In \cref{sec:baseline}, the baseline flow for elliptical cylinders with different \(Ar\) is described, and probe positions and distributions are designed using physical information from the flow field.
\Cref{sec:training} provides a detailed description of the controlled process for an elliptical cylinder with \(Ar=0.75\) during DRL training. We also illustrate the impact of synthetic jet actions, whether blowing or suction, on the streamlines around the elliptical cylinder using detailed streamline.
\Cref{sec:ellipses} systematically tests the flow control performance of elliptical cylinders with \(Ar\) ranging from 0 to 2 using DRL-based AFC, verifying the robustness of DRL-based AFC technology for elliptical cylinders with different geometric shapes.
\Cref{sec:controlled} analyzes the controlled flow to explore the physical mechanisms of control.
Finally, \cref{sec:Conclusions} concludes the paper by summarizing the key findings and discussing their implications.

\section{PROBLEM AND METHODOLOGY}\label{sec:Methodology} 

\subsection{Problem statement}\label{sec:numericalsimulation}

\paragraph{Model configuration}

\begin{figure*}[ht]
    \centering
    \begin{subfigure}{0.62\textwidth}
        \includegraphics{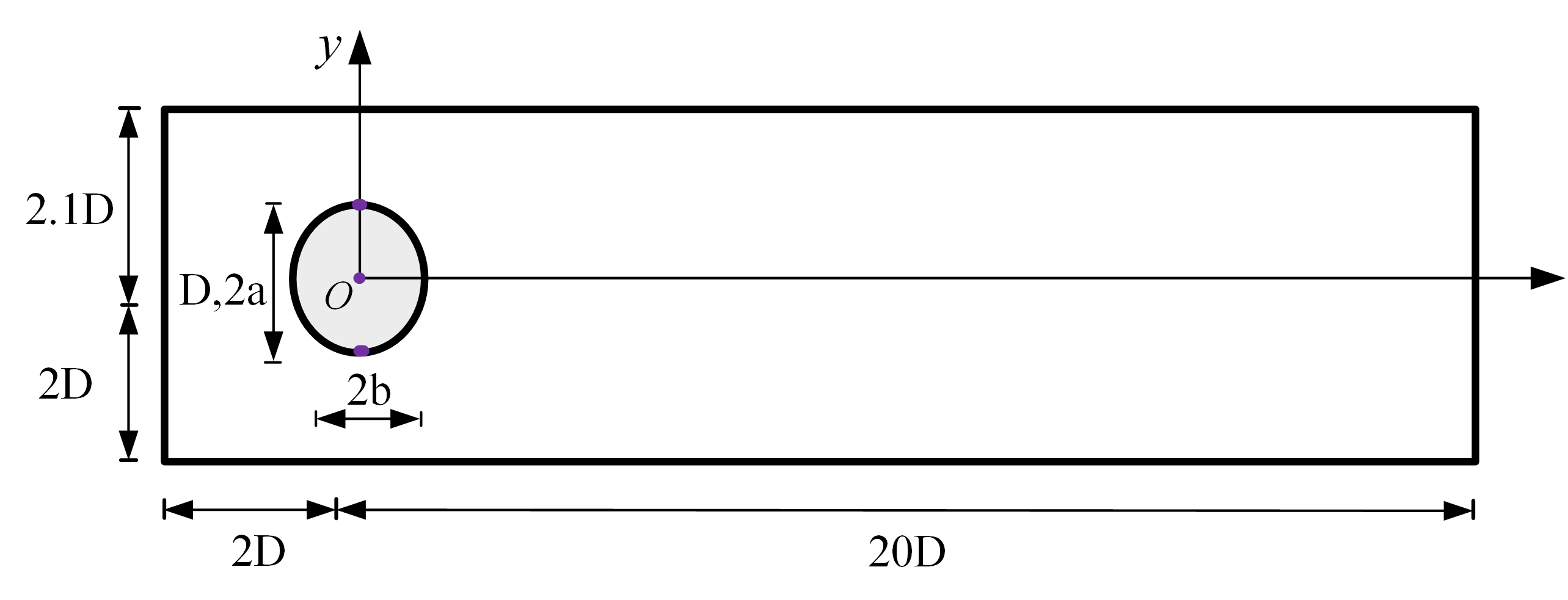}
        \caption{}
        \label{fig:set01}
    \end{subfigure}
    \begin{subfigure}{0.35\textwidth}
        \includegraphics{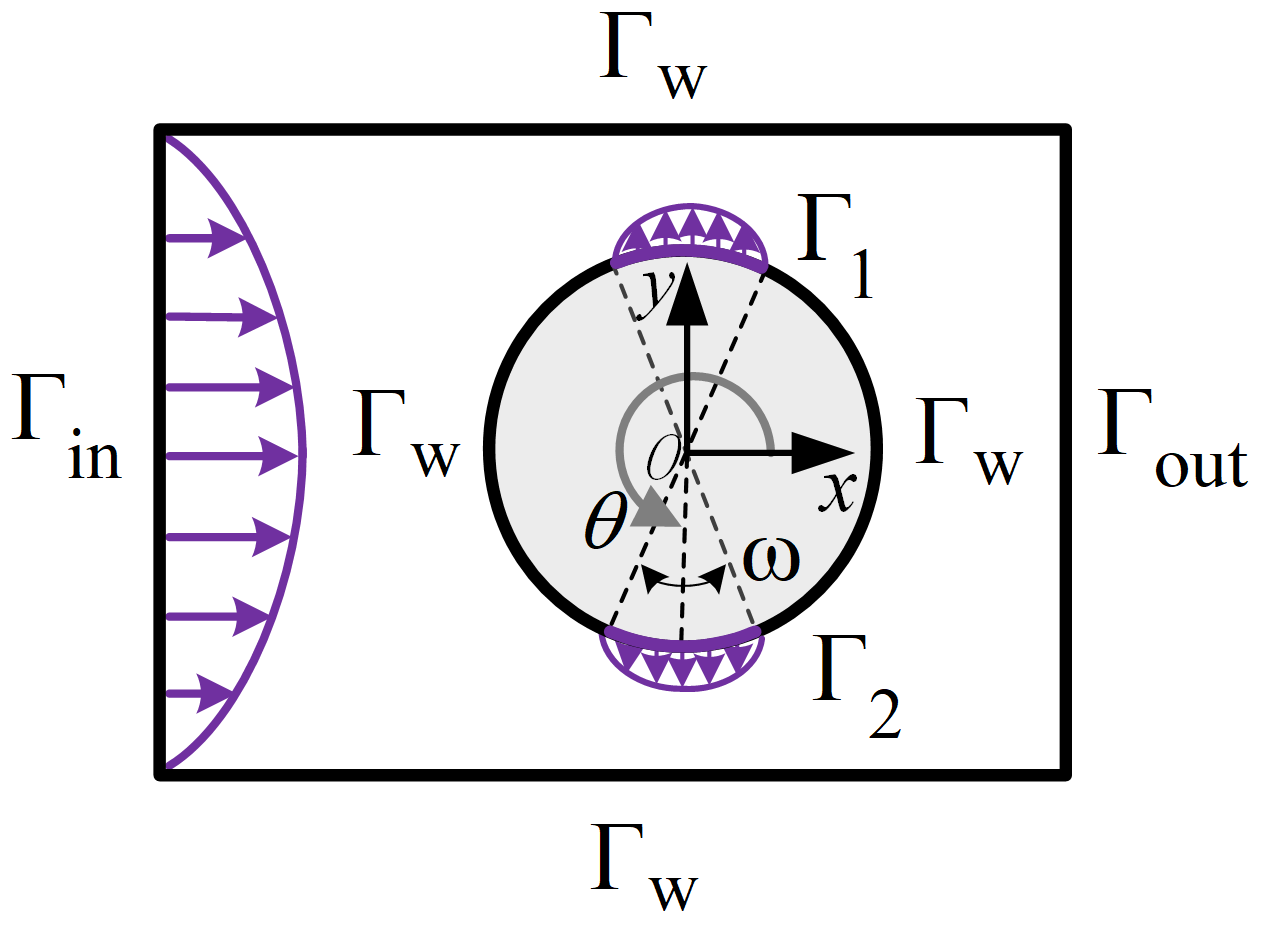}
        \caption{}
        \label{fig:set02}
    \end{subfigure}
    \begin{subfigure}{\textwidth}
        \centering
        \includegraphics{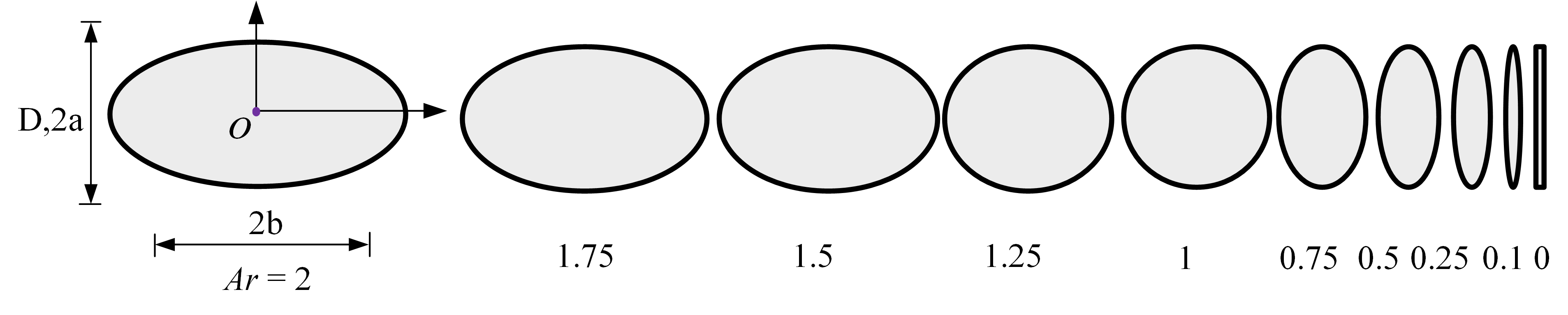}
        \caption{}
        \label{fig:set03}
    \end{subfigure}
    \caption{(a) Definition of the problem, including the dimensions of the computational domain and the establishment of a Cartesian coordinate system. (b) Establishment of boundary conditions; note that the dimensions of the elliptical cylinder and the computational domain in the figure are not to scale, focusing solely on the delineation of boundary conditions. (c) This study investigates elliptical cylinders with different aspect ratio: a normal flat plate corresponding to $Ar=0$, a circular cylinder with $Ar=1$, and an elliptical cylinder with $Ar=2$. The dimension $2a$ is fixed as $D$, and the $Ar$ is varied by changing the length of 2b.}
    \label{fig:set}
\end{figure*}

We investigate the flow characteristics around a two-dimensional elliptical cylinder. As illustrated in \Cref{fig:set01}, the flow is described using a coordinate system where the $x$-axis aligns with the flow direction, and the $y$-axis is the perpendicular direction. The origin of the coordinate system is set at the center of the elliptical cylinder, denoted as $O$. The computational domain extends from -2$D$ ahead of the elliptical cylinder to 20$D$ behind it in the flow direction, where $D$ represents a reference length. In the cross-flow direction, which is perpendicular to the flow, the domain spans from -2$D$ to 2.1$D$. The elliptical cylinder has one axis that remains fixed with a length of 2$a$ (equivalent to $D$), while the other axis has a length of 2$b$.
The synthetic jets, symmetrically positioned at $\theta_1 = 90^\circ$ and $\theta_2 = 270^\circ$ on the elliptical cylinder, have a fixed jet opening width of $\omega = 10^\circ$, as depicted in \Cref{fig:set02}. 
The aspect ratio ($Ar$) of the elliptical cylinder, defined as the ratio of the semi-axis $b$ to the semi-axis $a$, encompasses ten distinct geometrical shapes, ranging from a flat plate ($Ar=0$) with a thickness of $0.05D$ perpendicular to the flow, to a cylinder ($Ar=1$), and further to an elliptical cylinder ($Ar=2$), as depicted in \Cref{fig:set03}.

\paragraph{Governing equations}

Let $\Omega \subset \mathbb{R}^{nd}$ and $(0, T)$ be the spatial and temporal domains, respectively, where $nd$ is the number of space dimensions, and let $\Gamma$ denote the boundary of $\Omega$. The Navier-Stokes equations that describe the dynamics of this incompressible viscous fluid flow are given by:
\begin{subequations}
\begin{equation}
    \frac{\partial \mathbf{u}}{\partial t} + \mathbf{u} \cdot (\nabla \mathbf{u}) = -\nabla p + Re^{-1} \Delta \mathbf{u} \quad \text{on} \quad \Omega \times (0, T),
\end{equation}
\begin{equation}
    \nabla \cdot \mathbf{u} = 0 \quad \text{on} \quad \Omega \times (0, T),
\end{equation}
\end{subequations}
where $\mathbf{u}$, $t$ and $p$ are the velocity vector, time, and pressure, respectively. The Reynolds number, denoted by $Re$, is defined as $Re = \frac{\overline{U}D}{\nu}$, where $\overline{U} = 2U_m / 3 = 1$ represents the mean velocity magnitude, $U_m$ is the maximum velocity, and $\nu$ is the kinematic viscosity. In this study, $Re=100$ is utilized.

\paragraph{Boundary conditions}

The boundary conditions for the computational domain are as follows: The upstream boundary ($\Gamma_\text{in}$) is characterized by a parabolic velocity profile, the downstream boundary ($\Gamma_\text{out}$) is subjected to a Neumann-type boundary condition to nullify the stress vector, the upper and lower boundaries ($\Gamma_\text{w}$) employ a slip-wall boundary condition with zero velocity and tangential stress, and the boundaries of the synthetic jets ($\Gamma_{i}$, where $i = 1, 2$) are assigned a prescribed parabolic velocity profile representing fluid suction or injection. The synthetic jets feature a parabolic velocity distribution, where the sign of the jet velocity indicates fluid suction or injection. The design ensures mass flow balance through the condition $V_{\Gamma_1} = -V_{\Gamma_2}$, preserving mass conservation in the system. The non-slip solid wall boundary condition is applied to the rest of the elliptical cylinder except for the synthetic jets.

Mathematically, the boundary conditions are written as:
\begin{eqnarray}
\begin{array}{cl}
-\rho \mathbf{n} \cdot \mathbf{p} + Re^{-1} (\mathbf{n} \cdot \nabla \mathbf{u}) = 0  &\text{on\quad } \;\Gamma_{out}, \\
\mathbf{u} = 0  &\text{on \quad} \Gamma_{w}, \\
\;\mathbf{u} = U  &\text{on \quad} \Gamma_{in}, \\
\;\;\;\mathbf{u} = f_{Q_i}  & \text{on \quad} \Gamma_{i}, \quad i = 1, 2.
\end{array}
\end{eqnarray}
Here, $\mathbf{n}$ denotes the normal vector, $U$ is the inflow velocity, and $f_{Q_i}$ represents the radial velocity profiles simulating the suction or injection of fluid by the jets.
At the inlet $\Gamma_{\text{in}}$, the inflow velocity $U$ along the $x$-axis is prescribed by a parabolic velocity profile given by:
\begin{equation}\label{equ:1velocity}
U = U_m\frac{(H-2y)(H + 2y)}{H^2}
\end{equation}
where $H=4.1D$ represents the total height of the rectangular domain.

In particular, we choose $f_{Q_i}=A(\theta; Q_i)(x,y)$, where the modulation relies on the angular coordinate $\theta$ as illustrated in \cref{fig:set02}. For the jet, which has a width of $\omega$ and is centered at $\theta_0$ on the cylinder with radius $R$, the modulation is defined as follows:
\begin{equation}
A(\theta; Q) = Q \frac{\pi}{2 \omega R^2} \cos\left(\frac{\pi}{\omega} (\theta - \theta_0)\right).
\end{equation}
This modulation function ensures a smooth transition to the no-slip boundary conditions on the cylinder surfaces, enabling a comprehensive analysis of the flow characteristics and control challenges associated with the elliptical cylinder geometry.

\paragraph{Quantities of interest}
The lift coefficient ($C_L$) and drag coefficient ($C_D$) are physically important quantities in active flow control, and they are defined as follows:
\begin{align}
     C_L = \frac{F_L}{0.5\rho \overline{U}^2D}, \quad  C_D = \frac{F_D}{0.5\rho \overline{U}^2D}.
\end{align}
Here, $F_L$ and $F_D$ represent the lift and drag forces integrated over the surface of the elliptical cylinder, respectively, and $\rho$ is the fluid density. The Strouhal number ($St$) is used to characterize the characteristic frequency of oscillatory flow phenomena, and it is defined as follows:
\begin{equation}
St = \frac{f_s \cdot D}{\overline{U}}
\end{equation}
where $f_s$ is the shedding frequency calculated based on the periodic evolution of the $C_L$.

\subsection{Numerical simulation}

\paragraph{Solver details}
We utilize the open-source CFD package \texttt{OpenFOAM} as the incompressible flow solver, following the methodology described by Jasak \textit{et al.}\cite{jasakOpenFOAMOpenSource2009,jasakOpenFOAMLibraryComplex2013}. 
This approach is widely employed due to its ability to provide reliable numerical algorithms for solving the Navier-Stokes equations. 
The solver implemented in \texttt{OpenFOAM} uses the finite volume method to discretize the computational domain into a mesh consisting of control volumes. 
We choose the \texttt{pisoFoam} solver based on \texttt{OpenFOAM}, which is a transient solver for incompressible flow using the \texttt{PISO} algorithm. 
\texttt{PISO} algorithm is to iteratively correct the pressure and velocity fields to satisfy the incompressibility condition at each time step.
The time and spatial derivatives are discretized using the Crank-Nicolson scheme and the second-order central difference scheme, respectively. A non-dimensional time step of $\Delta t = 0.0005$ is selected. The CFL number, defined as $\left(\frac{\Delta t |U|}{\Delta x}\right)$—where $|U|$ represents the magnitude of the velocity vector per grid unit and $\Delta x$ is the grid length in the direction of the velocity—is less than 1, meeting the accuracy requirements.

\paragraph{Grid scheme}

\begin{figure*}[ht]
\centering
\includegraphics{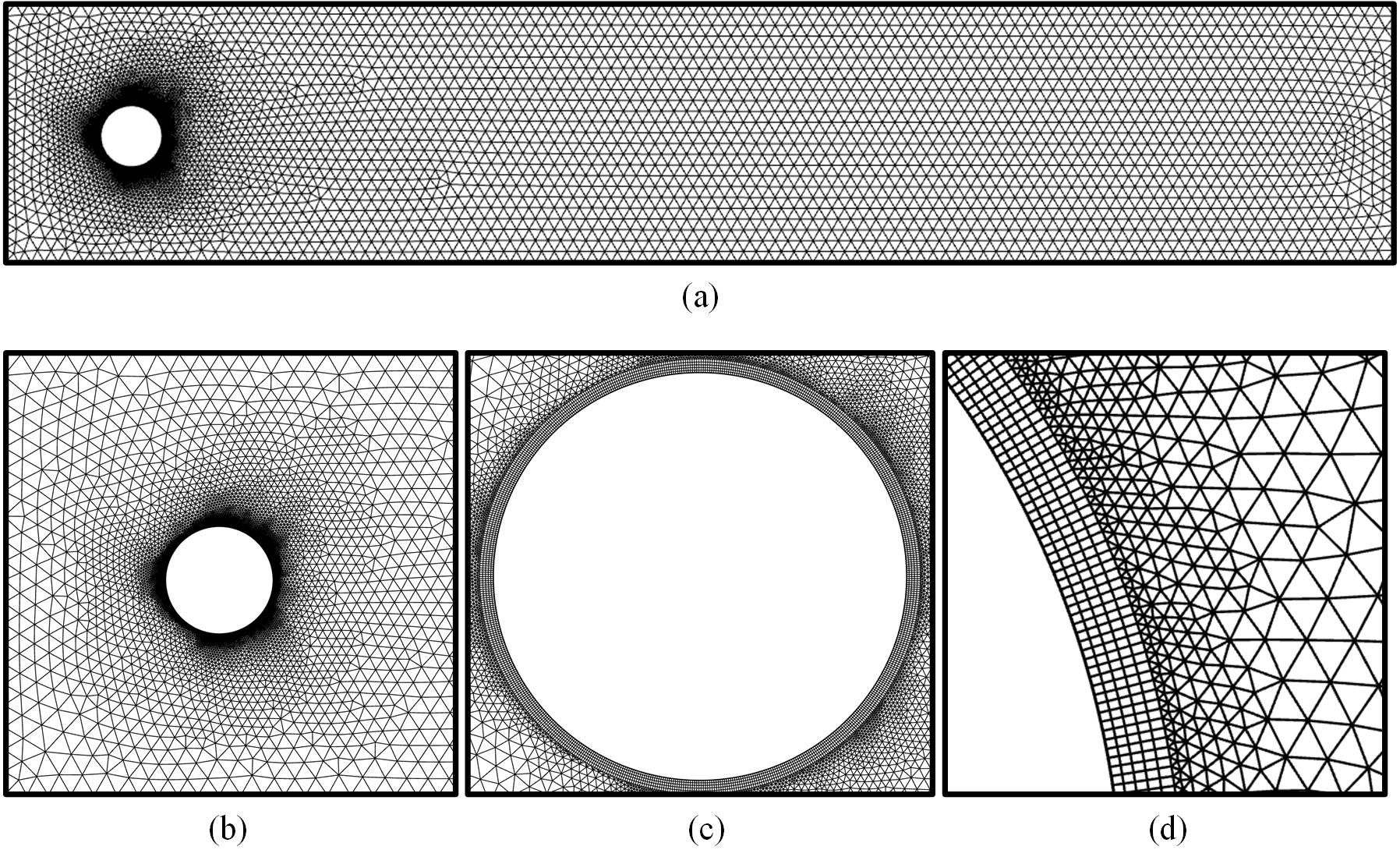}
\caption{For \(Ar=1\), the computational domain is discretized into 18,484 mesh cells, including both triangular and quadrilateral meshes. (a) Discretization of the computational domain is viewed from a global perspective. (b) A comprehensive display of the mesh around the elliptical cylinder. (c) Presentation of the multi-layered quadrilateral meshes surrounding the elliptical cylinder. (d) A magnified view of the quadrilateral meshes in the vicinity of the elliptical cylinder. }
\label{fig:mesh}
\end{figure*}

As an example, we examine a computational domain comprising an elliptical cylinder with an aspect ratio ($Ar=1$) of 1. To discretize this domain accurately, a hybrid meshing approach is adopted, integrating both structured (quadrilateral) and unstructured (triangular) meshes, as shown in \cref{fig:mesh}. The computational domain is discretized into 18,484 mesh cells, with a multi-layer quadrilateral mesh used for the area surrounding the elliptical cylinder and triangular meshes used for the remaining regions. In detail, \cref{fig:mesh}(a) provides an overview of the global discretization scheme, demonstrating the comprehensive approach to mesh division employed in the computational domain. \cref{fig:mesh}(b) offers a closer look at the mesh partitioning near the elliptical cylinder, highlighting the finer multi-layered quadrilateral meshes surrounding the cylinder compared to the triangular meshes. \cref{fig:mesh}(c) illustrates the division of the multi-layered quadrilateral meshes around the elliptical cylinder, emphasizing the precision in discretization near the structure. \cref{fig:mesh}(d) illustrates the mesh division strategy around the elliptical cylinder, demonstrating the integration between the quadrilateral and triangular meshes.
The method for mesh division for ellipses with \( Ar \) values other than 1 follows the same approach as that for \( Ar = 1 \), and is not separately demonstrated.

\paragraph{Grid independence}

Similarly, the case where the  $Ar$ value is 1 is also used as an example to explore grid convergence and validate the numerical method utilized. 
\Cref{tab:Grid} presents simulation results using three different mesh resolutions, allowing for a comparative analysis of the maximum drag coefficient \(C_{D,\max}\), the mean of the maximum drag coefficient \(C_{D,\text{mean}}\), the maximum lift coefficient \(C_{L,\max}\), and the $St$.
Furthermore, the results of the numerical simulations are compared with the findings of \citeauthor{rabault2019artificial} \cite{rabault2019artificial} and \citeauthor{schafer1996benchmark} \cite{schafer1996benchmark} for validation. The simulations conducted using a coarse mesh reveal significant deviations in the drag and lift coefficients when compared to the results obtained by \citeauthor{rabault2019artificial} \cite{rabault2019artificial}, which raises doubts about the accuracy of such numerical simulations. On the other hand, using a fine mesh with a high density of grid points significantly slows down the training speed for reinforcement learning, posing challenges for efficient simulation execution. It is always desirable to use the minimum number of grid elements for numerical simulation without affecting computational accuracy. As a result, the discretization strategy of the main mesh is adopted for subsequent numerical simulations and reinforcement learning training. The comparison of the computational results obtained from the main mesh with those obtained by \citeauthor{rabault2019artificial} \cite{rabault2019artificial} reveals minimal differences, indicating a high level of accuracy. Additionally, the accuracy of the main mesh's results falls within the upper and lower bounds provided by standard benchmarks \cite{schafer1996benchmark}, which confirms that the computational precision of the main mesh is sufficient for further training purposes. The methodology for examining mesh dependency for ellipses with an $Ar$ value other than 1 follows the same procedure established for $Ar=1$.

\begin{table}[htbp]
\centering
\caption{Grid independence test for an elliptical cylinder with \(Ar=1\) at \(Re=100\).}
\vspace{-\baselineskip}
\begin{tabular}{
  >{\centering\arraybackslash}p{0.2\textwidth}
  >{\centering\arraybackslash}p{0.13\textwidth}
  >{\centering\arraybackslash}p{0.10\textwidth}
  >{\centering\arraybackslash}p{0.11\textwidth}
  >{\centering\arraybackslash}p{0.10\textwidth}
  >{\centering\arraybackslash}p{0.11\textwidth}
  >{\centering\arraybackslash}p{0.11\textwidth} 
}
\toprule
Configuration & Mesh resolution & Mesh & \(C_{D, max}\) & \(C_{D, mean}\) & \(C_{L, max}\) & \(St\) \\
\midrule
\citeauthor{schafer1996benchmark}\cite{schafer1996benchmark} & - & - & 3.220–3.240 & - & 0.990–1.010 & 0.295–0.305 \\
\citeauthor{rabault2019artificial}\cite{rabault2019artificial} & - & 9262 &  & 3.205 & - &  \\
        & {Coarse}   & 10,540  & 3.242 & 3.224 & 1.052 & 0.304 \\
$Ar=1$    & {Medium}   & 18,484 & 3.225 & 3.205 & 0.990 & 0.300 \\
        & {Fine}     & 25,624 & 3.228 & 3.207 & 0.992 & 0.301 \\

\bottomrule
\end{tabular}
\label{tab:Grid}
\end{table}

\subsection{Deep reinforcement learning}\label{sec:DRL algorithm}

\begin{figure*}[htbp]
\centering
\includegraphics{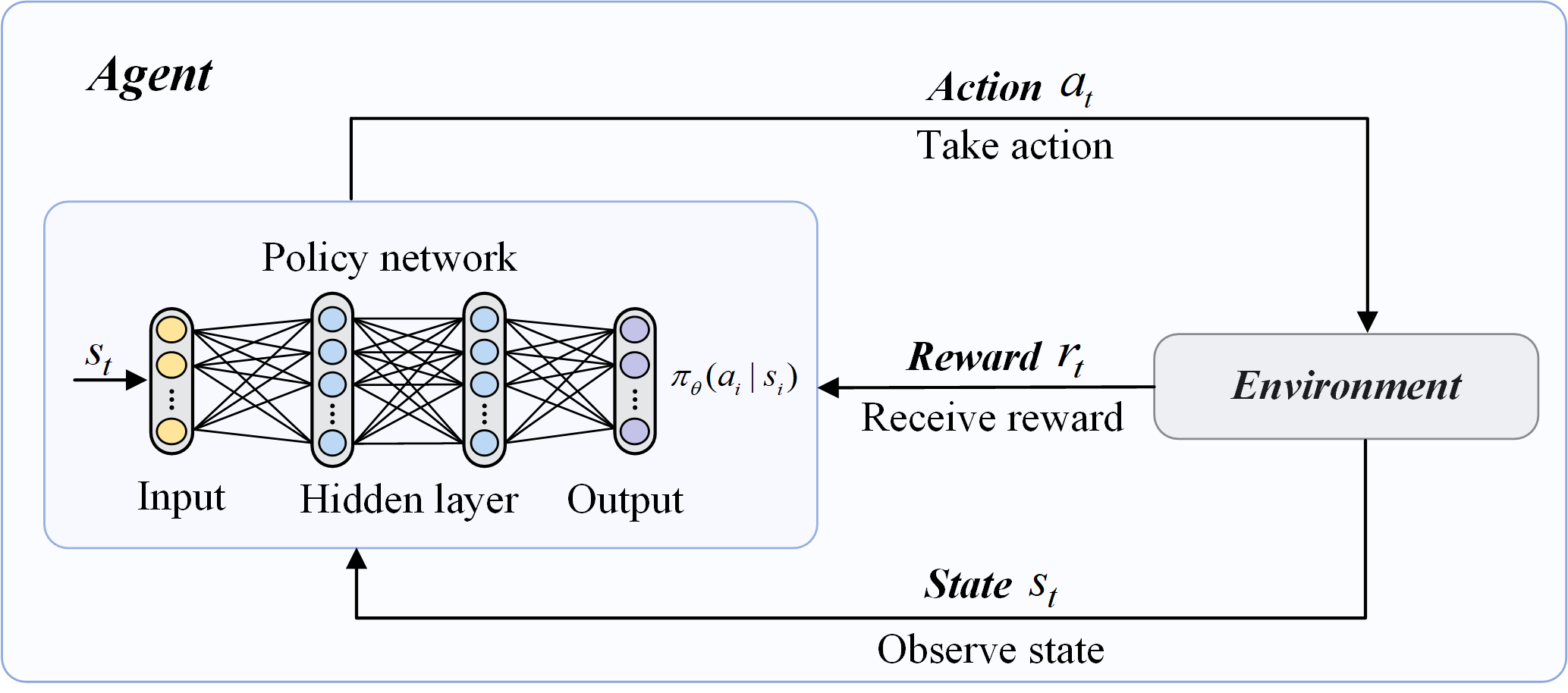}
\caption{DRL combines deep learning's capability to interpret complex data with reinforcement learning's strategy to make informed decisions.}
\label{fig:ML}
\end{figure*}

\paragraph{Model structure}
The structure of our employed DRL model is illustrated in \cref{fig:ML}. the DRL marries the perceptual power of deep learning with the decision-making and optimization capabilities of reinforcement learning, effectively overcoming their individual limitations and enhancing overall performance.
Reinforcement Learning serves as the core mechanism for facilitating learning through the interaction between an agent and its environment. At each time step $t$, the agent observes the current state $s_t$ of the environment, which contains all the necessary information for decision-making. Based on this observation, the agent selects and executes an action $a_t$ according to its policy $\pi$. The policy, serving as a strategy mapping states to actions, can be deterministic or probabilistic in nature. After the action is executed, the environment transitions to a new state $s_{t+1}$ and provides the agent with an immediate reward $r_{t+1}$. This reward, expressed as a numerical value, assesses the effectiveness of the action in progressing towards the agent's objectives. Consequently, the agent incorporates this feedback-comprising both the reward and the new state—to refine its policy, utilizing Reinforcement Learning algorithms designed to optimize the accumulation of long-term rewards.

\paragraph{Proximal policy optimization}

PPO algorithm algorithm is an improvement over the policy gradient methods. The core idea of the PPO algorithm is to update the policy through two crucial steps: data sampling and policy optimization.
In the data sampling phase, we interact with the environment using the current policy and collect a batch of trajectory data denoted as $\tau$, which can be mathematically expressed as:
\begin{equation}\label{eq:trajectory}
\tau = (s_0, a_0, r_1, s_1, a_1, r_2, s_2, \ldots, s_{T-1}, a_{T-1}, r_T, s_T).
\end{equation}
Here, $s_t$ represents the state at time step $t$, $a_t$ represents the action taken at time step $t$, and $r_t$ represents the reward received at time step $t$. The trajectory $\tau$ consists of a sequence of state-action-reward tuples observed during the interaction between the agent and the environment over a certain time horizon $T$. For a given trajectory $\tau$, at each interaction between the agent and the environment, the agent receives a reward value $r_t$ at time step $t$. Hence, the cumulative reward $R(\tau)$ of the trajectory $\tau$ is obtained by summing up the rewards obtained at each time step:
\begin{equation}
R(\tau) = \sum_{t=0}^{T} r_t.
\end{equation}
Here, $T$ represents the time horizon or the length of the trajectory, and $r_t$ represents the reward received at time step $t$. The cumulative reward $R(\tau)$ provides a measure of the total reward obtained by the agent throughout the trajectory $\tau$.

In the policy optimization phase, we use the sampled data obtained in the previous phase to update the policy. The method used to update the policy is gradient ascent.
To perform gradient ascent, we compute the gradient of the PPO objective function with respect to the policy parameters. The objective function of the PPO algorithm is a surrogate objective designed to strike a balance between exploration and exploitation. It can be expressed as follows:
\begin{equation}\label{eq:ppo}
\left.L^{C L I P}(\theta)=\hat{\mathbb{E}}_t\left[\min \left(r_t(\theta)\right) \hat{A}_t, \operatorname{clip}\left(r_t(\theta), 1-\varepsilon, 1+\varepsilon\right) \hat{A}_t\right)\right],
\end{equation}
where $\theta$ represents the policy parameters that are being optimized during the policy update process.
$\hat{\mathbb{E}}_t$ denotes the empirical expectation, indicating that the expectation is estimated using the samples collected from the environment. 
$r_t(\theta)$ represents the ratio of the probabilities between the new policy and the old policy, evaluated at time step $t$.
$\hat{A}_t$ corresponds to the advantage estimate at time step $t$. 
The advantage represents the advantage of taking a particular action compared to the expected value of the action under the current policy.
The $\operatorname{clip}(.)$ function performs clipping on the policy ratio to ensure that the update does not deviate too much from the old policy. It restricts the ratio within the range $[1-\varepsilon, 1+\varepsilon]$, where $\varepsilon$ is a small value.
$[\min \left(r_t(\theta)\right) \hat{A}_t, \operatorname{clip}\left(r_t(\theta), 1-\varepsilon, 1+\varepsilon\right) \hat{A}_t]$ expression represents the clipped surrogate objective. 
It consists of two terms: the first term is the minimum between $r_t(\theta) \hat{A}_t$ and $\operatorname{clip}\left(r_t(\theta), 1-\varepsilon, 1+\varepsilon\right) \hat{A}_t$, and the second term is the clipped version of $r_t(\theta) \hat{A}_t$. The purpose of this formulation is to balance policy improvement while ensuring a controlled update magnitude.

\subsection{DRL-enhanced active flow control}\label{DRL-Enhanced Active Flow Control}

\begin{figure*}[htbp]
    \centering
    \includegraphics{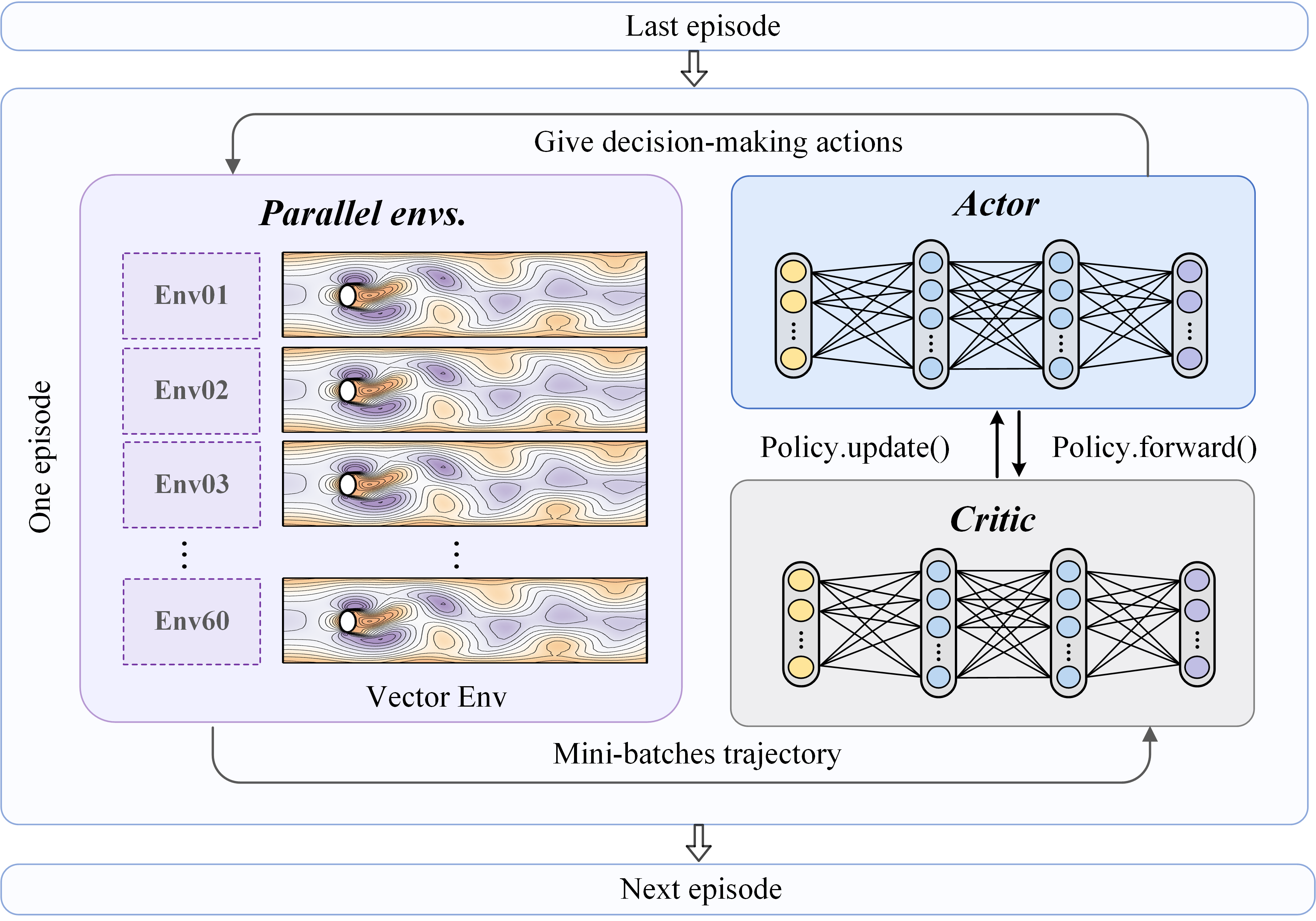}
    \caption{The Proximal Policy Optimization algorithm is trained using a parallel computing framework to simultaneously run multiple environments. In the interaction stage between agent and environment, an agent interacts with its environments by following the current policy $\pi$, where it executes actions $a_t$, receives states $s_t$, and gathers rewards $r_t$, identifying terminal states in the process. This interaction generates a set of trajectories $\{\tau_i\}_{i=1}^{n}$, each $\tau_i$ being a sequence of state-action-reward tuples ending at a terminal state $T$. The agent uses these trajectories to compute the policy gradient $\nabla_\theta J(\pi_\theta)$ and applies gradient ascent to update the policy parameters $\theta$, aiming to maximize cumulative rewards. Through repeated iterations, leveraging updated policies and recalculated gradients, the agent continuously enhances its policy effectiveness, optimizing its performance across various environments.}
    \label{fig:parall_drl}
\end{figure*}

\paragraph{Framework of DRL} 
Our AFC framework utilizes DRL algorithms, consisting of the agent and the environment. The agent, designed for continuous action control, employs algorithms like PPO or SAC. The environment chosen for interaction with the agent is the CFD simulation of the flow around an elliptical cylinder. The custom environment class, as presented by \citeauthor{rabault2020deep}\cite{rabault2020deep}, provides the main structure, and further details can be explored in their work. \cref{fig:parall_drl} illustrates the fundamental framework of DRL, including the computational process of a single training episode and the utilized formulas.
We refer to the work of \citeauthor{rabault2020deep}\cite{rabault2020deep} and use a multilayer perceptron with two fully connected layers of 512 neurons each as the policy network. This network structure is sufficient to handle the complexity of the flow control problem in this study. Simpler network structures lack the modeling capacity to capture the mapping relationships, while more complex network structures are harder to train. The architecture of the ANN needs to be designed based on the dimensions of the input and output data, which often relies on empirical knowledge.

\paragraph{Fundamental components configuration} 
In DRL-Enhanced active flow control,  it involves state ($s_t$), action ($a_t$), reward ($r_t$), and parameter, which are described as below.
The traditional state space $s_t$ poses challenges in learning policies due to its large size, emphasizing the reliance of the agent's performance in partially observable environments on the quality and relevance of observed data, as highlighted by \citeauthor{Viquerat}\cite{Viquerat}. To achieve the control objectives of reducing drag and lift and controlling vortex shedding in the wake field of the elliptical cylinder, strategically placed probes are utilized. The design of probe placement follows the principles proposed by \citeauthor{rabault2019artificial}\cite{rabault2019artificial}. This scheme was also adopted by \citeauthor{jia2024optimal}\cite{jia2024optimal} and \citeauthor{wangDRLinFluids}\cite{wangDRLinFluids} in their respective studies.

The control action $a_t$ involves blowing and suction of two synthetic jets on the elliptical cylinder. 
The flow rates of the upper and lower synthetic jets are constrained such that their sum is zero, i.e., $Q_{\text{lower}} + Q_{\text{upper}} = 0$. 
They also conducted an evaluation of the momentum injected into the flow field by the four jets.
The RL agent controls the velocity of the upper synthetic jet, while the lower synthetic jet has the same velocity but in the opposite direction. To ensure energy efficiency in AFC, the velocity of the synthetic jets is limited to less than 2\% of the inlet mean velocity, preventing the agent from utilizing excessively high jet velocities for drag reduction. 
Each control action, known as a `viscous action', has a duration of 50 time steps. To ensure smooth transitions between consecutive actions, a smoothing function $S$ is introduced:
 \begin{equation}
    S(V_{\Gamma_1,T_i}, a, V_{\Gamma_1,T_{i-1}}) = V_{\Gamma_1,T_i} + \beta \cdot (a - V_{\Gamma_1,T_{i-1}}),
    \end{equation}
where $V_{\Gamma_1,T_i}$ represents the current value at time step $i$, $\beta$ is a coefficient determining the extent of adjustment towards the target action $a$, and $V_{\Gamma_1,T_{i-1}}$ is the value at the previous time step $i$-1. This function ensures a smooth transition between consecutive actions, with the parameter $\beta$ controlling the smoothness of the transition.

The reward $r_t$ evaluates the agent based on the average drag coefficient, time-averaged drag coefficient, and time-averaged lift coefficient during the action, with the objective of minimizing drag while keeping the lift coefficient low.
The weight of the drag coefficient in the reward function is set to 1, indicating its importance, while the lift coefficient is assigned a weight $w$. The specific reward function takes the following form:
    \begin{equation}\label{eq:my reward}
        r_{T_i}=C_{D,0}-\left(C_D\right)_{T_i}-\omega\left|\left(C_L\right)_{T_i}\right|,
    \end{equation}
where $C_{D,0}$ represents the baseline drag coefficient, serving as a reference point for drag reduction. $(C_D)_{T_i}$ denotes the drag coefficient at time step $T_i$, with the objective of minimizing this value relative to the baseline. $(C_L)_{T_i}$ signifies the lift coefficient at time step $T_i$, whose absolute value is penalized to mitigate lift forces that may destabilize the flow around the bluff body. 
$\omega$ is a weighting factor that quantifies the trade-off between minimizing drag and controlling lift fluctuations.
\citeauthor{rabaultAccelerating}\cite{rabaultAccelerating} found that in the absence of a lift penalty, DRL produced a "cheating" strategy that resulted in reduced drag but was accompanied by a significant increase in lift. By incorporating a lift penalty in the reward function, this issue can be effectively mitigated. In previous work, the lift coefficient penalty factor is often set to 0.1\cite{jia2024optimal}, 0.2\cite{rabault2019artificial,tangRobustActiveFlow2020} or 1\cite{renApplying}. In this study, we use a penalty factor of 0.2, consistent with \citeauthor{rabault2019artificial}\cite{rabault2019artificial}

The time granularity parameter significantly affects the agent's performance and the complexity of the learning task. Precise adjustment to this hyperparameter, based on prior research and ongoing strategy refinement, is essential for achieving optimized control outcomes.
In particular, each action in our analysis is applied for a duration of 0.025 seconds, which corresponds to 25 numerical simulation time steps. 
Considering that the Strouhal number for the flow around the elliptical cylinder in this study is between 0.3 and 0.4, we refer to the work of \citeauthor{rabaultAccelerating}\cite{rabaultAccelerating}, where the duration of an episode corresponds to approximately 8 vortex shedding periods. Additionally, the work of \citeauthor{rabault2019artificial}\cite{rabault2019artificial} considers an episode duration equivalent to about 6.5 vortex shedding periods. 
Therefore, in this study, the duration of an episode for the DRL training for AFC on the elliptical cylinder corresponds to approximately 6.25 to 8.33 vortex shedding periods.
Each episode consists of 100 control actions, resulting in a maximum total time for an episode of $T_{\max} = T_{100} = 2.5$ non-dimensional time units.
The length of episodes is intentionally chosen to encompass multiple shedding cycles, which facilitates effective learning of the control algorithm by allowing the agent to observe and adapt to the flow dynamics over several cycles.
\citeauthor{rabaultAccelerating}\cite{rabaultAccelerating} studied the impact of action update frequency on control strategies and found that approximately \( T_a / T_K = 10\% \), where \( T_K \) represents the vortex shedding frequency and \( T_a \) represents the duration of the action. In this study, the ratio of action duration to vortex shedding frequency is around 6\% to 8\%. This ensures that the action update frequency is sufficiently high for the ANN to control the system effectively while allowing the system enough time to respond to the control actions.

\paragraph{Computing specifications} 

The DRL training and numerical simulation are conducted on a high-performance computing (HPC) system equipped with an \texttt{Intel}\textsuperscript{\textregistered} Xeon\textsuperscript{\textregistered}  Platinum 8358 CPU, operating at 2.60 GHz, and comprising a total of 64 cores evenly distributed across two sockets, each containing 32 cores. Specifically, the numerical simulation is performed using version 8 of the open-source software platform \texttt{OpenFOAM\textsuperscript{\textregistered}} version 8 \cite{jasakOpenFOAMLibraryComplex2013}. The deployment of DRL algorithms is based on the\texttt{Tensorforce} library \cite{schaarschmidt2017tensorforce}, while the definitions of artificial neural networks and gradient descent algorithms are facilitated using the  \texttt{TensorFlow} open-source library \cite{abadi2016tensorflow}.
For the DRL implementation, the agent is loaded using the Tensorforce platform with the PPO algorithm, and the environment is constructed using the  \texttt{Gym} interface \cite{brockman2016openai} . 
\cref{fig:parall_drl} illustrates the use of CFD calculations as the environment for DRL training and shows the effects of using 60 parallel CFD environments to accelerate the training process. 
This parallel strategy is inspired by the studies of \citeauthor{rabaultAccelerating}\cite{rabaultAccelerating} and \citeauthor{wangDRLinFluids}\cite{wangDRLinFluids}, and also incorporateds insights from the work of \citeauthor{wangDRLinFluids}\cite{wangDRLinFluids} 
For a more comprehensive understanding, readers are encouraged to consult their works and the references therein.

\section{RESULTS}\label{sec:results}

\subsection{Flow characteristics of baseline flow}\label{sec:baseline}

\paragraph{Vortex shedding phenomenon}

\begin{figure*}[ht]
    \centering
    \includegraphics{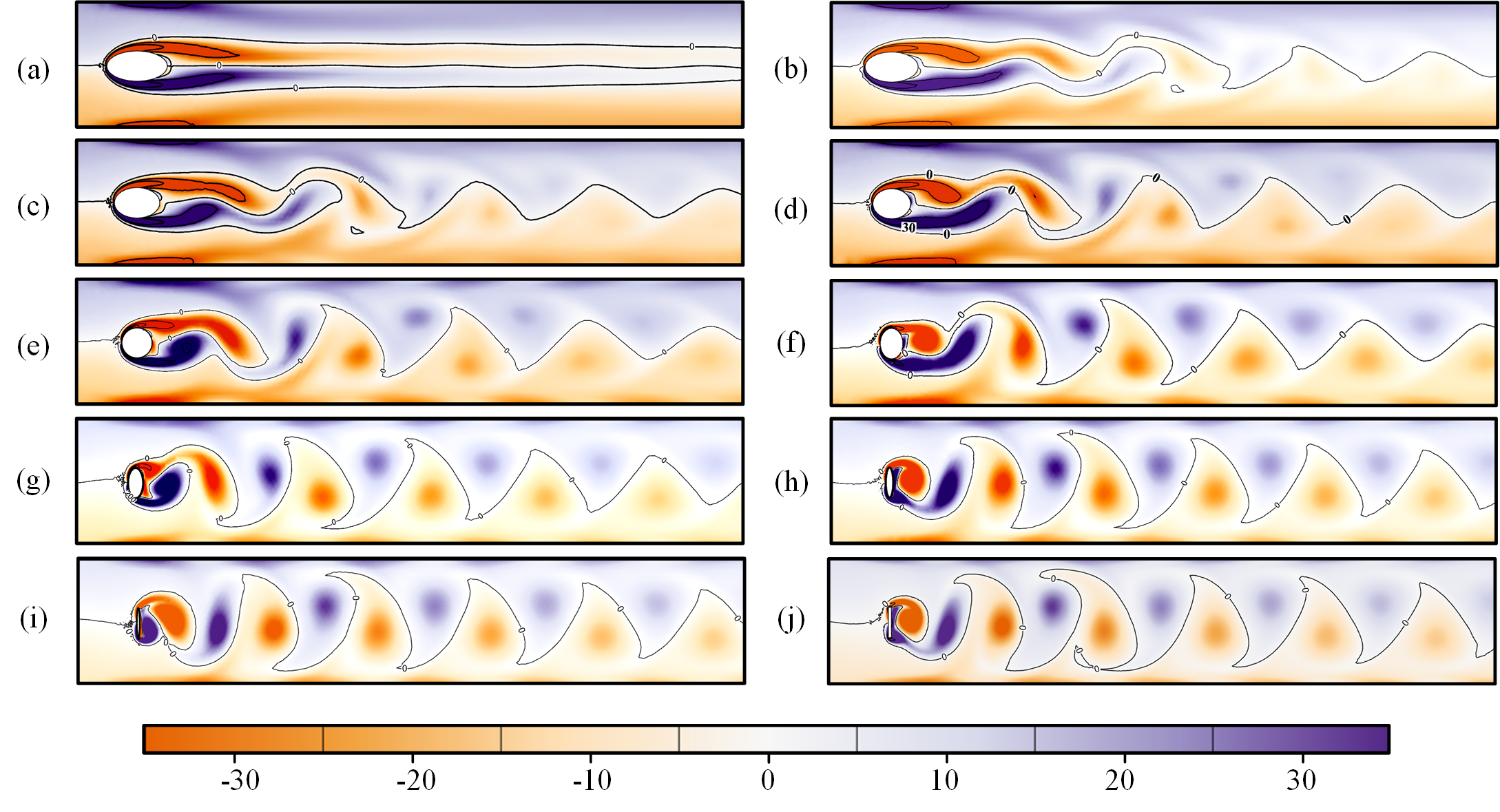}
    \caption{Contours of instantaneous vorticity for flow around an elliptic cylinder at \( Re = 100 \), varying from \( Ar = 2 \) to \( Ar = 0 \). Purple and orange indicate positive and negative vorticity, respectively. Purple represents the clockwise vortices shed from the upper side of the cylinder, while orange represents the counterclockwise vortices shed from the lower side. 
    (a) $Ar=2$; (b) $Ar=1.75$; (c) $Ar=1.5$; (d) $Ar=1.25$; (e) $Ar=1.0$; (f) $Ar=0.75$; (g) $Ar=0.5$; (h) $Ar=0.25$; (i) $Ar=0.1$; (j) $Ar=0$.}
    \label{fig05:vorticity}
\end{figure*}

\cref{fig05:vorticity} depicts the vortex shedding patterns within the wake field of elliptical cylinders with \(Ar\) from 0 to 2.
For an elliptical cylinder with $Ar=2$, the wake field does not exhibit vortex shedding. 
This is expected as the elongated elliptical cylinder results in weak adverse pressure gradients that allow the boundary layers to stay attached for a prolonged distance. The zero vorticity contour lines in the wake extend all the way to the outlet with slight fluctuations. This suggests that the wake is largely stable, with minor flow instability due to shear in the flow. 
It is worth highlighting that this test case of $Ar = 2$ serves as a special verification study for subsequent DRL control experiments. For a stable flow field with no vortex shedding, we expect the DRL-based flow control algorithm to conform to the underlying flow physics and to not introduce any actuation since the current baseline is essentially the optimal flow configuration. Based on our current knowledge, this sanity test that verifies minimal DRL-based control is often overlooked in the literature. We include it to emphasize the robustness of DRL-based control across a spectrum of flow physics. The detailed discussion of DRL control will be presented in \cref{sec:training}, where we will provide a comprehensive analysis of its principles and application.

As the $Ar$ decreases to 1.75, the recirculation region behind the cylinder starts to become unstable and begins to exhibit vortex shedding phenomena. As $Ar$ decreases further to 1.5 and 1.25, the vortices become more compact with increasing frequency, and there is increasing interaction between pairs of vortices through shearing, stretching, and detachment processes. It is also evident that the onset of vortex shedding shifts upstream with decreasing $Ar$.
As the $Ar$ decreases to 1, which transforms the elliptical cylinder into a perfect circle, we observe the typical von Kármán vortex street reported in many literature. This "standard aligned BvK street" is characterized by a series of regularly spaced vortices that alternately form on both sides of the cylinder.
As the $Ar$ decreases from $Ar = 0.75$ to $Ar = 0.25$, the recirculation area gradually diminishes, leading to a more compact vortex structure and an increase in shedding frequency.
At $Ar = 0.1$, for extremely low $Ar$, the recirculation area further decreases, resulting in an increase in both the frequency and intensity of vortex shedding.
At $Ar = 0$, when the cylinder degenerates into a vertical flat plate, flow separates right at the two sides of the flat plate and vortex shedding becomes more pronounced. This indicates an increase in flow instability and irregularity in vortex size and shape.

Across all test cases, persistent instabilities in the form of vortices are observed at the exit of the computational domain, indicating that the fluid flow remains unstable in the downstream direction. As the fluid advances downstream, there is a tendency for the peak vortex intensity to decrease. This attenuation of vorticity is primarily attributed to the viscous diffusion effects commonly present in low Reynolds number regions. This finding is consistent with the studies by \citeauthor{JOHNSON2004229} \cite{JOHNSON2004229}.
Between \(Ar\) of 2 and 1.25, in addition to the recirculation bubble downstream of the cylinder, two additional recirculation bubbles form near the upstream wall. As \(Ar\) is further reduced, the confined flow exhibits only one recirculation bubble. This observation is similar to that of \citeauthor{sahin2004numerical} \cite{sahin2004numerical}, where two additional recirculation bubbles appear on the solid sidewalls of the computational domain under the influence of blockage ratio and Reynolds number.
These phenomena in the wake field of elliptical cylinders represent some intriguing physical behaviors. 

This serves as a foundation for subsequent flow control studies.

\paragraph{Velocity fluctuation and probe layout optimization}

In addition to investigating the baseline flow field, we conduct an analysis of the velocity fluctuation component, which serves as a valuable guiding indicator for probe placement during control. By integrating our knowledge of fluid physics into the control agents, we can optimize their architecture while maintaining superior performance. An illustrative example of this approach is demonstrated in the work of  \citeauthor{pastoor2008feedback}\cite{pastoor2008feedback}, where phase control for drag reduction of a bluff body is achieved by using the decoupled development of shear layers and wake processes. Similarly, we utilize the physical information from the wake field to guide the flow control strategies of the agents, with a focus on the practical design of probe positions. Moreover, \citeauthor{protas2002drag} \cite{protas2002drag} highlighted the influence of oscillatory flow phenomena on the drag coefficient, represented by the adjustable component $C_D^0$. This understanding provides a strategic pathway for manipulating the von K\'{a}rm\'{a}n vortex street and achieving drag adjustment through targeted interventions. Expanding on this foundation, we tailor the physical information observed by the agents to specifically face the factors contributing to instability in the wake field.

\begin{figure*}[ht]
    \centering
    \includegraphics{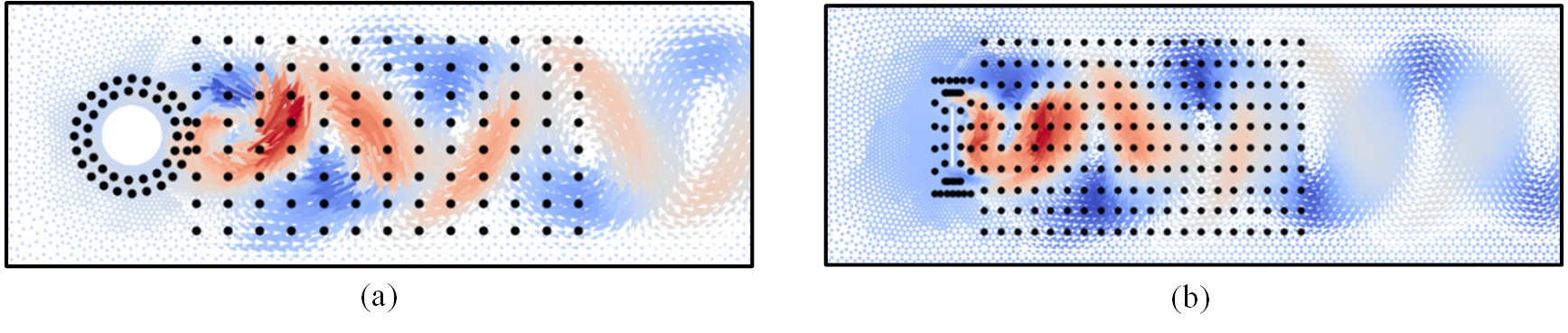} 
    \caption{The distribution of probes within the fluctuating velocity magnitude fields for different $Ar$ : (a) \(Ar = 1\); (b) \(Ar = 0\). To provide a clearer visualization of the probe distribution, we display to a specific region within the computational domain. The x-axis ranges from -2D to 10D, and the y-axis ranges from -2D to 2.1D, where D represents a reference length scale.}
    \label{fig:probes}
\end{figure*}

As depicted in \cref{fig:probes}, the fluctuating magnitude velocity fields around elliptical cylinders with $Ar$ of 0 and 1 are illustrated. The positions of the black dots on the diagram represent the locations of the probes. These probes are strategically placed around the elliptical cylinder and in the downstream wake region.
On one hand, the arrangement of probes around the elliptical cylinder is designed to capture physical information pertinent to the lift and drag forces, which are critical components of our control objectives. 
On the other hand, probes within the wake of the elliptical cylinder are positioned to encompass areas of maximum velocity fluctuations, facilitating the extraction of flow field data from regions where velocity fluctuations and vortex instabilities are most intense. The extraction of physical information from these locations enables the agents to acutely observe changes in the flow field.
In addition, \cref{fig:probes} illustrates two tests with different $Ar$, the configurations for other $Ar$ are similar and not further elaborated upon.

\subsection{Training of DRL-based flow control}\label{sec:training}

\begin{figure*}[htbp]
\centering
\includegraphics{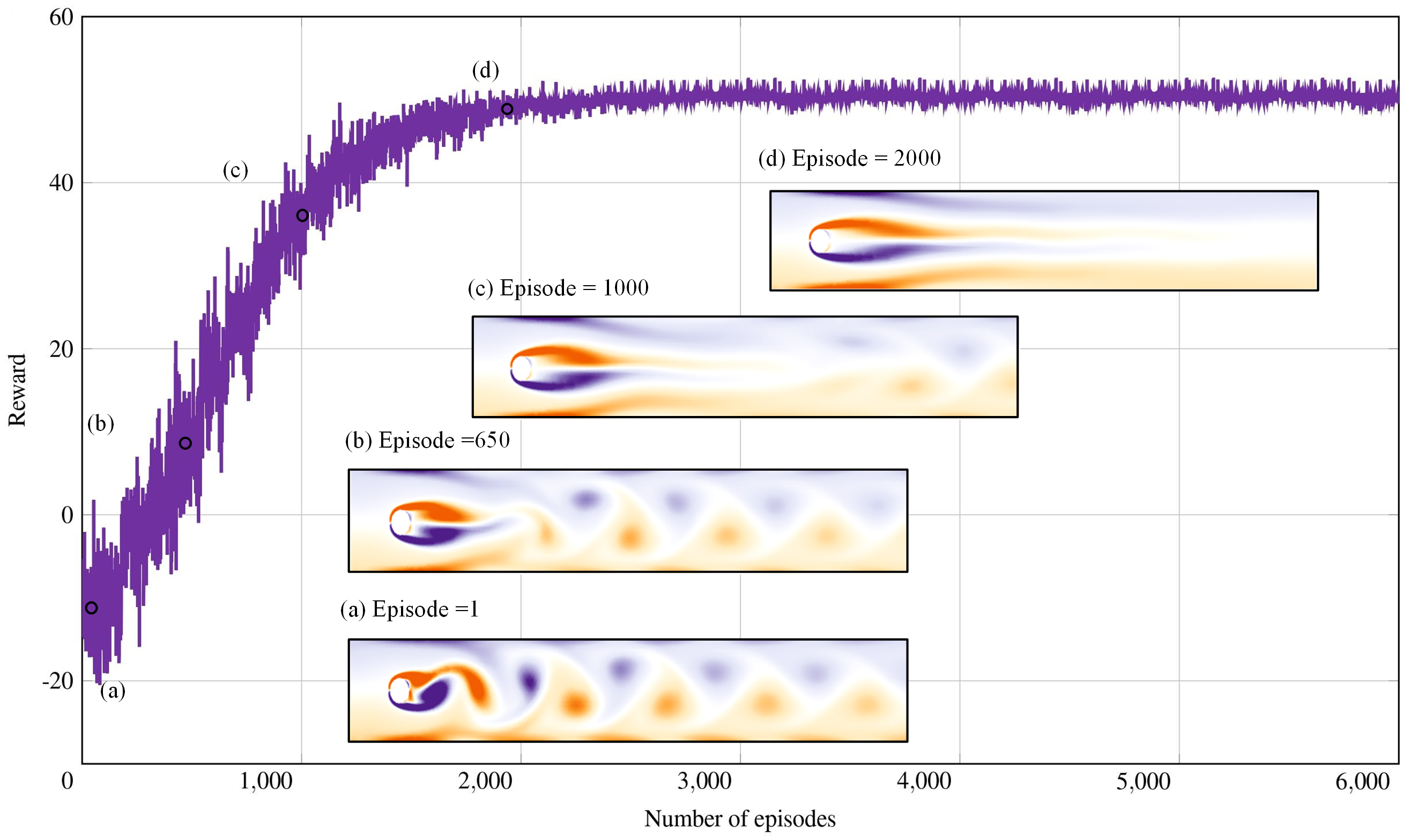}
\caption{For an elliptical cylinder with $Ar=0.75$, observed the evolution of the instantaneous vorticity snapshots when DRL training. Specifically, focused on four representative training episodes: (a) Episode = 1, (b) Episode = 650, (c) Episode = 1000, and (d) Episode = 2000. The variations in the instantaneous vorticity snapshots at these episodes describe the influence of flow control techniques on vortex shedding.}
\label{fig:reward01}
\end{figure*}

\begin{figure*}[htbp]
\centering
\includegraphics{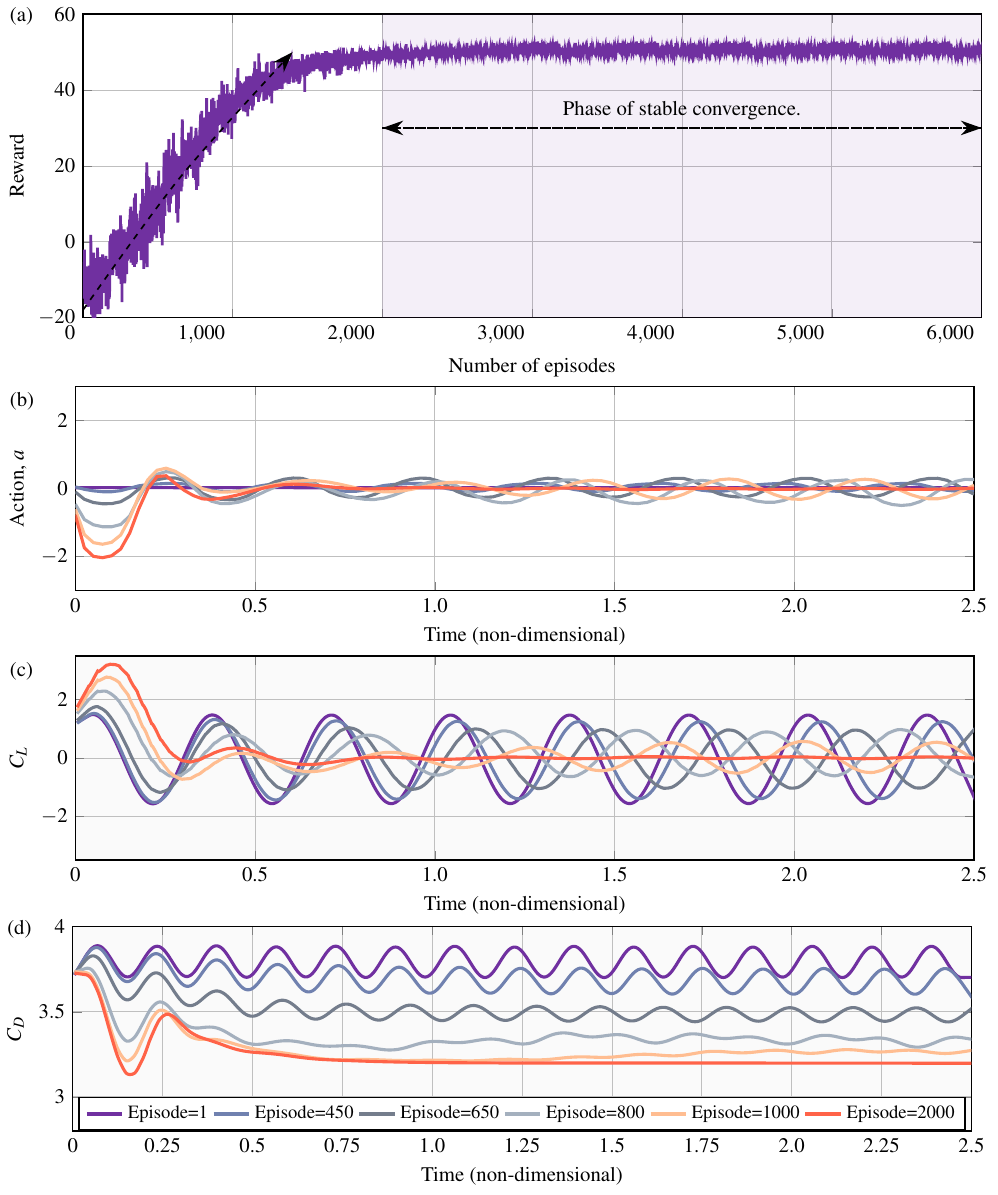}
\caption{For an elliptical cylinder with $Ar = 0.75$, the time series changes of the action, $C_L$, and $C_D$ during the characteristic episodes of DRL training. 
(a) The learning curve of DRL training for the elliptical cylinder with $Ar = 0.75$. The values of the synthetic jet actions and the time series of $C_L$ and $C_D$ for episodes 1, 450, 650, 800, 1000, and 2000 are summarized in (b), (c), and (d), respectively.
(b) The variation of the synthetic jet velocity with time.
(c) The time series of $C_L$ for the characteristic episodes.
(d) The time series of $C_D$ for the characteristic episodes. Note that the characteristic episodes in figures (b), (c), and (d) are consistent, with labels shown in (d).}
\label{fig:Ar075}
\end{figure*}

\begin{figure*}[htbp]
\centering
\includegraphics{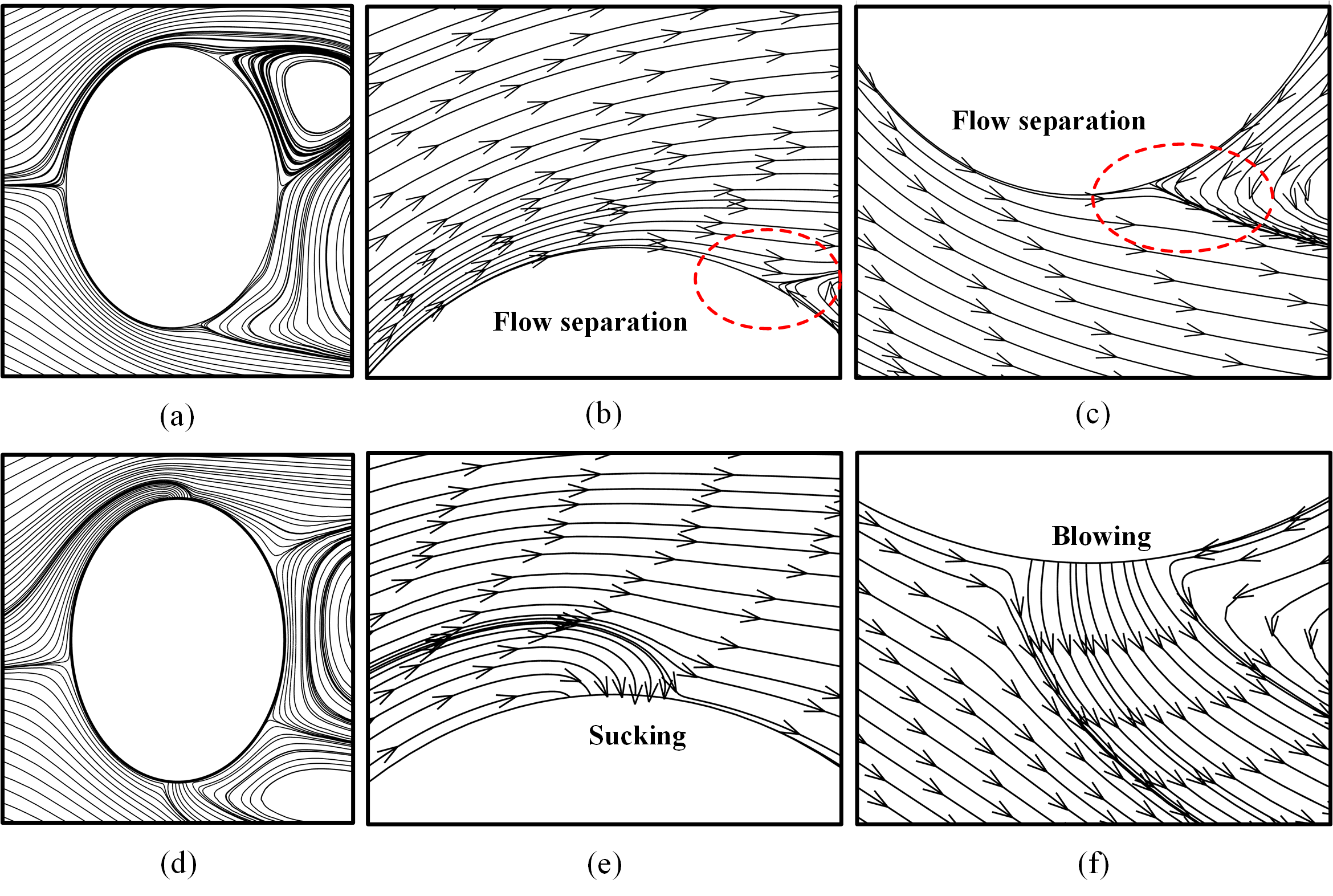}
\caption{Instantaneous streamlines of an elliptical cylinder with $ Ar = 0.75 $. (a) In the baseline flow, streamline around the elliptical cylinder. To highlight the changes in streamlines during the synthetic jet blowing or suction process, localized streamline near 90° and 270° of the elliptical cylinder are presented on the right side. (b) In the baseline flow, with the synthetic jets turned off, the streamlines flow along the elliptical cylinder's surface and separate near the aft side. (c) In the baseline flow, the fluid flows closely along the surface and separates near the aft side. (d) Instantaneous streamlines of the elliptical cylinder when flow control is activated. The synthetic jets perform suction on the upper side and blowing on the lower side of the elliptical cylinder. (e) and (f) display localized streamline near the synthetic jets. (e) Streamlines when the upper jet is performing suction. (f) Streamlines when the lower jet is performing blowing.}
\label{fig:jet}
\end{figure*}

In this section, we use an elliptical cylinder with \(Ar=0.75\) as an example to illustrate the interaction process between the agent and the environment during DRL training and the trial-and-error learning process to maximize the objective function. \Cref{fig:reward01} shows the evolution of the vorticity field of the elliptical cylinder throughout the DRL training process, clearly demonstrating how the vortices shed from the elliptical cylinder in the baseline flow are gradually expelled from the computational domain while suppressing the emergence of new vortices during the execution of active flow control.
During the training process, six characteristic episodes are marked, and the values of \(C_D\), \(C_L\), and the synthetic jet velocity for the elliptical cylinder are extracted for these episodes. \Cref{fig:Ar075} provides a more intuitive demonstration of the active flow control process and control strategies.
Finally, \Cref{fig:jet} offers a detailed analysis of the streamline patterns before and after the execution of synthetic jet actions, with a local magnification of the streamlines near the synthetic jet location. This clearly shows the changes in streamlines around the elliptical cylinder when the synthetic jet is in the blowing or suction state. 
This section provides a comprehensive description of the DRL training process and the operational mechanism of the synthetic jet.

To facilitate understanding of how the agent learns the optimal flow control strategy through trial and error, we use the learning curve from the training process and the changes in the vorticity field around the elliptical cylinder to describe it.
During DRL training, the agent first observes the state of the environment. We use several probes to provide physical information (velocity or pressure) of the flow field to characterize the state of the environment. The agent then calculates the initial reward function value based on the observed state and feeds the corresponding action (i.e., the jet velocity value) back to the CFD environment. Once the synthetic jets begin executing the action, they immediately influence the physical field around the elliptical cylinder, causing a change in its state.
Next, the agent observes the state of the CFD environment again and calculates the reward value corresponding to that action. If the reward value increases, it indicates that the control strategy is beneficial for the objective function. If the reward value decreases, the agent will adjust the control strategy in the opposite direction. Through this iterative interaction and feedback process, the objective function is gradually optimized to the best control strategy.

To demonstrate the suppression of wake vortices in an elliptical cylinder during DRL training, we provide an example using an elliptical cylinder with \(Ar=0.75\) in \Cref{fig:reward01}, observing the evolution of the instantaneous vorticity field throughout the DRL training process.
The DRL training consists of 6000 episodes, including a rapid growth phase and a convergence phase. After an initial rapid rise during the first 2000 episodes, the reward function value approaches its maximum. In the 2000-6000 episode phase, the reward function value fluctuates near its maximum, indicating that the training has stably converged.
We select four typical episodes during the surge phase to describe the changes in vortex shapes and positions. \Cref{fig:reward01}(a) shows the wake field of the elliptical cylinder at episode 1, where regular, periodic vortex shedding is observed. DRL training begins with control applied to an already fully developed flow around the elliptical cylinder.
In \Cref{fig:reward01}(b), no new vortices shed under the influence of synthetic jet blowing or suction, and the recirculation region behind the elliptical cylinder elongates. The initially unstable and oscillating recirculation region behind the elliptical cylinder becomes symmetric and stable during flow control execution.
\Cref{fig:reward01}(c) shows that the vortices shed in the baseline flow gradually move downstream, stretching and interacting with each other. At this stage, the shape of the recirculation bubble stabilizes, and no new vortices shed.
By episode 2000, \Cref{fig:reward01}(d) illustrates that no vortex shedding exists in the computational domain, the area of the recirculation bubble no longer expands, and the entire computational domain reaches a stable state. 
This indicates that when the DRL training reward value converges near its maximum, the agent has learned the optimal control strategy, effectively suppressing vortex shedding in the wake field through the actions of the synthetic jet.

To more clearly illustrate the variations in synthetic jet velocity, \( C_L \), and \( C_D \) during DRL training, we selected six characteristic episodes to extract the corresponding actions, \( C_L \), and \( C_D \). 
\cref{fig:Ar075}(a) divides the DRL training into two phases: the reward function value rapidly increases between episodes = 0 and episodes = 2000, and after episode = 2000, the reward function enters a plateau, indicating that the training has converged.
We selected episodes = 1, episodes = 450, episodes = 650, episodes = 800, episodes = 1000, and episodes = 2000 as characteristic episodes to observe the changes in \( C_D \), \( C_L \), and the control strategies for the elliptical cylinder. 
\cref{fig:Ar075}(b) describes the variation of the synthetic jet velocity over time, representing the development of the flow control strategy. It can be observed that at episode 1, the agent attempts to use a relatively low jet velocity. As the number of episodes increases, the amplitude of the jet velocity gradually increases. By episode = 2000, the control strategy involves using a high jet velocity in the initial phase to control the flow field, followed by maintaining stability with much lower jet velocities.
\cref{fig:Ar075}(c) shows the time series of \( C_L \) for the elliptical cylinder. At episode = 1, \( C_L \) exhibits periodic fluctuations. As the number of episodes increases, the amplitude of \( C_L \)'s periodic fluctuations decreases. By episode = 2000, \( C_L \) shows significant fluctuations in the initial phase of jet activation, then quickly reduces to near zero, maintaining zero for more than half the time, indicating that the lift around the elliptical cylinder is fully controlled.
\cref{fig:Ar075}(d) shows the time series of \( C_D \) for the elliptical cylinder. At episode = 1, \( C_D \) exhibits periodic oscillations. As the number of episodes increases, the amplitude and frequency of \( C_D \) fluctuations decrease. By episode = 2000, \( C_D \) only shows severe fluctuations at the beginning of control and then stabilizes near its minimum value.

In \cref{fig:jet}, we illustrate how the synthetic jets disrupt the streamlines along the sidewalls of an elliptical cylinder with \(Ar = 0.75\) during active flow control.
\cref{fig:jet}(a) shows the streamlines around the elliptical cylinder without active flow control. 
\cref{fig:jet}(b) and \cref{fig:jet}(c) provide zoomed-in views of the streamlines near the elliptical cylinder at the locations of the synthetic jets, which are not active at this point. 
Without control, the streamlines follow the contours of the elliptical cylinder and separate near the aft side.
\cref{fig:jet}(d) shows the streamlines around the elliptical cylinder with flow control activated. 
\cref{fig:jet}(e) and \cref{fig:jet}(f) again provide zoomed-in views of the streamlines at the synthetic jet locations, but this time with the jets active.
When control is activated, the synthetic jets either blow or suction relative to the sidewalls, disrupting the previously continuous streamlines that followed the solid boundary of the elliptical cylinder. 
The noticable difference between uncontrolled baseline and controlled flow is that the flow stays attached after applying jet control. In \cref{fig:jet}(b) and \cref{fig:jet}(c), distinct flow separation can be observed in the rear-end. With synthetic jet, the flow stays attached after the suction/blowing point in \cref{fig:jet}(e) and \cref{fig:jet}(f).
This illustrates the mechanism by which the synthetic jets influence the flow, breaking the streamlines and altering the flow pattern around the elliptical cylinder.

\subsection{DRL-based AFC for various ellipses}\label{sec:ellipses}
Using the DRL algorithm, we investigate the performance of actively controlling the magnitude of synthetic jets for targeted flow control throughout the process of elliptical cylinders transitioning from elongated bodies to circular cylinders, and finally to a normal flat plate, with aspect ratios ranging from 2 to 0 at $Re=100$. 
The evaluation of the control's effectiveness and performance is based on several criteria: intensity of vortex shedding after control, reductions in $C_D$, weakening of $C_L$, the actions (magnitude of the synthetic jets velocities), and the learning curves observed during the training process. 

\begin{figure*}[htbp]
\centering
\includegraphics{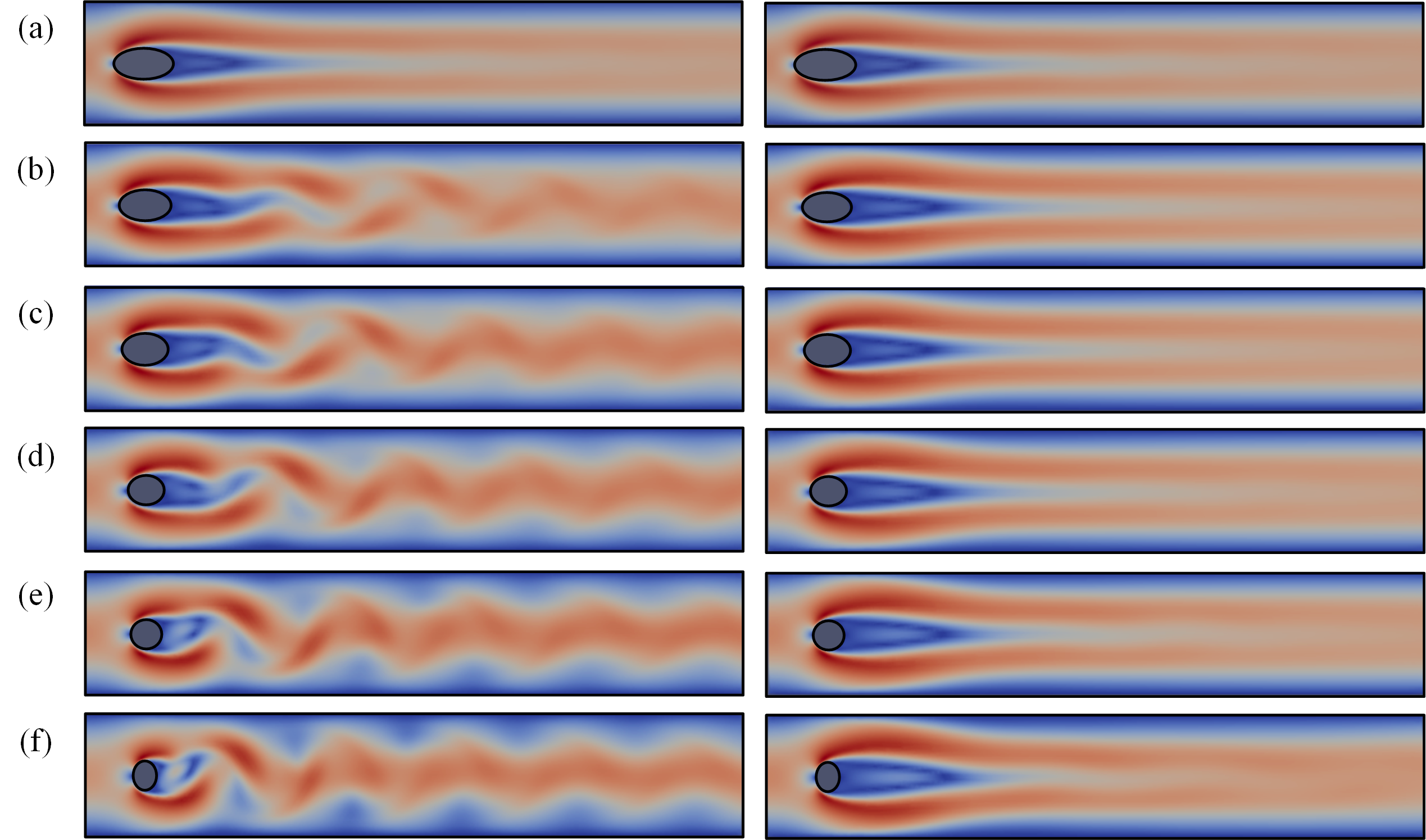}
\caption{Comparison of the velocity magnitude snapshots for an elliptical cylinder before and after control. Each row displays: on the left, the baseline simulation; on the right, the AFC results. (a) \(Ar = 2\); (b) \(Ar = 1.75\); (c) \(Ar = 1.5\); (d) \(Ar = 1.25\); (e) \(Ar = 1\); (f) \(Ar = 0.75\).}
\label{fig:ArU1}
\end{figure*}

\begin{figure*}[htbp]
    \centering
    \includegraphics{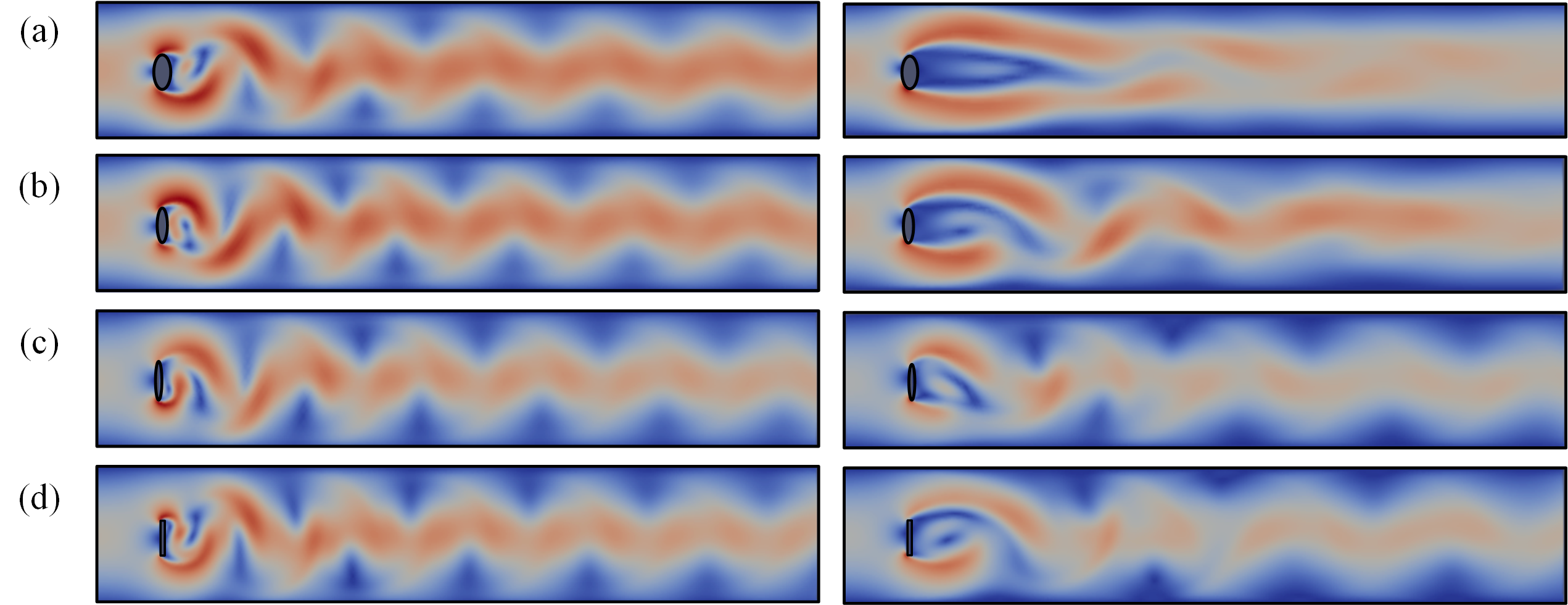}
    \caption{Comparison of the velocity magnitude snapshots for an elliptical cylinder before and after control. Each row displays: on the left, the baseline simulation; on the right, the AFC results. (a) $Ar = 0.5$; (b) $ Ar = 0.25$; (c) $Ar = 0.1$; (d) $Ar = 0$.}
    \label{fig:ArU2}
\end{figure*}

To demonstrate the suppression effect of AFC technology on vortex shedding phenomena, velocity magnitude snapshots of elliptical cylinders with \(Ar\) ranging from 2 to 0.75 and from 0.5 to 0 are displayed in \Cref{fig:ArU1} and \Cref{fig:ArU2}, respectively.
As shown in \Cref{fig:ArU1}, the DRL-based AFC approach demonstrates consistent and effective suppression of vortex shedding and enhancement of flow stability in the wake field. At $Ar$ = 2, the baseline flow field is stable without vortex shedding, and the controlled flow field remains stable and similar to the baseline. As $Ar$ decreases, the AFC successfully suppresses vortices that were previously formed on both sides of the elliptical cylinder. Despite increased vortex shedding and oscillatory behavior in the uncontrolled wake, AFC implementation at $Ar$ values of 1.5, 1.25, 1, and 0.75 transforms the alternating vortex streets into a stable wake field. 
Subsequently, as the $Ar$ continuously decreases, the baseline flow becomes more unstable, characterized by a diminishing recirculation region and increased vortex shedding intensity and frequency. This phenomenon is visually depicted in \Cref{fig:ArU2}.

In particular, at $Ar$ = 0.5, the application of flow control successfully suppresses the alternating vortex shedding, resulting in an elongated recirculation region. However, the wake flow still exhibits oscillations. For $Ar$ = 0.25, flow control elongates the recirculation bubble similar to the $Ar$ = 0.5 case, but with prominent oscillations. Vortex shedding persists in an alternating manner, resembling elongated and stretched vortex structures. Although complete suppression of vortex shedding is not achieved at $Ar$ = 0.1 or 0 due to the inherent chaotic nature of the wake behind vertical thin bodies, the recirculation bubbles are elongated, and the initiation point of shedding is shifted further downstream. The controlled flow exhibits reduced shedding frequency and intensity compared to the baseline conditions. 
\Cref{fig:ArU1} illustrates the effectiveness of AFC technology in eliminating vortex shedding and transitioning flow from unstable to stable modes within the range of $Ar$ from 2 to 0.75, while \Cref{fig:ArU2} indicates that at lower $Ar$ values (0.5 to 0), this technology can effectively alleviate vortex shedding and enhance flow stability to some extent.

\begin{figure*}[htbp]
\includegraphics[width=\textwidth]{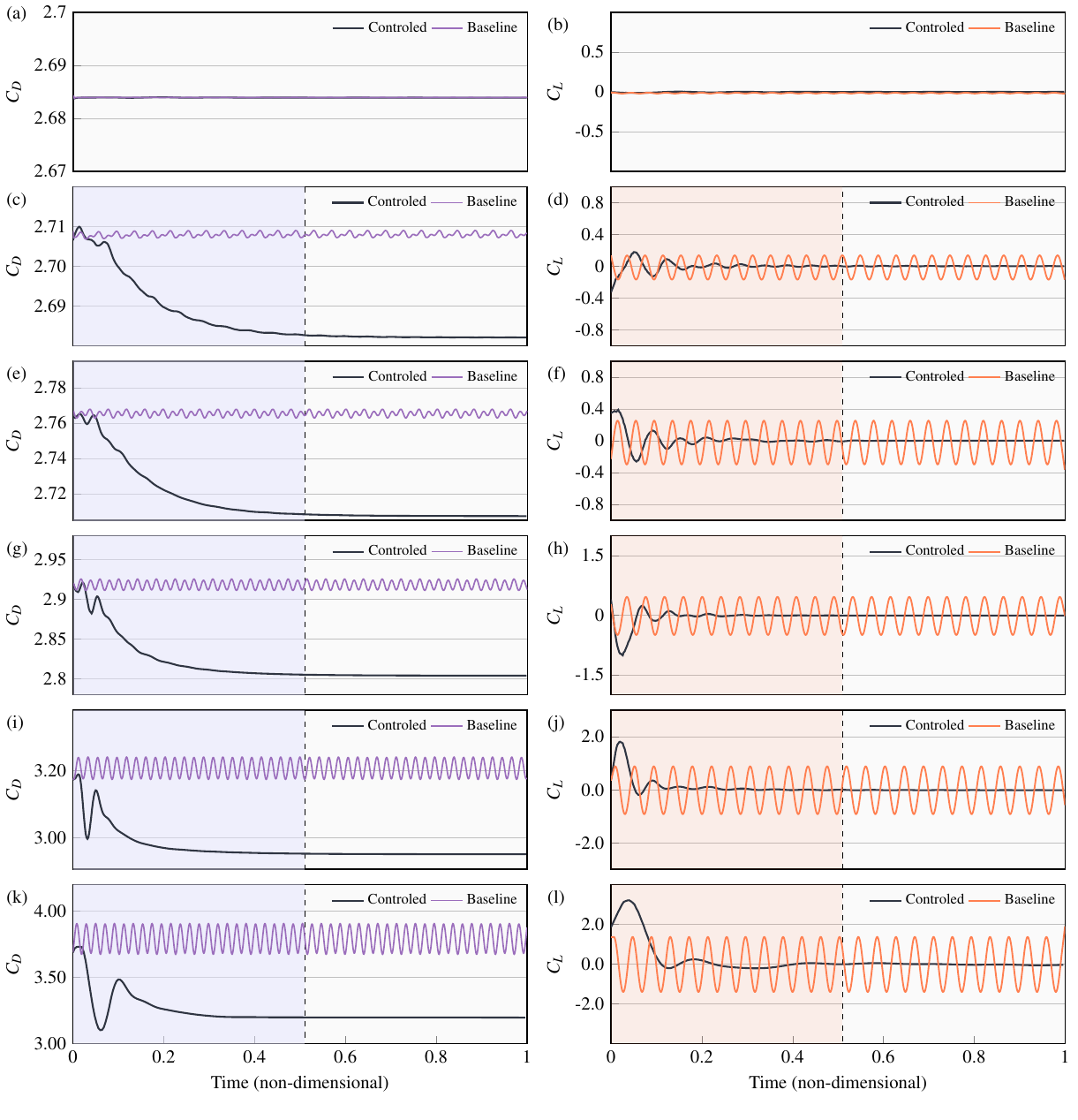}
\caption{Comparison of $C_D$ and $C_L$ before and after control. The background color represents the unstable stage of control, while the absence of background color indicates that the corresponding $C_D$ or $C_L$ remains in a stable stage during this period.
(a) and (b) $Ar=2$; (c) and (d) $Ar=1.75$; (e) and (f) $Ar=1.5$; (g) and (h) $Ar=1.25$; (i) and (j) $Ar=1$; (k) and (l) $Ar=0.75$.}
\label{fig:cdcl01control}
\end{figure*}

\begin{figure*}[htbp]
    \centering
    \includegraphics[width=\textwidth]{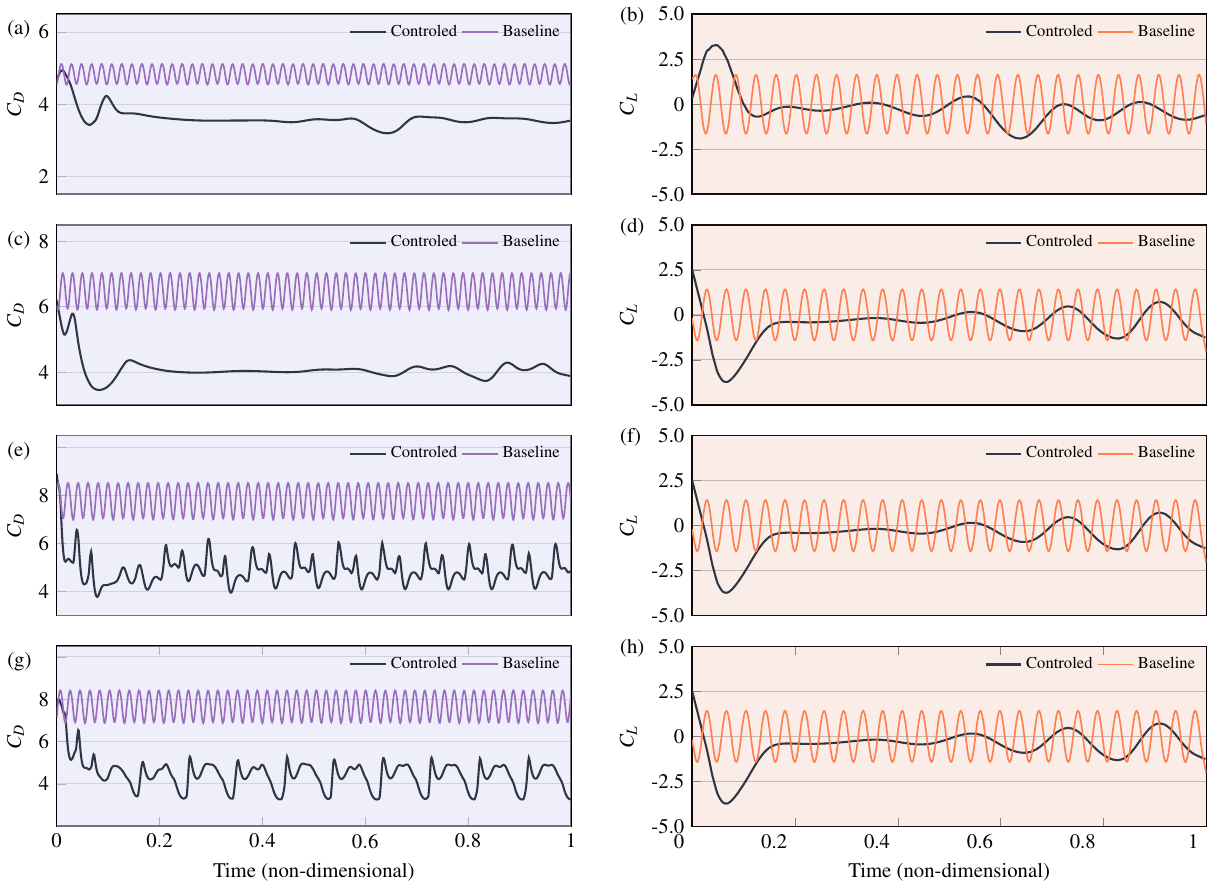}
    \caption{Comparison of $C_D$ and $C_L$ before and after control. The background color indicates that the drag or lift coefficient is still in an unstable stage after flow control. (a) and (b) $Ar = 0.5$; (c) and (d) $ Ar = 0.25$; (e) and (f) $Ar = 0.1$; (g) and (h) $Ar = 0$.}
    \label{fig:cdcl02control}
\end{figure*}

Further, \Cref{fig:cdcl01control} and \Cref{fig:cdcl02control} provide a comparative analysis of the $C_D$ and $C_L$ before and after implementing the DRL-based AFC strategy, showcasing its effectiveness across a range of $Ar$ from 2 to 0.75 and from 0.5 to 0, respectively. The training duration \( T_{\text{max}} \) is set to 2.5. Time is normalized using \( T_{\text{max}} \), with the control period starting at Time = 0 and ending at Time = 1.
As shown in \Cref{fig:cdcl01control}, the AFC strategy has no effect on the stability of both $C_D$ and $C_L$ at an $Ar$ of 2. However, when synthetic jets control is activated at $Ar$ values of 1.75, 1.5, 1.25, 1, and 0.75, there is an immediate and substantial reduction in $C_D$, which quickly reaches a minimum value. After half of the total control duration, $C_D$ stabilizes and remains constant at this minimum value. Similarly, $C_L$ experiences significant fluctuations initially but stabilizes near zero after half of the total control duration. The reductions in $C_D$ and $C_L$ at different $Ar$ values can be found in \Cref{tab:ratio}. Clearly, the periodic oscillations of \(C_D\) and \(C_L\) originally present in the baseline flow are significantly reduced, ultimately transitioning to a stable state. 

The impact of the AFC strategy on \(C_D\) and \(C_L\) at \(Ar=0.5\) and \(Ar=0.25\) is described in \Cref{fig:cdcl02control}.
While both coefficients experience reductions compared to the baseline flow, they do not achieve the same level of stability observed at $Ar=0.75$ and higher. Initially, there are abrupt changes in both $C_D$ and $C_L$, but they do not fully stabilize in the subsequent period. In the latter half of the control duration, there are fluctuations in both coefficients, but with lower amplitude and frequency compared to the baseline flow. 
At $Ar=0.1$ and $Ar=0$ are described in also in \Cref{fig:cdcl02control}, the controlled flow exhibits a sharp decrease in drag coefficient at the beginning of the control, followed by oscillations at a lower frequency and amplitude compared to the baseline flow. Although complete stability is not achieved, the average value of $C_D$ is significantly reduced. Similarly, $C_L$ initially decreases sharply and approaches zero, then exhibits oscillations at a lower frequency than the baseline $C_L$ oscillations.

\begin{table*}[htbp]
\centering
\caption{Quantified control effects of elliptical cylinders under DRL-based AFC.}
\begin{tabularx}{\textwidth}{
  >{\centering\arraybackslash}p{0.06\linewidth}
  >{\centering\arraybackslash}p{0.06\linewidth}
  >{\centering\arraybackslash}p{0.09\linewidth}
  >{\centering\arraybackslash}p{0.09\linewidth}
  >{\centering\arraybackslash}p{0.15\linewidth}
  >{\centering\arraybackslash}p{0.09\linewidth}
  >{\centering\arraybackslash}p{0.09\linewidth}
  >{\centering\arraybackslash}p{0.14\linewidth}
  >{\centering\arraybackslash}p{0.06\linewidth}
  >{\centering\arraybackslash}p{0.09\linewidth}  
}
\toprule
$Ar$ & Re & $\overline{C}_{D,\text{Baseline}}$ & $\overline{C}_{D,\text{Controlled}}$ & Drag Reduction (\%) & $\overline{C}_{L,\text{Baseline}}$ & $\overline{C}_{L,\text{Controlled}}$ & Lift Reduction (\%) & $\overline{a}$ & $\overline{a}$ ratio (\%) \\
\midrule
2    & 100 & 2.684 & 2.684 & 0.0   & 0.016 & 0.016 & 0  & 0 & 0  \\
1.75 & 100 & 2.707 & 2.684 & 0.9   & 0.165 & 0.001 & 99.7  & 0.007 & 0.7  \\
1.50  & 100 & 2.765 & 2.707 & 2.1   & 0.302 & 0.002 & 99.4  & 0.009 & 0.9  \\
1.25 & 100 & 2.918 & 2.804 & 3.9   & 0.541 & 0.006 & 98.9  & 0.006 & 0.6  \\
1.00    & 100 & 3.207 & 2.951 & 8.0   & 1.022 & 0.011 & 98.9  & 0.001 & 0.1  \\
0.75 & 100 & 3.792 & 3.197 & 15.7  & 1.603 & 0.076 & 95.2  & 0.010 & 1.0  \\
0.50  & 100 & 4.837 & 3.533 & 26.9  & 1.937 & 0.619 & 68.0  & 0.049 & 4.9  \\
0.25 & 100 & 6.503 & 4.167 & 35.9  & 1.707 & 0.579 & 66.1  & 0.126 & 12.6 \\
0.10  & 100 & 7.812 & 4.876 & 37.6  & 1.196 & 0.596 & 50.2  & 0.499 & 49.9 \\
0    & 100 & 7.679 & 4.331 & 43.6  & 1.513 & 0.605 & 60.0 & 0.521 & 52.1 \\
\bottomrule
\end{tabularx}
\label{tab:ratio}
\end{table*}

The DRL-based AFC implemented in our study is demonstrated across \Cref{fig:ArU1,fig:ArU2,fig:cdcl01control,fig:cdcl02control}. 
This control strategy not only achieves high control performance but also maintains robust effectiveness, as evidenced by the quantified variations in \(C_D\), \(C_L\), and the ratio of synthetic jet mass flow rate to incoming flow rate, presented in \cref{tab:ratio}.
$\overline{C}_{D,\text{Baseline}}$ and $\overline{C}_{L,\text{Baseline}}$ represents the time-averaged $C_D$ and $C_L$ of the elliptical cylinder once the baseline flow has fully developed.
$\overline{C}_{D,\text{Controlled}}$, $\overline{C}_{L,\text{Controlled}}$, and $\overline{a}$ denote the time-averaged values under stable conditions after AFC is applied, excluding the transient response phase.
At $Ar=2$, the results of AFC control still maintain a stable baseline flow, without applying external excitation to disrupt flow stability.
In the range of $Ar$ values from 1.75 to 0.75, the AFC control achieves substantial reductions in $C_D$ ranging from 0.9\% to 15.7\%, while the reduction rates in $C_L$ exceed 95\% in all cases.
Notably, the reduction in the coefficient of \(C_L\) for elliptical cylinders, ranging from 99.7\% to 95.2\%, aligns with the achievement of a completely stable state (free from oscillations and fluctuations) for both \(C_D\) and \(C_L\) as illustrated in \cref{fig:cdcl01control}. 
This consistency is also mirrored in the suppression of alternating vortex shedding in the Baseline, as depicted in \cref{fig:ArU1}.

\begin{figure*}[htbp]
\centering
\includegraphics[width=0.95\textwidth]{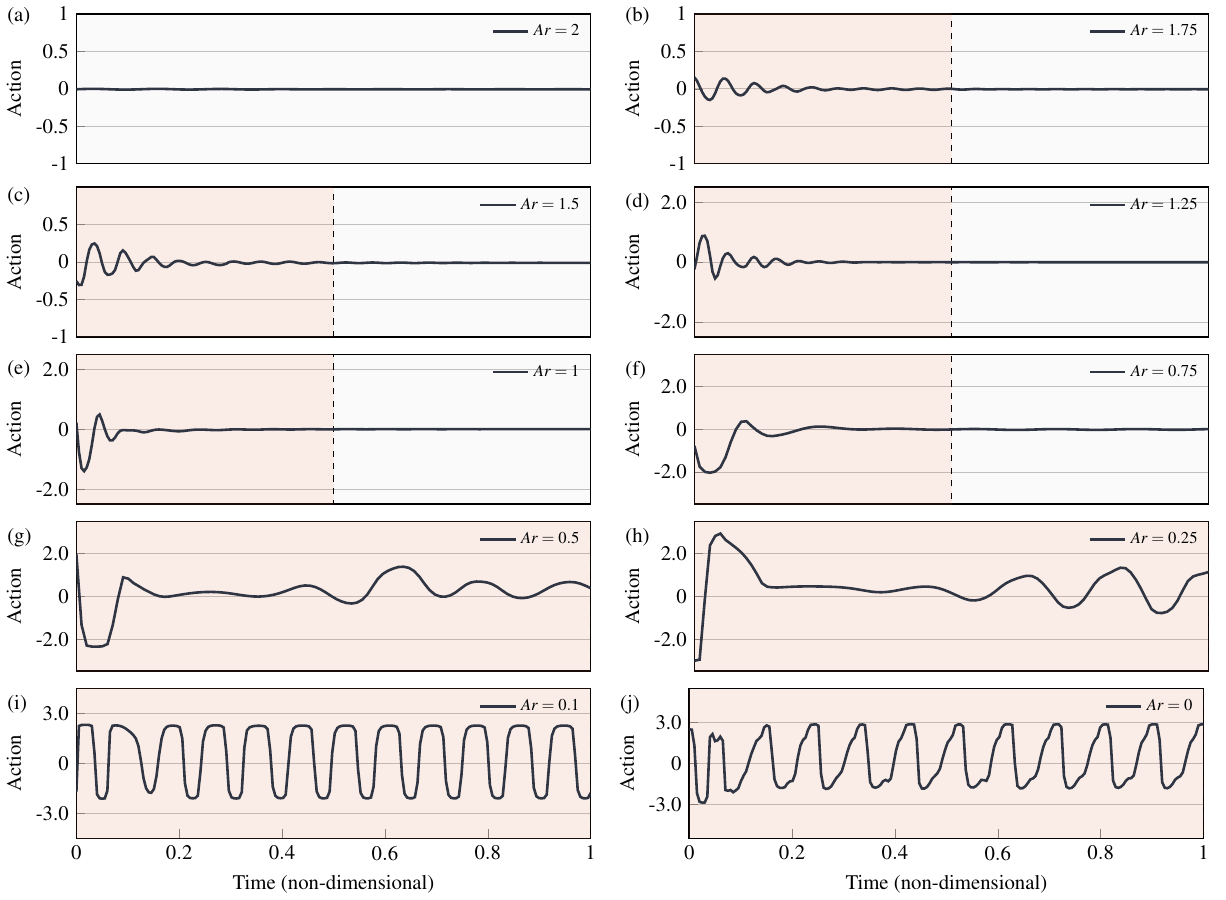}
\caption{Action, which is the mass flow rate of the synthetic jets. The background color represents that the value of the action is in an unstable stage, and the absence of a background color means that the action has been in a stable stage during this period. (a) $Ar=2$; (b) $Ar=1.75$; (c) $Ar=1.5$; (d) $Ar=1.25$; (e) $Ar=1$; (f) $Ar=0.75$; (g) $Ar=0.5$; (h) $Ar=0.25$; (i) $Ar=0.1$; (j) $Ar=0$.}
\label{fig:action}
\end{figure*}

Interestingly, as the $Ar$ decreases to 0.5, 0.25, 0.1, and 0, the reduction rates of the \(C_D\) significantly increase to 26.9\%, 35.9\%, 37.6\%, and 43.6\% respectively, while the reduction rates of the \(C_L\) are observed at 68.0\%, 66.1\%, 50.2\%, and 60.0\%. \citeauthor{protas2002drag}\cite{protas2002drag} delineated the instantaneous drag coefficient into two components: the invariant drag (\(C_D^b\)) and the amendable drag (\(C_D^0\)). 
The \(C_D^b\) remains constant at a given Reynolds number, while the \(C_D^0\) is related to oscillatory flow phenomena, representing the additional resistance caused by fluid oscillations. 
This amendable part can be adjusted through external interventions, demonstrating the potential for flow control.
To mitigate the \(C_D^0\) caused by flow instabilities, we utilize external excitations such as synthetic jets to manipulate the flow field, thereby optimizing the overall fluid dynamics performance. 
As the \(Ar\) of the elliptical cylinder decreases, the pulsations, oscillations, and instabilities within the wake intensify, enhancing the controllable component of the drag. 
This explains why the drag reduction ratio gradually increases with the decrease in $Ar$ as shown in \Cref{tab:ratio}.
However, as the \(Ar\) decreases to 0.5, 0.25, 0.1, and 0, the reduction rates of the $C_L$ are significantly lower than when \(Ar\) is greater than or equal to 0.75. 
Within this range of \(Ar\), only 50.2\% to 68\% of the $C_L$ is reduced, consistent with the fact that neither the $C_L$ nor the $C_D$ achieve the complete stability observed in \cref{fig:cdcl02control}. Additionally, the baseline instability is not fully mitigated as shown in \cref{fig:ArU2}.
This highlights the fact that there is still room for improvement in the current DRL-based AFC framework for controlling the flow instability of elliptical cylinders when $Ar$ is less than or equal to 0.5. Addressing this challenge remains a significant hurdle in the field of active flow control technology for slender structures.


\Cref{fig:action} illustrates the time history curve of the action (the mass flow rate of synthetic jets) during the application of DRL for AFC in the flow around an elliptical cylinder with $Ar$ ranging from 0 to 2. At $Ar=2$, the AFC has minimal impact, with the action value close to zero, indicating stable flow control without nonphysical perturbations. From $Ar=1.75$ to $Ar=0.75$, the agent employs a deliberate strategy of selecting larger action values for higher mass flow rates, aiming to assert flow control through significant energy inputs. As the control progresses, the action values gradually decrease, reducing the mass flow rate. In the later stages, the action values converge towards zero, optimizing for energy efficiency while maintaining flow characteristics. However, at $Ar=0.5$ and $Ar=0.25$, stable flow control is challenging, as the mass flow rate does not stabilize near zero. The non-zero action values indicate the difficulty in suppressing the chaotic wake for vertical slender bodies under these conditions. For $Ar=0.1$ and $Ar=0$, the action values exhibit periodic fluctuations, showcasing the DRL agent's ability to recognize and exploit dominant flow frequencies. 

Furthermore, in \cref{tab:ratio}, we normalize the mass flow rates of the synthetic jets against the incoming far-field flow rate, reflecting their relative strength compared to the incoming flow.
$\overline{a}$ ratio (\%) represents the ratio of the mass flow rate of the synthetic jets to the flow rate from the far-field inflow boundary.
For $Ar$ greater than 0.75, the ratio of synthetic jets to the inflow is less than 1\%, indicating that the mass flow rate utilized in the flow control strategy proposed by the PPO agent is minimal, in line with the energy efficiency requirements of AFC technologies.
For $Ar$ ranging from 0.75 to 1.75, the elliptical cylinder achieves a reduction in $C_D$ and attenuates up to 95\% of $C_L$ with very low energy consumption from external excitation. However, as $Ar$ decreases from 0.5 to 0.1, the ratio of synthetic jets to the inflow increases significantly. Despite this increase in energy input, the capability to attenuate $C_L$ decreases. This indicates that at lower $Ar$ values, the instability in the wake field cannot be fully controlled, even with higher energy from external excitation.
The control strategies derived from DRL-based AFC technology exhibit greater complexity compared to conventional harmonic forcing implemented in previous works (\citeauthor{bergmann2005optimal}\cite{bergmann2005optimal}), highlighting the value of employing ANN as controllers. Typically, approximately 93\% of the drag induced by the shedding of vortices is suppressed by the discovered control laws, with minimal jet intensities required for drag reduction.
In our research, the control strategy derived from the DRL-based AFC system dynamically responds in real time to variations in the flow state. 
This emphasizes its capability to adaptively adjust to changing conditions, demonstrating the robustness and responsiveness of the AFC framework. Furthermore, for elliptical cylinders with \(Ar\) of 1.75, 1.5, 1.25, 1, and 0.75, the magnitude of the mass flow rate injected by the jets normalized by the mass flow rate of the mainstream intersecting with the cylinder is within 1\%.


\begin{figure*}[htbp]
\centering
\includegraphics[width=0.87\textwidth]{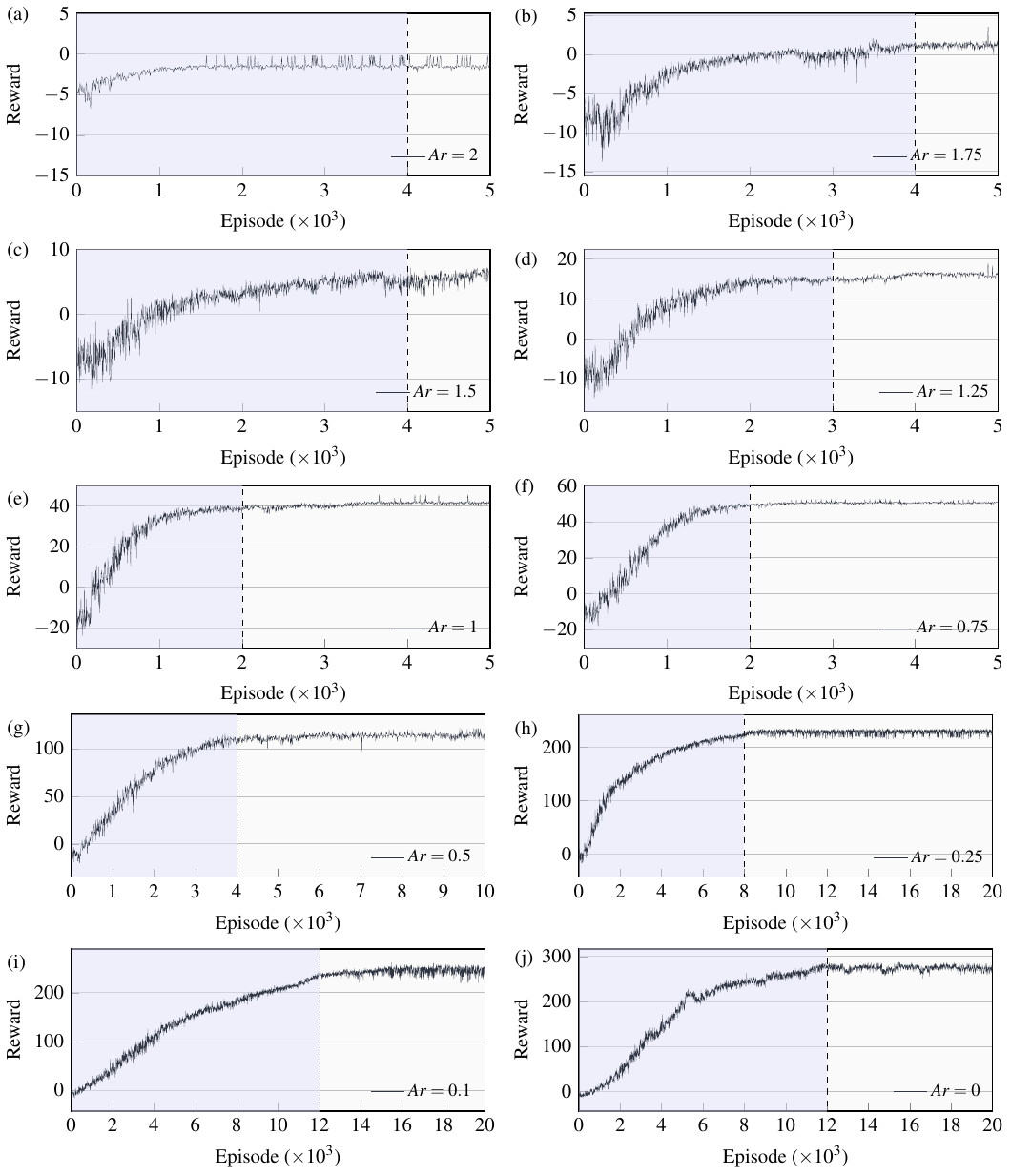}
\caption{When $Ar$ of the elliptical cylinder is in the range of 2-0, the learning curve of PPO algorithm training. (a) $Ar=2$; (b) $Ar=1.75$; (c) $Ar=1.5$; (d) $Ar=1.25$; (e) $Ar=1$. (f) $Ar=0.75$; (g) $Ar=0.5$; (h) $Ar=0.25$; (i) $Ar=0.1$; (j) $Ar=0$.}
\label{fig:reward}
\end{figure*}

The above content details the effectiveness of the flow control strategy, with the corresponding DRL training process curves for elliptical cylinders with different \(Ar\) values presented in \cref{fig:reward}.
First, we observed the overall trend of the reward function. For \(Ar\) values ranging from 1.75 to 0, we noted that the reward function increases with the number of training iterations, eventually converging to a stable value. However, for \(Ar=2\), after 5,000 training iterations, the reward function reaches a plateau but remains below zero, indicating that although the reward function has converged, the agent is unable to achieve the desired objective.
Next, we examined the convergence performance of the reward function. For \(Ar\) values ranging from 1.75 to 0, the reward function exhibits clear convergence points, stabilizing after a plateau phase. This indicates that the agent has learned effective strategies, achieving stable returns under these conditions. 
However, the convergence speed of DRL training varies significantly across different \(Ar\) values. For \(Ar\) values of 1.75, 1.5, and 1.25, signs of convergence appear around 2,000 iterations. For \(Ar\) values of 1 and 0.75, signs of convergence emerge around 1,000 iterations.

Despite setting the total training iterations to 5,000 for \(Ar\) values between 1.75 and 0.75, we observed that the reward function converges much earlier for \(Ar\) values of 1 and 0.75, indicating a faster convergence speed and the agent’s ability to quickly learn effective strategies.
For \(Ar=0.5\), convergence appears after 4,000 iterations. For \(Ar=0.25\), convergence is reached after 8,000 iterations. For \(Ar=0.1\), convergence can be observed after 10,000 iterations. For \(Ar=0\), convergence appears after 12,000 iterations. 
In the range of \(Ar=0.5\) to 0, the DRL training convergence speed is significantly slower compared to \(Ar\) values greater than 0.5. This reflects a lower training efficiency and greater difficulty in exploring effective strategies for the agent when \(Ar\) is less than or equal to 0.5.
Furthermore, in the case of different \(Ar\), DRL training was conducted in parallel using 60 CFD environments, allowing the DRL algorithm to collect data simultaneously from multiple independent simulations. The parameters of the ANN were updated every 60 iterations, which means that the agent and the environment only interacted without updating the policy network for the first 60 episodes. After 60 episodes, the ANN network parameters started to update, resulting in an initial learning curve that exhibits a noticeable step-like increase. This parallel acceleration effect has also been observed in the research conducted by \citeauthor{rabaultAccelerating}\cite{rabaultAccelerating} and \citeauthor{jia2024optimal}\cite{jia2024optimal}.

\subsection{Analysis of the Controlled Flow}\label{sec:controlled}

\begin{figure*}[htbp]
    \centering
    \includegraphics{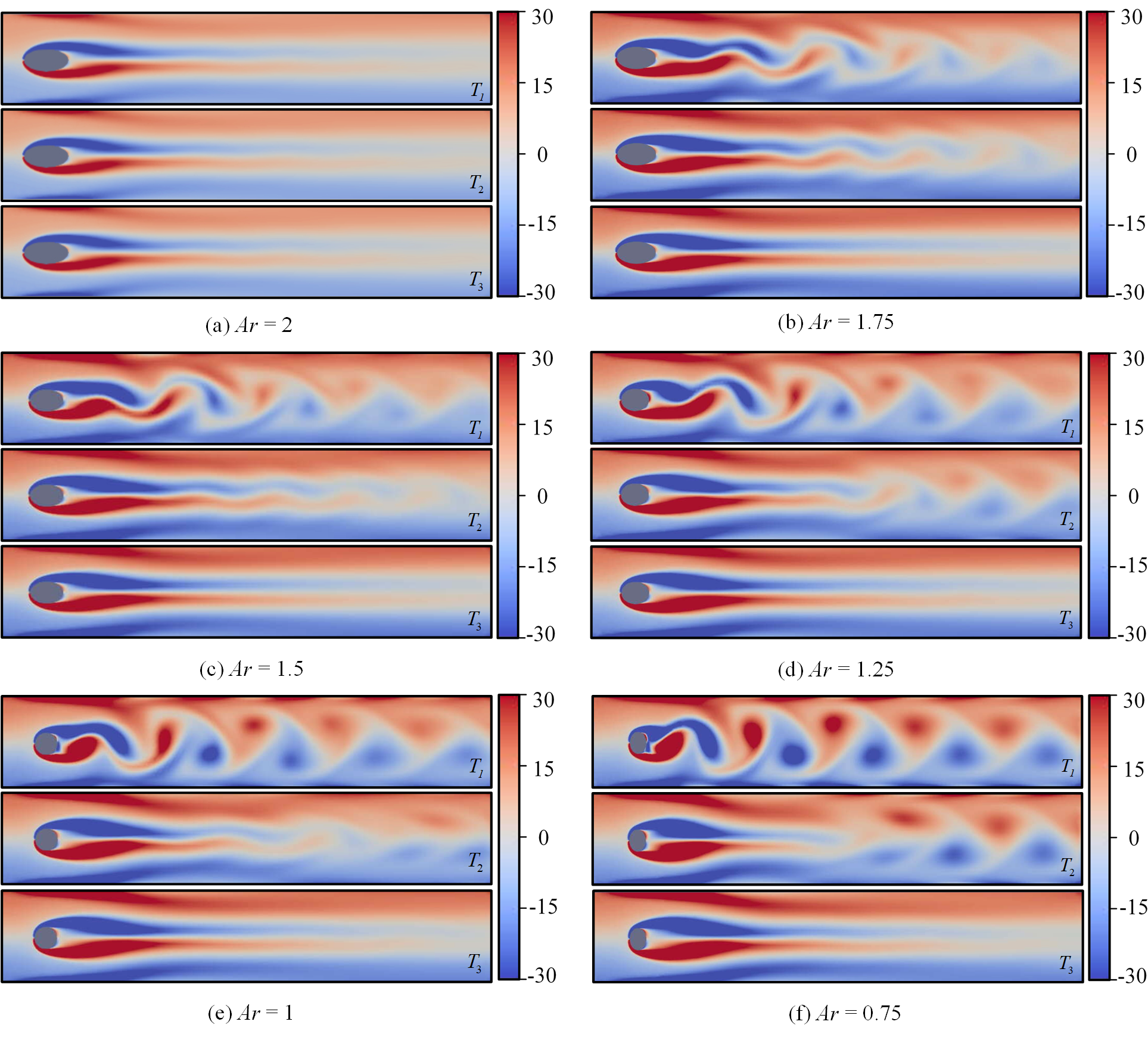}
    \caption{When $Ar=2$, 1.75, 1.5, 1.25, 1, 0.75, the vortex shedding of the elliptical cylinder can be completely suppressed. The instantaneous vorticity snapshots of the wake flow field behind an elliptical cylinder, where the vortex shedding process is suppressed, feature key moments designated as $T_1$, $T_2$, and $T_3$. These represent the instances of the onset of control (when the vortex is fully developed), during the control process, and at the result of control, respectively. (a) $Ar=2$, (b) $Ar=1.75$, (c) $Ar=1.5$, (d) $Ar=1.25$, (e) $Ar=1.0$, (f) $Ar=0.75$.}
    \label{fig:T3Ar111}
\end{figure*}

\cref{fig:T3Ar111} illustrates the suppression of vortex shedding in the wake of elliptical cylinders under the influence of AFC based on DRL. The elliptical cylinders examined have $Ar$ of 2, 1.75, 1.5, 1.25, 1.0, and 0.75, scenarios in which vortex shedding can be completely suppressed. At the initial moment of control ($T_1$), vortices in the wake of the elliptical cylinders are fully developed. During this phase, vortices are observed detaching from the cylinder at a consistent frequency and intensity, forming a Kármán vortex street in the downstream flow. As the control process advances ($T_2$), external excitation methods, such as synthetic jets, are introduced to suppress the vortices in the wake of the elliptical cylinders. The objective of the control is to alter the characteristics of the flow field to weaken or eliminate the formation and persistence of vortices. With the progression of the control measures, a snapshot of the flow field is captured at the result of the control ($T_3$). At this juncture, vortices within the wake of the elliptical cylinders are observed to be suppressed, demonstrating the effectiveness of the DRL-based AFC strategy in stabilizing complex fluid dynamics.

In the study of elliptical cylinders, the phenomenon of vortex shedding and its control are crucial for understanding the structure of flow fields. For elliptical cylinders with different $Ar$, the behavior of the flow fields shows significant differences. 
At $Ar = 2$, both before and after control, the instantaneous vorticity maps at moments $T_1$, $T_2$, and $T_3$ demonstrate a high degree of stability in the flow field, with no evident vortex shedding observed. 
For elliptical cylinders with \(Ar\) values of 1.75, 1.5, and 1.25, at $T_1$, alternating vortices form behind the cylinder, influenced by the elliptical shape effect, which results in lower intensity and frequency of vortex shedding. By $T_2$, with the initiation of control, synthetic jets quickly intervene in the regular vortex shedding process, leading to a delay in vortex detachment and the formation of "elongated vortex structures." The original vortices are significantly elongated and narrowed, and the instantaneous recirculation bubbles are also elongated longitudinally and tend towards stability. By $T_3$, the original vortex shedding has been completely suppressed, with no new vortices forming, and the instantaneous recirculation bubbles reach their maximum extent and remain stable, demonstrating the effectiveness of the control strategy.

\begin{figure*}[htbp]
    \includegraphics{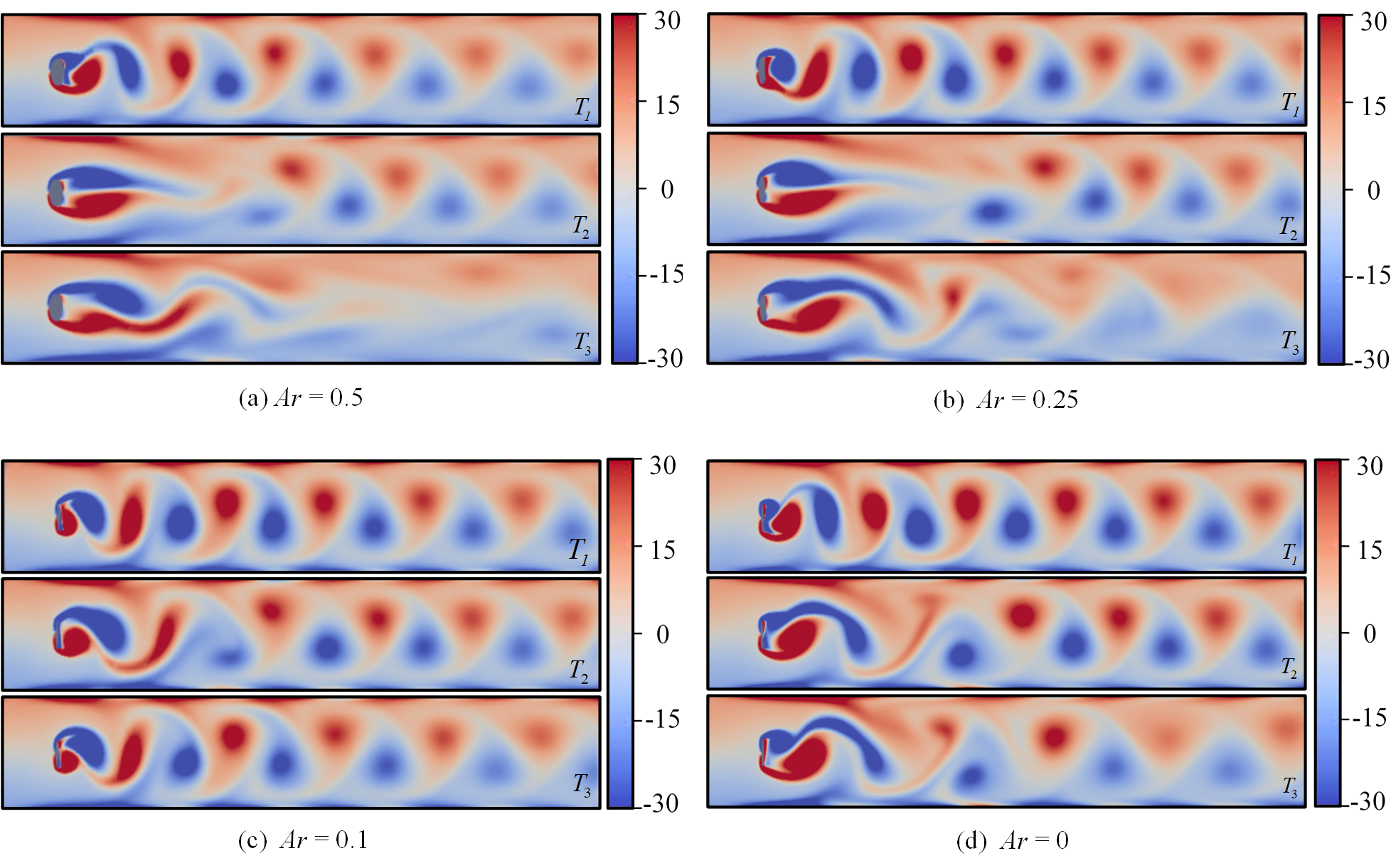}
    \label{fig:T3Ar05}
    \caption{When $Ar = 0.5$, 0.25, 0.1 and 0, vortex shedding from the elliptical cylinder is not entirely suppressed. The instantaneous vorticity snapshots of the wake flow field behind the elliptical cylinder capture key moments designated as $T_1$, $T_2$, and $T_3$. These correspond to the initiation of control (when the vortex is fully developed), the early phase of the control process, the mid-to-late phase of the control process, and the termination phase of control, respectively. (a) $Ar=0.5$, (b) $Ar=0.25$, (c) $Ar=0.1$, (d) $Ar=0$.}
    \label{fig:T3Ar_222}
\end{figure*}

At \(Ar=1\) and \(0.75\), during the $T_1$ phase, a standard Bénard–von Kármán vortex street forms behind the elliptical cylinder, characterized by a series of regularly spaced vortices. These vortices alternately shed from both sides of the cylinder, creating a regular pattern of vortex arrangement. 
Moving into the second phase ($T_2$), similar to cases with higher aspect ratios ($Ar=1.75$, 1.5, 1.25), the recirculation bubbles behind the cylinder are elongated longitudinally, and the originally shed vortices are also elongated and narrowed, with shearing interactions occurring between them. By the third time point ($T_3$), the vortices behind the elliptical cylinder are completely suppressed, and the regions of positive and negative vorticity in the flow field nearly achieve symmetry, demonstrating high stability.
\cref{fig:T3Ar_222} delineates scenarios where the suppression of vortex shedding from elliptical cylinders is not entirely effective for \(Ar\) of 0.5, 0.25, 0.1, and 0. At $T_1$, the vortex shedding behind the cylinder is fully developed, $T_2$ represents the control phase, and $T_3$ provides a snapshot of the flow field after control cessation.

For \(Ar=0.5\) and \(Ar=0.25\), During the $T_1$ phase, vortices are observed detaching from the cylinder with certain frequency and intensity, forming a Kármán vortex street in the downstream wake.
In the $T_2$ phase, as control commences, the recirculation bubble is noted to elongate longitudinally to some extent, with the recirculation area appearing relatively symmetric and stable. The intervention of synthetic jets delays the detachment of vortices, which are stretched and deformed due to shear with the side walls.
By $T_3$, the initially generated vortices are significantly elongated and deformed, exiting the computational domain, yet the recirculation bubble behind the cylinder remains unstable and continues to oscillate, leading to new vortex shedding. The emergence of new irregular vortex shedding highlights persistent instability behind the elliptical cylinder.
For elliptical cylinders with \(Ar=0.5\) and \(Ar=0.25\), the DRL-based control strategy does not completely suppress vortex shedding. The instability of the wake field is mitigated, demonstrating partial efficacy of the control strategy in manipulating flow dynamics, although full stability is not achieved.

For \(Ar = 0.1\) and \(Ar = 0\), during $T_1$, the vortex shedding frequency and intensity from the cylinder are more severe compared to \(Ar = 0.5\), with enhanced flow instability. Multi-level vortex shedding occurs, with primary and secondary vortices differing in shape, frequency, and intensity.
At $T_2$, vortices behind the cylinder are elongated and delayed, and the recirculation area significantly enlarges. The instability of the recirculation area manifests as asymmetry, swinging, and oscillation phenomena.
By $T_3$, although the original shedding vortices are elongated and expelled from the computational domain, new, unstable, and irregular shedding occurs. Shed vortices swinging to the side walls are sheared, stretched, and undergo deformation, resulting in irregularly shaped vortices. This highlights the persistent challenges in managing flow dynamics and instability at lower $Ar$.

\section{CONCLUSIONS}\label{sec:Conclusions} 

In this study, we investigated the use of DRL-based AFC to suppress vortex shedding, reduce drag, and mitigate lift fluctuations of elliptical cylinders at a $Re=100$. The $Ar$ of the elliptical cylinder ranged from an elliptical cylinder ($Ar=2.0$) to a circle cylinder ($Ar=1.0$), and finally to a flat plate ($Ar=0$). 
We employed the PPO algorithm to precisely control the mass flow rates of synthetic jets located on the upper and lower surfaces of the elliptical cylinder. 
The primary objective of this work was to gain further insights into the control capabilities of DRL algorithms for complex flow systems.
The main findings of this study can be summarized as follows:

\begin{itemize}
    \item At \(Ar=2\), the baseline flow is already in a stable state. When AFC is applied, the agent does not perform any actions that would disturb the flow field. This indicates that the DRL agent is capable of recognizing a stable flow condition and making intelligent decisions accordingly.  
    However, for $Ar$ of 1.75, 1.5, 1.25, 1, and 0.75, the reduction rates in the drag coefficient around the cylinder are respectively 0.9\%, 2.1\%, 3.9\%, 8.0\%, and 15.7\%, while the reduction rates in lift coefficient are 99.7\%, 99.4\%, 98.9\%, 98.9\%, and 95.2\%. 
    The originally periodic oscillations of the drag coefficient are reduced to their minimum value and stabilized. Simultaneously, over 95\% of the lift coefficient is suppressed, ultimately stabilizing close to zero.
    With the activation of synthetic jets, the oscillating recirculation bubble in the baseline flow is gradually elongated and transformed into a stable and symmetrical recirculation bubble. Simultaneously, the existing vortices in the wake region become progressively elongated, ceasing the generation of new vortices.
    The originally periodic shedding of vortices transitions into a stable flow field, eliminating the phenomenon of vortex shedding.
    These results underscore the robust adaptability of DRL-based AFC across different elliptical cylinder shapes ($Ar$ between 0.75 and 1.75), and its capability of learning optimal control strategies directly through interaction with the environment to achieve a reduction in drag and lift coefficients around the cylinder. 
    Moreover, for elliptical cylinders with $Ar$ of 1.75, 1.5, 1.25, 1, and 0.75, the mass flow rate ratios of synthetic jets relative to the inlet flow rate are 0.7\%, 0.9\%, 0.6\%, 0.1\%, and 1.0\% respectively. 
    This indicates the effectiveness of the flow control strategy, which utilizes less than 1\% of external excitation energy, showcasing its significant energy-saving attribute.
    \item  When the $Ar$ of the elliptical cylinder decreases to 0.5, 0.25, 0.1, and 0, the reduction rates of the drag coefficient and lift coefficient are 26.9\%, 35.9\%, 37.6\%, and 43.6\%, respectively, while the corresponding reduction rates for the lift coefficient are 68.0\%, 66.1\%, 50.2\%, and 60.1\%. Despite the effective reduction in both drag and lift coefficients achieved by the DRL-based AFC strategy, vortex shedding phenomena have not yet stabilized compared to cases with larger $Ar$ values. 
    The drag and lift coefficients exhibit significant fluctuations during the initial stages of control, eventually settling into oscillatory states with smaller amplitudes and frequencies compared to the baseline flow. In particular, for elliptical cylinders with $Ar$ values of 0.1 and 0, the lift and drag coefficients continue to exhibit periodic oscillations.
    This indicates that for elliptical cylinders within the $Ar$ range of 0.5 to 0, the AFC control strategy can partially suppress lift and reduce drag.
    For $Ar$ of 0.5 and 0.25, the recirculation region behind the elliptical cylinder initially elongates, exhibiting symmetric recirculation bubbles. 
    As control progresses, the recirculation region becomes unstable, displaying lateral oscillations and the shedding of long, narrow vortices. 
    For $Ar$ values of 0.1 and 0, the recirculation region remains unstable throughout the control phase, accompanied by the shedding of new vortices.
    However, compared to the baseline flow, the frequency and intensity of vortex shedding decrease.
    Correspondingly, for elliptical cylinders with $Ar$ values of 0.5, 0.25, 0.1, and 0, the ratios of the mass flow rates of the synthetic jets relative to the incoming flow rate are 4.9\%, 12.6\%, 49.9\%, and 52.1\%, respectively. 
    This indicates that despite using higher external energy, it is still not possible to suppress the oscillations and fluctuations in the lift coefficient.
\end{itemize}

To summarize, for elliptical cylinders with $Ar$ ranging from 1.75 to 0.75, intelligent agents trained using the PPO algorithm have adeptly learned to adopt an energy-efficient flow control strategy. 
The control policies based on the PPO algorithm significantly reduce $C_L$ around the elliptical cylinder, greatly diminish $C_D$, and completely suppress vortex shedding in the wake.
Regarding elliptical cylinder with $Ar$ values ranging from 0.5 to 0, the control strategies obtained from agents trained with the PPO algorithm can moderately alleviate vortex shedding and reduce the amplitude and frequency of fluctuations in $C_D$ and $C_L$. 
These control results demonstrate the adaptability of DRL-based AFC strategies to various geometric shapes, their robustness against uncertainties, and their ability to handle complex problems.

\section*{ACKNOWLEDGMENTS}

The authors would like to express their gratitude to Dr. Jean Rabault (University of Oslo, Oslo, Norway), Dr. Jichao Li (National University of Singapore, Singapore), and Mr. Qiulei Wang (The University of Hong Kong, Hong Kong SAR, China) for making their open-source codes for DRL and numerical simulation available online. These resources have significantly contributed to the research presented in this paper. 
The relevant code repositories can be accessed at \url{https://github.com/jerabaul29/Cylinder2DFlowControlDRLParallel} \cite{rabault2024cylinder2dflowcontroldrlparallel,rabault2019artificial},
\url{https://github.com/npuljc/RL_control_Nek5000} \cite{Nek5000,liReinforcementlearning},
and \url{https://github.com/venturi123/DRLinFluids} \cite{DRLinFluids,wangDRLinFluids}.

\section*{AUTHOR DECLARATIONS}

\subsection{Conflict of Interest}

The authors report no conflict of interest.

\subsection{Author Contributions}

\textbf{Wang Jia:} Conceptualization (equal); Data curation (lead); Formal analysis (lead); Methodology(lead); Validation (lead); Visualization (lead); Writing – original draft (lead). 
\textbf{Hang Xu:}
Conceptualization (equal); Investigation (equal); Supervision (lead); Validation (equal); Writing –review \& editing (lead). 

\section*{DATA AVAILABILITY}

The data that support the findings of this study are available from the corresponding author upon reasonable request.

\section*{AUTHOR ORCIDs}

\noindent Wang Jia \href{https://orcid.org/0009-0008-2786-397X}{https://orcid.org/0009-0008-2786-397X}\\
Hang Xu \href{https://orcid.org/0000-0003-4176-0738}{https://orcid.org/0000-0003-4176-0738}.


\section*{References}
\bibliography{aipsamp}

\begin{thebibliography}{65}%
\makeatletter
\providecommand \@ifxundefined [1]{%
 \@ifx{#1\undefined}
}%
\providecommand \@ifnum [1]{%
 \ifnum #1\expandafter \@firstoftwo
 \else \expandafter \@secondoftwo
 \fi
}%
\providecommand \@ifx [1]{%
 \ifx #1\expandafter \@firstoftwo
 \else \expandafter \@secondoftwo
 \fi
}%
\providecommand \natexlab [1]{#1}%
\providecommand \enquote  [1]{``#1''}%
\providecommand \bibnamefont  [1]{#1}%
\providecommand \bibfnamefont [1]{#1}%
\providecommand \citenamefont [1]{#1}%
\providecommand \href@noop [0]{\@secondoftwo}%
\providecommand \href [0]{\begingroup \@sanitize@url \@href}%
\providecommand \@href[1]{\@@startlink{#1}\@@href}%
\providecommand \@@href[1]{\endgroup#1\@@endlink}%
\providecommand \@sanitize@url [0]{\catcode `\\12\catcode `\$12\catcode
  `\&12\catcode `\#12\catcode `\^12\catcode `\_12\catcode `\%12\relax}%
\providecommand \@@startlink[1]{}%
\providecommand \@@endlink[0]{}%
\providecommand \url  [0]{\begingroup\@sanitize@url \@url }%
\providecommand \@url [1]{\endgroup\@href {#1}{\urlprefix }}%
\providecommand \urlprefix  [0]{URL }%
\providecommand \Eprint [0]{\href }%
\providecommand \doibase [0]{http://dx.doi.org/}%
\providecommand \selectlanguage [0]{\@gobble}%
\providecommand \bibinfo  [0]{\@secondoftwo}%
\providecommand \bibfield  [0]{\@secondoftwo}%
\providecommand \translation [1]{[#1]}%
\providecommand \BibitemOpen [0]{}%
\providecommand \bibitemStop [0]{}%
\providecommand \bibitemNoStop [0]{.\EOS\space}%
\providecommand \EOS [0]{\spacefactor3000\relax}%
\providecommand \BibitemShut  [1]{\csname bibitem#1\endcsname}%
\let\auto@bib@innerbib\@empty
\bibitem [{\citenamefont {Collis}\ and\ \citenamefont
  {Joslin}(2004)}]{collis2004issues}%
  \BibitemOpen
  \bibfield  {author} {\bibinfo {author} {\bibfnamefont {S.~S.}\ \bibnamefont
  {Collis}}\ and\ \bibinfo {author} {\bibfnamefont {R.~D.}\ \bibnamefont
  {Joslin}},\ }\bibfield  {title} {\enquote {\bibinfo {title} {Issues in active
  flow control: theory, control, simulation, and experiment},}\ }\href
  {\doibase https://doi.org/10.1016/j.paerosci.2004.06.001} {\bibfield
  {journal} {\bibinfo  {journal} {Progress in aerospace sciences}\ }\textbf
  {\bibinfo {volume} {40}},\ \bibinfo {pages} {237--289} (\bibinfo {year}
  {2004})}\BibitemShut {NoStop}%
\bibitem [{\citenamefont {Jahanmiri}(2010)}]{jahanmiri2010active}%
  \BibitemOpen
  \bibfield  {author} {\bibinfo {author} {\bibfnamefont {M.}~\bibnamefont
  {Jahanmiri}},\ }\bibfield  {title} {\enquote {\bibinfo {title} {Active flow
  control: a review},}\ }\href {\doibase 10.1016/j.flowmeasinst.2009.11.001}
  {\bibfield  {journal} {\bibinfo  {journal} {Flow Measurement and
  Instrumentation}\ }\textbf {\bibinfo {volume} {21}},\ \bibinfo {pages}
  {7--28} (\bibinfo {year} {2010})}\BibitemShut {NoStop}%
\bibitem [{\citenamefont {Awada}, \citenamefont {Younes},\ and\ \citenamefont
  {Ilinca}(2021)}]{en14113058}%
  \BibitemOpen
  \bibfield  {author} {\bibinfo {author} {\bibfnamefont {A.}~\bibnamefont
  {Awada}}, \bibinfo {author} {\bibfnamefont {R.}~\bibnamefont {Younes}}, \
  and\ \bibinfo {author} {\bibfnamefont {A.}~\bibnamefont {Ilinca}},\
  }\bibfield  {title} {\enquote {\bibinfo {title} {Review of vibration control
  methods for wind turbines},}\ }\href {\doibase 10.3390/en14113058} {\bibfield
   {journal} {\bibinfo  {journal} {Energies}\ }\textbf {\bibinfo {volume} {14}}
  (\bibinfo {year} {2021}),\ 10.3390/en14113058}\BibitemShut {NoStop}%
\bibitem [{\citenamefont {Wang}\ \emph {et~al.}(2018)\citenamefont {Wang},
  \citenamefont {Hu}, \citenamefont {Li},\ and\ \citenamefont {Fu}}]{Detached}%
  \BibitemOpen
  \bibfield  {author} {\bibinfo {author} {\bibfnamefont {L.}~\bibnamefont
  {Wang}}, \bibinfo {author} {\bibfnamefont {R.}~\bibnamefont {Hu}}, \bibinfo
  {author} {\bibfnamefont {L.}~\bibnamefont {Li}}, \ and\ \bibinfo {author}
  {\bibfnamefont {S.}~\bibnamefont {Fu}},\ }\bibfield  {title} {\enquote
  {\bibinfo {title} {Detached-eddy simulations for active flow control},}\
  }\href {\doibase 10.2514/1.J055891} {\bibfield  {journal} {\bibinfo
  {journal} {AIAA Journal}\ }\textbf {\bibinfo {volume} {56}},\ \bibinfo
  {pages} {1447--1462} (\bibinfo {year} {2018})}\BibitemShut {NoStop}%
\bibitem [{\citenamefont {Aram}\ \emph {et~al.}(2018)\citenamefont {Aram},
  \citenamefont {Lee}, \citenamefont {Shan},\ and\ \citenamefont
  {Vargas}}]{Aram2018}%
  \BibitemOpen
  \bibfield  {author} {\bibinfo {author} {\bibfnamefont {S.}~\bibnamefont
  {Aram}}, \bibinfo {author} {\bibfnamefont {Y.-T.}\ \bibnamefont {Lee}},
  \bibinfo {author} {\bibfnamefont {H.}~\bibnamefont {Shan}}, \ and\ \bibinfo
  {author} {\bibfnamefont {A.}~\bibnamefont {Vargas}},\ }\bibfield  {title}
  {\enquote {\bibinfo {title} {Computational fluid dynamic analysis of fluidic
  actuator for active flow control applications},}\ }\href {\doibase
  10.2514/1.J056255} {\bibfield  {journal} {\bibinfo  {journal} {AIAA Journal}\
  }\textbf {\bibinfo {volume} {56}},\ \bibinfo {pages} {111--120} (\bibinfo
  {year} {2018})}\BibitemShut {NoStop}%
\bibitem [{\citenamefont {Jain}, \citenamefont {Yeo},\ and\ \citenamefont
  {Chopra}(2010)}]{jain2010computational}%
  \BibitemOpen
  \bibfield  {author} {\bibinfo {author} {\bibfnamefont {R.}~\bibnamefont
  {Jain}}, \bibinfo {author} {\bibfnamefont {H.}~\bibnamefont {Yeo}}, \ and\
  \bibinfo {author} {\bibfnamefont {I.}~\bibnamefont {Chopra}},\ }\bibfield
  {title} {\enquote {\bibinfo {title} {Computational fluid
  dynamics—computational structural dynamics analysis of active control of
  helicopter rotor for performance improvement},}\ }\href {\doibase
  10.4050/JAHS.55.042004} {\bibfield  {journal} {\bibinfo  {journal} {Journal
  of the American Helicopter Society}\ }\textbf {\bibinfo {volume} {55}},\
  \bibinfo {pages} {42004} (\bibinfo {year} {2010})}\BibitemShut {NoStop}%
\bibitem [{\citenamefont {Li}\ \emph {et~al.}(2022)\citenamefont {Li},
  \citenamefont {Chang}, \citenamefont {Kong},\ and\ \citenamefont
  {Bao}}]{LI202214}%
  \BibitemOpen
  \bibfield  {author} {\bibinfo {author} {\bibfnamefont {Y.}~\bibnamefont
  {Li}}, \bibinfo {author} {\bibfnamefont {J.}~\bibnamefont {Chang}}, \bibinfo
  {author} {\bibfnamefont {C.}~\bibnamefont {Kong}}, \ and\ \bibinfo {author}
  {\bibfnamefont {W.}~\bibnamefont {Bao}},\ }\bibfield  {title} {\enquote
  {\bibinfo {title} {Recent progress of machine learning in flow modeling and
  active flow control},}\ }\href {\doibase
  https://doi.org/10.1016/j.cja.2021.07.027} {\bibfield  {journal} {\bibinfo
  {journal} {Chinese Journal of Aeronautics}\ }\textbf {\bibinfo {volume}
  {35}},\ \bibinfo {pages} {14--44} (\bibinfo {year} {2022})}\BibitemShut
  {NoStop}%
\bibitem [{\citenamefont {Pozorski}\ and\ \citenamefont
  {Wacławczyk}(2020)}]{pr8111379}%
  \BibitemOpen
  \bibfield  {author} {\bibinfo {author} {\bibfnamefont {J.}~\bibnamefont
  {Pozorski}}\ and\ \bibinfo {author} {\bibfnamefont {M.}~\bibnamefont
  {Wacławczyk}},\ }\bibfield  {title} {\enquote {\bibinfo {title} {Mixing in
  turbulent flows: An overview of physics and modelling},}\ }\href {\doibase
  10.3390/pr8111379} {\bibfield  {journal} {\bibinfo  {journal} {Processes}\
  }\textbf {\bibinfo {volume} {8}} (\bibinfo {year} {2020}),\
  10.3390/pr8111379}\BibitemShut {NoStop}%
\bibitem [{\citenamefont {Reichstein}\ \emph {et~al.}(2019)\citenamefont
  {Reichstein}, \citenamefont {Camps-Valls}, \citenamefont {Stevens},
  \citenamefont {Jung}, \citenamefont {Denzler}, \citenamefont {Carvalhais},\
  and\ \citenamefont {Prabhat}}]{reichstein2019deep}%
  \BibitemOpen
  \bibfield  {author} {\bibinfo {author} {\bibfnamefont {M.}~\bibnamefont
  {Reichstein}}, \bibinfo {author} {\bibfnamefont {G.}~\bibnamefont
  {Camps-Valls}}, \bibinfo {author} {\bibfnamefont {B.}~\bibnamefont
  {Stevens}}, \bibinfo {author} {\bibfnamefont {M.}~\bibnamefont {Jung}},
  \bibinfo {author} {\bibfnamefont {J.}~\bibnamefont {Denzler}}, \bibinfo
  {author} {\bibfnamefont {N.}~\bibnamefont {Carvalhais}}, \ and\ \bibinfo
  {author} {\bibnamefont {Prabhat}},\ }\bibfield  {title} {\enquote {\bibinfo
  {title} {Deep learning and process understanding for data-driven earth system
  science},}\ }\href {\doibase 10.1038/s41586-019-0912-1} {\bibfield  {journal}
  {\bibinfo  {journal} {Nature}\ }\textbf {\bibinfo {volume} {566}},\ \bibinfo
  {pages} {195--204} (\bibinfo {year} {2019})}\BibitemShut {NoStop}%
\bibitem [{\citenamefont {Vinuesa}(2024)}]{vinuesa2024perspectives}%
  \BibitemOpen
  \bibfield  {author} {\bibinfo {author} {\bibfnamefont {R.}~\bibnamefont
  {Vinuesa}},\ }\bibfield  {title} {\enquote {\bibinfo {title} {Perspectives on
  predicting and controlling turbulent flows through deep learning},}\ }\href
  {\doibase 10.1063/5.0190452} {\bibfield  {journal} {\bibinfo  {journal}
  {Physics of Fluids}\ }\textbf {\bibinfo {volume} {36}},\ \bibinfo {pages}
  {031401} (\bibinfo {year} {2024})}\BibitemShut {NoStop}%
\bibitem [{\citenamefont {Ren}, \citenamefont {Rabault},\ and\ \citenamefont
  {Tang}(2021)}]{renApplying}%
  \BibitemOpen
  \bibfield  {author} {\bibinfo {author} {\bibfnamefont {F.}~\bibnamefont
  {Ren}}, \bibinfo {author} {\bibfnamefont {J.}~\bibnamefont {Rabault}}, \ and\
  \bibinfo {author} {\bibfnamefont {H.}~\bibnamefont {Tang}},\ }\bibfield
  {title} {\enquote {\bibinfo {title} {Applying deep reinforcement learning to
  active flow control in turbulent conditions},}\ }\href {\doibase
  10.1063/5.0037371} {\bibfield  {journal} {\bibinfo  {journal} {Physics of
  Fluids}\ }\textbf {\bibinfo {volume} {33}},\ \bibinfo {pages} {037121}
  (\bibinfo {year} {2021})},\ \Eprint {http://arxiv.org/abs/2006.10683}
  {2006.10683 [physics]} \BibitemShut {NoStop}%
\bibitem [{\citenamefont {Subramaniam}(2020)}]{Subramaniam}%
  \BibitemOpen
  \bibfield  {author} {\bibinfo {author} {\bibfnamefont {S.}~\bibnamefont
  {Subramaniam}},\ }\bibfield  {title} {\enquote {\bibinfo {title} {Multiphase
  flows: Rich physics, challenging theory, and big simulations},}\ }\href
  {\doibase 10.1103/PhysRevFluids.5.110520} {\bibfield  {journal} {\bibinfo
  {journal} {Physical Review Fluids}\ }\textbf {\bibinfo {volume} {5}},\
  \bibinfo {pages} {110520} (\bibinfo {year} {2020})}\BibitemShut {NoStop}%
\bibitem [{\citenamefont {Brunton}, \citenamefont {Noack},\ and\ \citenamefont
  {Koumoutsakos}(2020)}]{annurevfluid}%
  \BibitemOpen
  \bibfield  {author} {\bibinfo {author} {\bibfnamefont {S.~L.}\ \bibnamefont
  {Brunton}}, \bibinfo {author} {\bibfnamefont {B.~R.}\ \bibnamefont {Noack}},
  \ and\ \bibinfo {author} {\bibfnamefont {P.}~\bibnamefont {Koumoutsakos}},\
  }\bibfield  {title} {\enquote {\bibinfo {title} {Machine learning for fluid
  mechanics},}\ }\href {\doibase 10.1146/annurev-fluid-010719-060214}
  {\bibfield  {journal} {\bibinfo  {journal} {Annual Review of Fluid
  Mechanics}\ }\textbf {\bibinfo {volume} {52}},\ \bibinfo {pages} {477--508}
  (\bibinfo {year} {2020})}\BibitemShut {NoStop}%
\bibitem [{\citenamefont {Brunton}\ and\ \citenamefont
  {Noack}(2015)}]{bruntonClosed}%
  \BibitemOpen
  \bibfield  {author} {\bibinfo {author} {\bibfnamefont {S.~L.}\ \bibnamefont
  {Brunton}}\ and\ \bibinfo {author} {\bibfnamefont {B.~R.}\ \bibnamefont
  {Noack}},\ }\bibfield  {title} {\enquote {\bibinfo {title} {Closed-{{Loop
  Turbulence Control}}: {{Progress}} and {{Challenges}}},}\ }\href {\doibase
  10.1115/1.4031175} {\bibfield  {journal} {\bibinfo  {journal} {Applied
  Mechanics Reviews}\ }\textbf {\bibinfo {volume} {67}},\ \bibinfo {pages}
  {050801} (\bibinfo {year} {2015})}\BibitemShut {NoStop}%
\bibitem [{\citenamefont {Ren}, \citenamefont {Hu},\ and\ \citenamefont
  {Tang}(2020)}]{ren2020active}%
  \BibitemOpen
  \bibfield  {author} {\bibinfo {author} {\bibfnamefont {F.}~\bibnamefont
  {Ren}}, \bibinfo {author} {\bibfnamefont {H.-b.}\ \bibnamefont {Hu}}, \ and\
  \bibinfo {author} {\bibfnamefont {H.}~\bibnamefont {Tang}},\ }\bibfield
  {title} {\enquote {\bibinfo {title} {Active flow control using machine
  learning: A brief review},}\ }\href {\doibase 10.1007/s42241-020-0026-0}
  {\bibfield  {journal} {\bibinfo  {journal} {Journal of Hydrodynamics}\
  }\textbf {\bibinfo {volume} {32}},\ \bibinfo {pages} {247--253} (\bibinfo
  {year} {2020})}\BibitemShut {NoStop}%
\bibitem [{\citenamefont {Arulkumaran}\ \emph {et~al.}(2017)\citenamefont
  {Arulkumaran}, \citenamefont {Deisenroth}, \citenamefont {Brundage},\ and\
  \citenamefont {Bharath}}]{arulkumaran2017deep}%
  \BibitemOpen
  \bibfield  {author} {\bibinfo {author} {\bibfnamefont {K.}~\bibnamefont
  {Arulkumaran}}, \bibinfo {author} {\bibfnamefont {M.~P.}\ \bibnamefont
  {Deisenroth}}, \bibinfo {author} {\bibfnamefont {M.}~\bibnamefont
  {Brundage}}, \ and\ \bibinfo {author} {\bibfnamefont {A.~A.}\ \bibnamefont
  {Bharath}},\ }\bibfield  {title} {\enquote {\bibinfo {title} {Deep
  reinforcement learning: A brief survey},}\ }\href {\doibase
  10.1109/MSP.2017.2743240} {\bibfield  {journal} {\bibinfo  {journal} {IEEE
  Signal Processing Magazine}\ }\textbf {\bibinfo {volume} {34}},\ \bibinfo
  {pages} {26--38} (\bibinfo {year} {2017})}\BibitemShut {NoStop}%
\bibitem [{\citenamefont {Fran\c{c}ois-Lavet}\ \emph
  {et~al.}(2018)\citenamefont {Fran\c{c}ois-Lavet}, \citenamefont {Henderson},
  \citenamefont {Islam}, \citenamefont {Bellemare},\ and\ \citenamefont
  {Pineau}}]{franccois2018introduction}%
  \BibitemOpen
  \bibfield  {author} {\bibinfo {author} {\bibfnamefont {V.}~\bibnamefont
  {Fran\c{c}ois-Lavet}}, \bibinfo {author} {\bibfnamefont {P.}~\bibnamefont
  {Henderson}}, \bibinfo {author} {\bibfnamefont {R.}~\bibnamefont {Islam}},
  \bibinfo {author} {\bibfnamefont {M.~G.}\ \bibnamefont {Bellemare}}, \ and\
  \bibinfo {author} {\bibfnamefont {J.}~\bibnamefont {Pineau}},\ }\bibfield
  {title} {\enquote {\bibinfo {title} {An introduction to deep reinforcement
  learning},}\ }\href {\doibase 10.1561/2200000071} {\bibfield  {journal}
  {\bibinfo  {journal} {Found. Trends Mach. Learn.}\ }\textbf {\bibinfo
  {volume} {11}},\ \bibinfo {pages} {219–354} (\bibinfo {year}
  {2018})}\BibitemShut {NoStop}%
\bibitem [{\citenamefont {Wang}\ \emph {et~al.}(2020)\citenamefont {Wang},
  \citenamefont {Liu}, \citenamefont {Zhang} \emph {et~al.}}]{9904958}%
  \BibitemOpen
  \bibfield  {author} {\bibinfo {author} {\bibfnamefont {H.}~\bibnamefont
  {Wang}}, \bibinfo {author} {\bibfnamefont {N.}~\bibnamefont {Liu}}, \bibinfo
  {author} {\bibfnamefont {Y.}~\bibnamefont {Zhang}},  \emph {et~al.},\
  }\bibfield  {title} {\enquote {\bibinfo {title} {Deep reinforcement learning:
  a survey},}\ }\href {\doibase 10.1631/FITEE.1900533} {\bibfield  {journal}
  {\bibinfo  {journal} {Frontiers of Information Technology \& Electronic
  Engineering}\ }\textbf {\bibinfo {volume} {21}},\ \bibinfo {pages}
  {1726--1744} (\bibinfo {year} {2020})}\BibitemShut {NoStop}%
\bibitem [{\citenamefont {Janiesch}, \citenamefont {Zschech},\ and\
  \citenamefont {Heinrich}(2021)}]{Janiesch2021}%
  \BibitemOpen
  \bibfield  {author} {\bibinfo {author} {\bibfnamefont {C.}~\bibnamefont
  {Janiesch}}, \bibinfo {author} {\bibfnamefont {P.}~\bibnamefont {Zschech}}, \
  and\ \bibinfo {author} {\bibfnamefont {K.}~\bibnamefont {Heinrich}},\
  }\bibfield  {title} {\enquote {\bibinfo {title} {Machine learning and deep
  learning},}\ }\href {\doibase 10.1007/s12525-021-00475-2} {\bibfield
  {journal} {\bibinfo  {journal} {Electronic Markets}\ }\textbf {\bibinfo
  {volume} {31}},\ \bibinfo {pages} {685--695} (\bibinfo {year}
  {2021})}\BibitemShut {NoStop}%
\bibitem [{\citenamefont {LeCun}, \citenamefont {Bengio},\ and\ \citenamefont
  {Hinton}(2015)}]{lecun2015deep}%
  \BibitemOpen
  \bibfield  {author} {\bibinfo {author} {\bibfnamefont {Y.}~\bibnamefont
  {LeCun}}, \bibinfo {author} {\bibfnamefont {Y.}~\bibnamefont {Bengio}}, \
  and\ \bibinfo {author} {\bibfnamefont {G.}~\bibnamefont {Hinton}},\
  }\bibfield  {title} {\enquote {\bibinfo {title} {Deep learning},}\ }\href
  {\doibase 10.1038/nature14539} {\bibfield  {journal} {\bibinfo  {journal}
  {Nature}\ }\textbf {\bibinfo {volume} {521}},\ \bibinfo {pages} {436--444}
  (\bibinfo {year} {2015})}\BibitemShut {NoStop}%
\bibitem [{\citenamefont {Dargazany}(2021)}]{dargazany2021drl}%
  \BibitemOpen
  \bibfield  {author} {\bibinfo {author} {\bibfnamefont {A.}~\bibnamefont
  {Dargazany}},\ }\href@noop {} {\enquote {\bibinfo {title} {Drl: Deep
  reinforcement learning for intelligent robot control -- concept, literature,
  and future},}\ } (\bibinfo {year} {2021}),\ \Eprint
  {http://arxiv.org/abs/2105.13806} {arXiv:2105.13806 [cs.LG]} \BibitemShut
  {NoStop}%
\bibitem [{\citenamefont {Granter}, \citenamefont {Beck},\ and\ \citenamefont
  {Papke}(2017)}]{granter2017alphago}%
  \BibitemOpen
  \bibfield  {author} {\bibinfo {author} {\bibfnamefont {S.~R.}\ \bibnamefont
  {Granter}}, \bibinfo {author} {\bibfnamefont {A.~H.}\ \bibnamefont {Beck}}, \
  and\ \bibinfo {author} {\bibfnamefont {D.~J.}\ \bibnamefont {Papke}},\
  }\bibfield  {title} {\enquote {\bibinfo {title} {Alphago, deep learning, and
  the future of the human microscopist},}\ }\href {\doibase
  https://doi.org/10.5858/arpa.2016-0471-ED} {\bibfield  {journal} {\bibinfo
  {journal} {Arch Pathol Lab Med}\ }\textbf {\bibinfo {volume} {141}},\
  \bibinfo {pages} {619--621} (\bibinfo {year} {2017})}\BibitemShut {NoStop}%
\bibitem [{\citenamefont {Zhao}, \citenamefont {Queralta},\ and\ \citenamefont
  {Westerlund}(2020)}]{9308468}%
  \BibitemOpen
  \bibfield  {author} {\bibinfo {author} {\bibfnamefont {W.}~\bibnamefont
  {Zhao}}, \bibinfo {author} {\bibfnamefont {J.~P.}\ \bibnamefont {Queralta}},
  \ and\ \bibinfo {author} {\bibfnamefont {T.}~\bibnamefont {Westerlund}},\
  }\bibfield  {title} {\enquote {\bibinfo {title} {Sim-to-real transfer in deep
  reinforcement learning for robotics: a survey},}\ }in\ \href {\doibase
  10.1109/SSCI47803.2020.9308468} {\emph {\bibinfo {booktitle} {2020 IEEE
  Symposium Series on Computational Intelligence (SSCI)}}}\ (\bibinfo {year}
  {2020})\ pp.\ \bibinfo {pages} {737--744}\BibitemShut {NoStop}%
\bibitem [{\citenamefont {Sallab}\ \emph {et~al.}(2017)\citenamefont {Sallab},
  \citenamefont {Abdou}, \citenamefont {Perot},\ and\ \citenamefont
  {Yogamani}}]{Sallab_2017}%
  \BibitemOpen
  \bibfield  {author} {\bibinfo {author} {\bibfnamefont {A.~E.}\ \bibnamefont
  {Sallab}}, \bibinfo {author} {\bibfnamefont {M.}~\bibnamefont {Abdou}},
  \bibinfo {author} {\bibfnamefont {E.}~\bibnamefont {Perot}}, \ and\ \bibinfo
  {author} {\bibfnamefont {S.}~\bibnamefont {Yogamani}},\ }\bibfield  {title}
  {\enquote {\bibinfo {title} {Deep reinforcement learning framework for
  autonomous driving},}\ }\href {\doibase
  10.2352/issn.2470-1173.2017.19.avm-023} {\bibfield  {journal} {\bibinfo
  {journal} {Electronic Imaging}\ }\textbf {\bibinfo {volume} {29}},\ \bibinfo
  {pages} {70–76} (\bibinfo {year} {2017})}\BibitemShut {NoStop}%
\bibitem [{\citenamefont {Garnier}\ \emph {et~al.}(2021)\citenamefont
  {Garnier}, \citenamefont {Viquerat}, \citenamefont {Rabault}, \citenamefont
  {Larcher}, \citenamefont {Kuhnle},\ and\ \citenamefont
  {Hachem}}]{GARNIER2021104973}%
  \BibitemOpen
  \bibfield  {author} {\bibinfo {author} {\bibfnamefont {P.}~\bibnamefont
  {Garnier}}, \bibinfo {author} {\bibfnamefont {J.}~\bibnamefont {Viquerat}},
  \bibinfo {author} {\bibfnamefont {J.}~\bibnamefont {Rabault}}, \bibinfo
  {author} {\bibfnamefont {A.}~\bibnamefont {Larcher}}, \bibinfo {author}
  {\bibfnamefont {A.}~\bibnamefont {Kuhnle}}, \ and\ \bibinfo {author}
  {\bibfnamefont {E.}~\bibnamefont {Hachem}},\ }\bibfield  {title} {\enquote
  {\bibinfo {title} {A review on deep reinforcement learning for fluid
  mechanics},}\ }\href {\doibase
  https://doi.org/10.1016/j.compfluid.2021.104973} {\bibfield  {journal}
  {\bibinfo  {journal} {Computers \& Fluids}\ }\textbf {\bibinfo {volume}
  {225}},\ \bibinfo {pages} {104973} (\bibinfo {year} {2021})}\BibitemShut
  {NoStop}%
\bibitem [{\citenamefont {Razdan}\ and\ \citenamefont {Shah}(2022)}]{Popat}%
  \BibitemOpen
  \bibfield  {author} {\bibinfo {author} {\bibfnamefont {S.}~\bibnamefont
  {Razdan}}\ and\ \bibinfo {author} {\bibfnamefont {S.}~\bibnamefont {Shah}},\
  }\bibfield  {title} {\enquote {\bibinfo {title} {Optimization of fluid
  modeling and flow control processes using machine learning: A brief
  review},}\ }in\ \href {\doibase 10.1007/978-981-19-0676-3_6} {\emph {\bibinfo
  {booktitle} {Advances in Mechanical Engineering and Material Science}}},\
  \bibinfo {editor} {edited by\ \bibinfo {editor} {\bibfnamefont {K.~C.}\
  \bibnamefont {Popat}}, \bibinfo {editor} {\bibfnamefont {S.}~\bibnamefont
  {Kanagaraj}}, \bibinfo {editor} {\bibfnamefont {P.~S.~R.}\ \bibnamefont
  {Sreekanth}}, \ and\ \bibinfo {editor} {\bibfnamefont {V.~M.~R.}\
  \bibnamefont {Kumar}}}\ (\bibinfo  {publisher} {Springer Nature Singapore},\
  \bibinfo {address} {Singapore},\ \bibinfo {year} {2022})\ pp.\ \bibinfo
  {pages} {63--85}\BibitemShut {NoStop}%
\bibitem [{\citenamefont {Xie}\ \emph {et~al.}(2023)\citenamefont {Xie},
  \citenamefont {Zheng}, \citenamefont {Ji}, \citenamefont {Zhang},
  \citenamefont {Bi}, \citenamefont {Zhou},\ and\ \citenamefont
  {Zheng}}]{Aerospace2023}%
  \BibitemOpen
  \bibfield  {author} {\bibinfo {author} {\bibfnamefont {F.}~\bibnamefont
  {Xie}}, \bibinfo {author} {\bibfnamefont {C.}~\bibnamefont {Zheng}}, \bibinfo
  {author} {\bibfnamefont {T.}~\bibnamefont {Ji}}, \bibinfo {author}
  {\bibfnamefont {X.}~\bibnamefont {Zhang}}, \bibinfo {author} {\bibfnamefont
  {R.}~\bibnamefont {Bi}}, \bibinfo {author} {\bibfnamefont {H.}~\bibnamefont
  {Zhou}}, \ and\ \bibinfo {author} {\bibfnamefont {Y.}~\bibnamefont {Zheng}},\
  }\bibfield  {title} {\enquote {\bibinfo {title} {Deep reinforcement learning:
  A new beacon for intelligent active flow control},}\ }\href {\doibase
  10.3389/arc.2023.11130} {\bibfield  {journal} {\bibinfo  {journal} {Aerospace
  Research Communications}\ }\textbf {\bibinfo {volume} {1}} (\bibinfo {year}
  {2023}),\ 10.3389/arc.2023.11130}\BibitemShut {NoStop}%
\bibitem [{\citenamefont {Zhao}\ \emph {et~al.}(2024)\citenamefont {Zhao},
  \citenamefont {Zhao}, \citenamefont {Deng}, \citenamefont {Wang},
  \citenamefont {Zhang}, \citenamefont {Zheng}, \citenamefont {Cao},
  \citenamefont {Nan}, \citenamefont {Lian},\ and\ \citenamefont
  {Burke}}]{ZHAO2024122836}%
  \BibitemOpen
  \bibfield  {author} {\bibinfo {author} {\bibfnamefont {J.}~\bibnamefont
  {Zhao}}, \bibinfo {author} {\bibfnamefont {W.}~\bibnamefont {Zhao}}, \bibinfo
  {author} {\bibfnamefont {B.}~\bibnamefont {Deng}}, \bibinfo {author}
  {\bibfnamefont {Z.}~\bibnamefont {Wang}}, \bibinfo {author} {\bibfnamefont
  {F.}~\bibnamefont {Zhang}}, \bibinfo {author} {\bibfnamefont
  {W.}~\bibnamefont {Zheng}}, \bibinfo {author} {\bibfnamefont
  {W.}~\bibnamefont {Cao}}, \bibinfo {author} {\bibfnamefont {J.}~\bibnamefont
  {Nan}}, \bibinfo {author} {\bibfnamefont {Y.}~\bibnamefont {Lian}}, \ and\
  \bibinfo {author} {\bibfnamefont {A.~F.}\ \bibnamefont {Burke}},\ }\bibfield
  {title} {\enquote {\bibinfo {title} {Autonomous driving system: A
  comprehensive survey},}\ }\href {\doibase
  https://doi.org/10.1016/j.eswa.2023.122836} {\bibfield  {journal} {\bibinfo
  {journal} {Expert Systems with Applications}\ }\textbf {\bibinfo {volume}
  {242}},\ \bibinfo {pages} {122836} (\bibinfo {year} {2024})}\BibitemShut
  {NoStop}%
\bibitem [{\citenamefont {Husen}, \citenamefont {Chaudary},\ and\ \citenamefont
  {Ahmad}(2022)}]{Husen2023}%
  \BibitemOpen
  \bibfield  {author} {\bibinfo {author} {\bibfnamefont {A.}~\bibnamefont
  {Husen}}, \bibinfo {author} {\bibfnamefont {M.~H.}\ \bibnamefont {Chaudary}},
  \ and\ \bibinfo {author} {\bibfnamefont {F.}~\bibnamefont {Ahmad}},\
  }\bibfield  {title} {\enquote {\bibinfo {title} {A survey on requirements of
  future intelligent networks: Solutions and future research directions},}\
  }\href {\doibase 10.1145/3524106} {\bibfield  {journal} {\bibinfo  {journal}
  {ACM Comput. Surv.}\ }\textbf {\bibinfo {volume} {55}} (\bibinfo {year}
  {2022}),\ 10.1145/3524106}\BibitemShut {NoStop}%
\bibitem [{\citenamefont {Bhola}\ \emph {et~al.}(2023)\citenamefont {Bhola},
  \citenamefont {Pawar}, \citenamefont {Balaprakash},\ and\ \citenamefont
  {Maulik}}]{BHOLA2023112018}%
  \BibitemOpen
  \bibfield  {author} {\bibinfo {author} {\bibfnamefont {S.}~\bibnamefont
  {Bhola}}, \bibinfo {author} {\bibfnamefont {S.}~\bibnamefont {Pawar}},
  \bibinfo {author} {\bibfnamefont {P.}~\bibnamefont {Balaprakash}}, \ and\
  \bibinfo {author} {\bibfnamefont {R.}~\bibnamefont {Maulik}},\ }\bibfield
  {title} {\enquote {\bibinfo {title} {Multi-fidelity reinforcement learning
  framework for shape optimization},}\ }\href {\doibase
  https://doi.org/10.1016/j.jcp.2023.112018} {\bibfield  {journal} {\bibinfo
  {journal} {Journal of Computational Physics}\ }\textbf {\bibinfo {volume}
  {482}},\ \bibinfo {pages} {112018} (\bibinfo {year} {2023})}\BibitemShut
  {NoStop}%
\bibitem [{\citenamefont {Vinuesa}\ \emph {et~al.}(2022)\citenamefont
  {Vinuesa}, \citenamefont {Lehmkuhl}, \citenamefont {Lozano-Dur{\'a}n},\ and\
  \citenamefont {Rabault}}]{vinuesa2022flow}%
  \BibitemOpen
  \bibfield  {author} {\bibinfo {author} {\bibfnamefont {R.}~\bibnamefont
  {Vinuesa}}, \bibinfo {author} {\bibfnamefont {O.}~\bibnamefont {Lehmkuhl}},
  \bibinfo {author} {\bibfnamefont {A.}~\bibnamefont {Lozano-Dur{\'a}n}}, \
  and\ \bibinfo {author} {\bibfnamefont {J.}~\bibnamefont {Rabault}},\
  }\bibfield  {title} {\enquote {\bibinfo {title} {Flow control in wings and
  discovery of novel approaches via deep reinforcement learning},}\ }\href
  {\doibase 10.3390/fluids7020062} {\bibfield  {journal} {\bibinfo  {journal}
  {Fluids}\ }\textbf {\bibinfo {volume} {7}},\ \bibinfo {pages} {62} (\bibinfo
  {year} {2022})}\BibitemShut {NoStop}%
\bibitem [{\citenamefont {Vignon}, \citenamefont {Rabault},\ and\ \citenamefont
  {Vinuesa}(2023)}]{Vignon2023}%
  \BibitemOpen
  \bibfield  {author} {\bibinfo {author} {\bibfnamefont {C.}~\bibnamefont
  {Vignon}}, \bibinfo {author} {\bibfnamefont {J.}~\bibnamefont {Rabault}}, \
  and\ \bibinfo {author} {\bibfnamefont {R.}~\bibnamefont {Vinuesa}},\
  }\bibfield  {title} {\enquote {\bibinfo {title} {{Recent advances in applying
  deep reinforcement learning for flow control: Perspectives and future
  directions}},}\ }\href {\doibase 10.1063/5.0143913} {\bibfield  {journal}
  {\bibinfo  {journal} {Physics of Fluids}\ }\textbf {\bibinfo {volume} {35}},\
  \bibinfo {pages} {031301} (\bibinfo {year} {2023})}\BibitemShut {NoStop}%
\bibitem [{\citenamefont {Viquerat}\ \emph {et~al.}(2022)\citenamefont
  {Viquerat}, \citenamefont {Meliga}, \citenamefont {Larcher},\ and\
  \citenamefont {Hachem}}]{Viquerat}%
  \BibitemOpen
  \bibfield  {author} {\bibinfo {author} {\bibfnamefont {J.}~\bibnamefont
  {Viquerat}}, \bibinfo {author} {\bibfnamefont {P.}~\bibnamefont {Meliga}},
  \bibinfo {author} {\bibfnamefont {A.}~\bibnamefont {Larcher}}, \ and\
  \bibinfo {author} {\bibfnamefont {E.}~\bibnamefont {Hachem}},\ }\bibfield
  {title} {\enquote {\bibinfo {title} {A review on deep reinforcement learning
  for fluid mechanics: An update},}\ }\href {\doibase 10.1063/5.0128446}
  {\bibfield  {journal} {\bibinfo  {journal} {Physics of Fluids}\ }\textbf
  {\bibinfo {volume} {34}} (\bibinfo {year} {2022}),\
  10.1063/5.0128446}\BibitemShut {NoStop}%
\bibitem [{\citenamefont {Rabault}\ \emph {et~al.}(2019)\citenamefont
  {Rabault}, \citenamefont {Kuchta}, \citenamefont {Jensen}, \citenamefont
  {R{\'e}glade},\ and\ \citenamefont {Cerardi}}]{rabault2019artificial}%
  \BibitemOpen
  \bibfield  {author} {\bibinfo {author} {\bibfnamefont {J.}~\bibnamefont
  {Rabault}}, \bibinfo {author} {\bibfnamefont {M.}~\bibnamefont {Kuchta}},
  \bibinfo {author} {\bibfnamefont {A.}~\bibnamefont {Jensen}}, \bibinfo
  {author} {\bibfnamefont {U.}~\bibnamefont {R{\'e}glade}}, \ and\ \bibinfo
  {author} {\bibfnamefont {N.}~\bibnamefont {Cerardi}},\ }\bibfield  {title}
  {\enquote {\bibinfo {title} {Artificial neural networks trained through deep
  reinforcement learning discover control strategies for active flow
  control},}\ }\href {\doibase 10.1017/jfm.2019.62} {\bibfield  {journal}
  {\bibinfo  {journal} {Journal of fluid mechanics}\ }\textbf {\bibinfo
  {volume} {865}},\ \bibinfo {pages} {281--302} (\bibinfo {year}
  {2019})}\BibitemShut {NoStop}%
\bibitem [{\citenamefont {Tang}\ \emph {et~al.}(2020)\citenamefont {Tang},
  \citenamefont {Rabault}, \citenamefont {Kuhnle}, \citenamefont {Wang},\ and\
  \citenamefont {Wang}}]{tangRobustActiveFlow2020}%
  \BibitemOpen
  \bibfield  {author} {\bibinfo {author} {\bibfnamefont {H.}~\bibnamefont
  {Tang}}, \bibinfo {author} {\bibfnamefont {J.}~\bibnamefont {Rabault}},
  \bibinfo {author} {\bibfnamefont {A.}~\bibnamefont {Kuhnle}}, \bibinfo
  {author} {\bibfnamefont {Y.}~\bibnamefont {Wang}}, \ and\ \bibinfo {author}
  {\bibfnamefont {T.}~\bibnamefont {Wang}},\ }\bibfield  {title} {\enquote
  {\bibinfo {title} {Robust active flow control over a range of {{Reynolds}}
  numbers using an artificial neural network trained through deep reinforcement
  learning},}\ }\href {\doibase 10.1063/5.0006492} {\bibfield  {journal}
  {\bibinfo  {journal} {Physics of Fluids}\ }\textbf {\bibinfo {volume} {32}},\
  \bibinfo {pages} {053605} (\bibinfo {year} {2020})}\BibitemShut {NoStop}%
\bibitem [{\citenamefont {Heess}\ \emph {et~al.}(2017)\citenamefont {Heess},
  \citenamefont {Dhruva}, \citenamefont {Sriram}, \citenamefont {Lemmon},
  \citenamefont {Merel}, \citenamefont {Wayne}, \citenamefont {Tassa},
  \citenamefont {Erez}, \citenamefont {Wang}, \citenamefont {Eslami},
  \citenamefont {Riedmiller},\ and\ \citenamefont
  {Silver}}]{heess2017emergence}%
  \BibitemOpen
  \bibfield  {author} {\bibinfo {author} {\bibfnamefont {N.~M.~O.}\
  \bibnamefont {Heess}}, \bibinfo {author} {\bibfnamefont {T.}~\bibnamefont
  {Dhruva}}, \bibinfo {author} {\bibfnamefont {S.}~\bibnamefont {Sriram}},
  \bibinfo {author} {\bibfnamefont {J.}~\bibnamefont {Lemmon}}, \bibinfo
  {author} {\bibfnamefont {J.}~\bibnamefont {Merel}}, \bibinfo {author}
  {\bibfnamefont {G.}~\bibnamefont {Wayne}}, \bibinfo {author} {\bibfnamefont
  {Y.}~\bibnamefont {Tassa}}, \bibinfo {author} {\bibfnamefont
  {T.}~\bibnamefont {Erez}}, \bibinfo {author} {\bibfnamefont {Z.}~\bibnamefont
  {Wang}}, \bibinfo {author} {\bibfnamefont {S.~M.~A.}\ \bibnamefont {Eslami}},
  \bibinfo {author} {\bibfnamefont {M.~A.}\ \bibnamefont {Riedmiller}}, \ and\
  \bibinfo {author} {\bibfnamefont {D.}~\bibnamefont {Silver}},\ }\bibfield
  {title} {\enquote {\bibinfo {title} {Emergence of locomotion behaviours in
  rich environments},}\ }\href
  {https://api.semanticscholar.org/CorpusID:30099687} {\bibfield  {journal}
  {\bibinfo  {journal} {ArXiv}\ }\textbf {\bibinfo {volume} {abs/1707.02286}}
  (\bibinfo {year} {2017})}\BibitemShut {NoStop}%
\bibitem [{\citenamefont {Wang}\ and\ \citenamefont
  {Xu}(2024{\natexlab{a}})}]{jia2024optimal}%
  \BibitemOpen
  \bibfield  {author} {\bibinfo {author} {\bibfnamefont {J.}~\bibnamefont
  {Wang}}\ and\ \bibinfo {author} {\bibfnamefont {H.}~\bibnamefont {Xu}},\
  }\bibfield  {title} {\enquote {\bibinfo {title} {Optimal parallelization
  strategies for active flow control in deep reinforcement learning-based
  computational fluid dynamics},}\ }\href {\doibase 10.1063/5.0204237}
  {\bibfield  {journal} {\bibinfo  {journal} {Physics of Fluids}\ }\textbf
  {\bibinfo {volume} {36}},\ \bibinfo {pages} {043623} (\bibinfo {year}
  {2024}{\natexlab{a}})}\BibitemShut {NoStop}%
\bibitem [{\citenamefont {Li}\ and\ \citenamefont
  {Zhang}(2022)}]{liReinforcementlearning}%
  \BibitemOpen
  \bibfield  {author} {\bibinfo {author} {\bibfnamefont {J.}~\bibnamefont
  {Li}}\ and\ \bibinfo {author} {\bibfnamefont {M.}~\bibnamefont {Zhang}},\
  }\bibfield  {title} {\enquote {\bibinfo {title} {Reinforcement-learning-based
  control of confined cylinder wakes with stability analyses},}\ }\href
  {\doibase 10.1017/jfm.2021.1045} {\bibfield  {journal} {\bibinfo  {journal}
  {Journal of Fluid Mechanics}\ }\textbf {\bibinfo {volume} {932}},\ \bibinfo
  {pages} {A44} (\bibinfo {year} {2022})},\ \Eprint
  {http://arxiv.org/abs/2111.07498} {2111.07498 [physics]} \BibitemShut
  {NoStop}%
\bibitem [{\citenamefont {Wang}\ \emph
  {et~al.}(2022{\natexlab{a}})\citenamefont {Wang}, \citenamefont {Yan},
  \citenamefont {Hu}, \citenamefont {Li}, \citenamefont {Xiao}, \citenamefont
  {Xiong}, \citenamefont {Rabault},\ and\ \citenamefont
  {Noack}}]{wangDRLinFluids}%
  \BibitemOpen
  \bibfield  {author} {\bibinfo {author} {\bibfnamefont {Q.}~\bibnamefont
  {Wang}}, \bibinfo {author} {\bibfnamefont {L.}~\bibnamefont {Yan}}, \bibinfo
  {author} {\bibfnamefont {G.}~\bibnamefont {Hu}}, \bibinfo {author}
  {\bibfnamefont {C.}~\bibnamefont {Li}}, \bibinfo {author} {\bibfnamefont
  {Y.}~\bibnamefont {Xiao}}, \bibinfo {author} {\bibfnamefont {H.}~\bibnamefont
  {Xiong}}, \bibinfo {author} {\bibfnamefont {J.}~\bibnamefont {Rabault}}, \
  and\ \bibinfo {author} {\bibfnamefont {B.~R.}\ \bibnamefont {Noack}},\
  }\bibfield  {title} {\enquote {\bibinfo {title} {{{DRLinFluids}}: {{An}}
  open-source {{Python}} platform of coupling deep reinforcement learning and
  {{OpenFOAM}}},}\ }\href {\doibase 10.1063/5.0103113} {\bibfield  {journal}
  {\bibinfo  {journal} {Physics of Fluids}\ }\textbf {\bibinfo {volume} {34}},\
  \bibinfo {pages} {081801} (\bibinfo {year} {2022}{\natexlab{a}})}\BibitemShut
  {NoStop}%
\bibitem [{\citenamefont {Wang}\ and\ \citenamefont
  {Xu}(2024{\natexlab{b}})}]{jia2024robust}%
  \BibitemOpen
  \bibfield  {author} {\bibinfo {author} {\bibfnamefont {J.}~\bibnamefont
  {Wang}}\ and\ \bibinfo {author} {\bibfnamefont {H.}~\bibnamefont {Xu}},\
  }\bibfield  {title} {\enquote {\bibinfo {title} {Robust and adaptive deep
  reinforcement learning for enhancing flow control around a square cylinder
  with varying reynolds numbers},}\ }\href {\doibase 10.1063/5.0207879}
  {\bibfield  {journal} {\bibinfo  {journal} {Physics of Fluids}\ }\textbf
  {\bibinfo {volume} {36}},\ \bibinfo {pages} {054103} (\bibinfo {year}
  {2024}{\natexlab{b}})}\BibitemShut {NoStop}%
\bibitem [{\citenamefont {Wang}\ \emph
  {et~al.}(2022{\natexlab{b}})\citenamefont {Wang}, \citenamefont {Mei},
  \citenamefont {Aubry}, \citenamefont {Chen}, \citenamefont {Wu},\ and\
  \citenamefont {Wu}}]{wang2022deep}%
  \BibitemOpen
  \bibfield  {author} {\bibinfo {author} {\bibfnamefont {Y.-Z.}\ \bibnamefont
  {Wang}}, \bibinfo {author} {\bibfnamefont {Y.-F.}\ \bibnamefont {Mei}},
  \bibinfo {author} {\bibfnamefont {N.}~\bibnamefont {Aubry}}, \bibinfo
  {author} {\bibfnamefont {Z.}~\bibnamefont {Chen}}, \bibinfo {author}
  {\bibfnamefont {P.}~\bibnamefont {Wu}}, \ and\ \bibinfo {author}
  {\bibfnamefont {W.-T.}\ \bibnamefont {Wu}},\ }\bibfield  {title} {\enquote
  {\bibinfo {title} {Deep reinforcement learning based synthetic jet control on
  disturbed flow over airfoil},}\ }\href {\doibase 10.1063/5.0080922}
  {\bibfield  {journal} {\bibinfo  {journal} {Physics of Fluids}\ }\textbf
  {\bibinfo {volume} {34}},\ \bibinfo {pages} {033606} (\bibinfo {year}
  {2022}{\natexlab{b}})}\BibitemShut {NoStop}%
\bibitem [{\citenamefont {He}\ \emph {et~al.}(2023)\citenamefont {He},
  \citenamefont {Wang}, \citenamefont {Hua}, \citenamefont {Chen},
  \citenamefont {Li},\ and\ \citenamefont {Wu}}]{he2023policy}%
  \BibitemOpen
  \bibfield  {author} {\bibinfo {author} {\bibfnamefont {X.-J.}\ \bibnamefont
  {He}}, \bibinfo {author} {\bibfnamefont {Y.-Z.}\ \bibnamefont {Wang}},
  \bibinfo {author} {\bibfnamefont {Y.}~\bibnamefont {Hua}}, \bibinfo {author}
  {\bibfnamefont {Z.-H.}\ \bibnamefont {Chen}}, \bibinfo {author}
  {\bibfnamefont {Y.-B.}\ \bibnamefont {Li}}, \ and\ \bibinfo {author}
  {\bibfnamefont {W.-T.}\ \bibnamefont {Wu}},\ }\bibfield  {title} {\enquote
  {\bibinfo {title} {Policy transfer of reinforcement learning-based flow
  control: From two- to three-dimensional environment},}\ }\href {\doibase
  10.1063/5.0147190} {\bibfield  {journal} {\bibinfo  {journal} {Physics of
  Fluids}\ }\textbf {\bibinfo {volume} {35}},\ \bibinfo {pages} {055116}
  (\bibinfo {year} {2023})}\BibitemShut {NoStop}%
\bibitem [{\citenamefont {Fan}\ \emph {et~al.}(2020)\citenamefont {Fan},
  \citenamefont {Yang}, \citenamefont {Wang}, \citenamefont {Triantafyllou},\
  and\ \citenamefont {Karniadakis}}]{Dixiapnas}%
  \BibitemOpen
  \bibfield  {author} {\bibinfo {author} {\bibfnamefont {D.}~\bibnamefont
  {Fan}}, \bibinfo {author} {\bibfnamefont {L.}~\bibnamefont {Yang}}, \bibinfo
  {author} {\bibfnamefont {Z.}~\bibnamefont {Wang}}, \bibinfo {author}
  {\bibfnamefont {M.~S.}\ \bibnamefont {Triantafyllou}}, \ and\ \bibinfo
  {author} {\bibfnamefont {G.~E.}\ \bibnamefont {Karniadakis}},\ }\bibfield
  {title} {\enquote {\bibinfo {title} {Reinforcement learning for bluff body
  active flow control in experiments and simulations},}\ }\href {\doibase
  10.1073/pnas.2004939117} {\bibfield  {journal} {\bibinfo  {journal}
  {Proceedings of the National Academy of Sciences}\ }\textbf {\bibinfo
  {volume} {117}},\ \bibinfo {pages} {26091--26098} (\bibinfo {year}
  {2020})}\BibitemShut {NoStop}%
\bibitem [{\citenamefont {Ren}, \citenamefont {Wang},\ and\ \citenamefont
  {Tang}(2021)}]{Bluffbody}%
  \BibitemOpen
  \bibfield  {author} {\bibinfo {author} {\bibfnamefont {F.}~\bibnamefont
  {Ren}}, \bibinfo {author} {\bibfnamefont {C.}~\bibnamefont {Wang}}, \ and\
  \bibinfo {author} {\bibfnamefont {H.}~\bibnamefont {Tang}},\ }\bibfield
  {title} {\enquote {\bibinfo {title} {{Bluff body uses
  deep-reinforcement-learning trained active flow control to achieve
  hydrodynamic stealth}},}\ }\href {\doibase 10.1063/5.0060690} {\bibfield
  {journal} {\bibinfo  {journal} {Physics of Fluids}\ }\textbf {\bibinfo
  {volume} {33}},\ \bibinfo {pages} {093602} (\bibinfo {year}
  {2021})}\BibitemShut {NoStop}%
\bibitem [{\citenamefont {Paris}, \citenamefont {Beneddine},\ and\
  \citenamefont {Dandois}(2021)}]{paris}%
  \BibitemOpen
  \bibfield  {author} {\bibinfo {author} {\bibfnamefont {R.}~\bibnamefont
  {Paris}}, \bibinfo {author} {\bibfnamefont {S.}~\bibnamefont {Beneddine}}, \
  and\ \bibinfo {author} {\bibfnamefont {J.}~\bibnamefont {Dandois}},\
  }\bibfield  {title} {\enquote {\bibinfo {title} {Robust flow control and
  optimal sensor placement using deep reinforcement learning},}\ }\href
  {\doibase 10.1017/jfm.2020.1170} {\bibfield  {journal} {\bibinfo  {journal}
  {Journal of Fluid Mechanics}\ }\textbf {\bibinfo {volume} {913}},\ \bibinfo
  {pages} {A25} (\bibinfo {year} {2021})}\BibitemShut {NoStop}%
\bibitem [{\citenamefont {Chen}\ \emph {et~al.}(2023)\citenamefont {Chen},
  \citenamefont {Wang}, \citenamefont {Yan}, \citenamefont {Hu},\ and\
  \citenamefont {Noack}}]{chen2023deep}%
  \BibitemOpen
  \bibfield  {author} {\bibinfo {author} {\bibfnamefont {W.}~\bibnamefont
  {Chen}}, \bibinfo {author} {\bibfnamefont {Q.}~\bibnamefont {Wang}}, \bibinfo
  {author} {\bibfnamefont {L.}~\bibnamefont {Yan}}, \bibinfo {author}
  {\bibfnamefont {G.}~\bibnamefont {Hu}}, \ and\ \bibinfo {author}
  {\bibfnamefont {B.~R.}\ \bibnamefont {Noack}},\ }\bibfield  {title} {\enquote
  {\bibinfo {title} {Deep reinforcement learning-based active flow control of
  vortex-induced vibration of a square cylinder},}\ }\href {\doibase
  10.1063/5.0152777} {\bibfield  {journal} {\bibinfo  {journal} {Physics of
  Fluids}\ }\textbf {\bibinfo {volume} {35}},\ \bibinfo {pages} {053610}
  (\bibinfo {year} {2023})}\BibitemShut {NoStop}%
\bibitem [{\citenamefont {Xia}\ \emph {et~al.}(2024)\citenamefont {Xia},
  \citenamefont {Zhang}, \citenamefont {Kerrigan},\ and\ \citenamefont
  {Rigas}}]{Xia2024}%
  \BibitemOpen
  \bibfield  {author} {\bibinfo {author} {\bibfnamefont {C.}~\bibnamefont
  {Xia}}, \bibinfo {author} {\bibfnamefont {J.}~\bibnamefont {Zhang}}, \bibinfo
  {author} {\bibfnamefont {E.}~\bibnamefont {Kerrigan}}, \ and\ \bibinfo
  {author} {\bibfnamefont {G.}~\bibnamefont {Rigas}},\ }\bibfield  {title}
  {\enquote {\bibinfo {title} {Active flow control for bluff body drag
  reduction using reinforcement learning with partial measurements},}\ }\href
  {\doibase 10.1017/jfm.2024.69} {\bibfield  {journal} {\bibinfo  {journal}
  {Journal of Fluid Mechanics}\ }\textbf {\bibinfo {volume} {981}},\ \bibinfo
  {pages} {A17} (\bibinfo {year} {2024})}\BibitemShut {NoStop}%
\bibitem [{\citenamefont {Jia}\ and\ \citenamefont {Xu}(2024)}]{jia2024effect}%
  \BibitemOpen
  \bibfield  {author} {\bibinfo {author} {\bibfnamefont {W.}~\bibnamefont
  {Jia}}\ and\ \bibinfo {author} {\bibfnamefont {H.}~\bibnamefont {Xu}},\
  }\href@noop {} {\enquote {\bibinfo {title} {Effect of synthetic jets actuator
  parameters on deep reinforcement learning-based flow control performance in a
  square cylinder},}\ } (\bibinfo {year} {2024}),\ \Eprint
  {http://arxiv.org/abs/2405.12834} {arXiv:2405.12834 [physics.flu-dyn]}
  \BibitemShut {NoStop}%
\bibitem [{\citenamefont {Yan}\ \emph {et~al.}(2023)\citenamefont {Yan},
  \citenamefont {Li}, \citenamefont {Hu}, \citenamefont {Chen}, \citenamefont
  {Zhong},\ and\ \citenamefont {Noack}}]{yan2023stabilizing}%
  \BibitemOpen
  \bibfield  {author} {\bibinfo {author} {\bibfnamefont {L.}~\bibnamefont
  {Yan}}, \bibinfo {author} {\bibfnamefont {Y.}~\bibnamefont {Li}}, \bibinfo
  {author} {\bibfnamefont {G.}~\bibnamefont {Hu}}, \bibinfo {author}
  {\bibfnamefont {W.-l.}\ \bibnamefont {Chen}}, \bibinfo {author}
  {\bibfnamefont {W.}~\bibnamefont {Zhong}}, \ and\ \bibinfo {author}
  {\bibfnamefont {B.~R.}\ \bibnamefont {Noack}},\ }\bibfield  {title} {\enquote
  {\bibinfo {title} {Stabilizing the square cylinder wake using deep
  reinforcement learning for different jet locations},}\ }\href {\doibase
  10.1063/5.0171188} {\bibfield  {journal} {\bibinfo  {journal} {Physics of
  Fluids}\ }\textbf {\bibinfo {volume} {35}},\ \bibinfo {pages} {115104}
  (\bibinfo {year} {2023})}\BibitemShut {NoStop}%
\bibitem [{\citenamefont {Rabault}\ \emph {et~al.}(2020)\citenamefont
  {Rabault}, \citenamefont {Ren}, \citenamefont {Zhang}, \citenamefont {Tang},\
  and\ \citenamefont {Xu}}]{rabault2020deep}%
  \BibitemOpen
  \bibfield  {author} {\bibinfo {author} {\bibfnamefont {J.}~\bibnamefont
  {Rabault}}, \bibinfo {author} {\bibfnamefont {F.}~\bibnamefont {Ren}},
  \bibinfo {author} {\bibfnamefont {W.}~\bibnamefont {Zhang}}, \bibinfo
  {author} {\bibfnamefont {H.}~\bibnamefont {Tang}}, \ and\ \bibinfo {author}
  {\bibfnamefont {H.}~\bibnamefont {Xu}},\ }\bibfield  {title} {\enquote
  {\bibinfo {title} {Deep reinforcement learning in fluid mechanics: A
  promising method for both active flow control and shape optimization},}\
  }\href {\doibase https://doi.org/10.1007/s42241-020-0028-y} {\bibfield
  {journal} {\bibinfo  {journal} {Journal of Hydrodynamics}\ }\textbf {\bibinfo
  {volume} {32}},\ \bibinfo {pages} {234--246} (\bibinfo {year}
  {2020})}\BibitemShut {NoStop}%
\bibitem [{\citenamefont {Jasak}(2009)}]{jasakOpenFOAMOpenSource2009}%
  \BibitemOpen
  \bibfield  {author} {\bibinfo {author} {\bibfnamefont {H.}~\bibnamefont
  {Jasak}},\ }\bibfield  {title} {\enquote {\bibinfo {title} {{{OpenFOAM}}:
  {{Open}} source {{CFD}} in research and industry},}\ }\href {\doibase
  10.2478/IJNAOE-2013-0011} {\bibfield  {journal} {\bibinfo  {journal}
  {International Journal of Naval Architecture and Ocean Engineering}\ }\textbf
  {\bibinfo {volume} {1}},\ \bibinfo {pages} {89--94} (\bibinfo {year}
  {2009})}\BibitemShut {NoStop}%
\bibitem [{\citenamefont {Jasak}, \citenamefont {Jemcov},\ and\ \citenamefont
  {Tukovic}(2007)}]{jasakOpenFOAMLibraryComplex2013}%
  \BibitemOpen
  \bibfield  {author} {\bibinfo {author} {\bibfnamefont {H.}~\bibnamefont
  {Jasak}}, \bibinfo {author} {\bibfnamefont {A.}~\bibnamefont {Jemcov}}, \
  and\ \bibinfo {author} {\bibfnamefont {Z.}~\bibnamefont {Tukovic}},\
  }\bibfield  {title} {\enquote {\bibinfo {title} {Openfoam: A c++ library for
  complex physics simulations},}\ }in\ \href
  {http://csabai.web.elte.hu/http/simulationLab/jasakEtAlOpenFoam.pdf} {\emph
  {\bibinfo {booktitle} {International Workshop on Coupled Methods in Numerical
  Dynamics}}}\ (\bibinfo  {publisher} {Inter-University Centre, Dubrovnik},\
  \bibinfo {address} {Croatia},\ \bibinfo {year} {2007})\ pp.\ \bibinfo {pages}
  {1--20}\BibitemShut {NoStop}%
\bibitem [{\citenamefont {Schäfer}\ \emph {et~al.}(1996)\citenamefont
  {Schäfer}, \citenamefont {Turek}, \citenamefont {Durst}, \citenamefont
  {Krause},\ and\ \citenamefont {Rannacher}}]{schafer1996benchmark}%
  \BibitemOpen
  \bibfield  {author} {\bibinfo {author} {\bibfnamefont {M.}~\bibnamefont
  {Schäfer}}, \bibinfo {author} {\bibfnamefont {S.}~\bibnamefont {Turek}},
  \bibinfo {author} {\bibfnamefont {F.}~\bibnamefont {Durst}}, \bibinfo
  {author} {\bibfnamefont {E.}~\bibnamefont {Krause}}, \ and\ \bibinfo {author}
  {\bibfnamefont {R.}~\bibnamefont {Rannacher}},\ }\bibfield  {title} {\enquote
  {\bibinfo {title} {Benchmark computations of laminar flow around a
  cylinder},}\ }in\ \href {\doibase 10.1007/978-3-322-89849-4_39} {\emph
  {\bibinfo {booktitle} {Flow Simulation with High-Performance Computers
  II}}},\ \bibinfo {series} {Notes on Numerical Fluid Mechanics (NNFM)},
  Vol.~\bibinfo {volume} {48},\ \bibinfo {editor} {edited by\ \bibinfo {editor}
  {\bibfnamefont {E.~H.}\ \bibnamefont {Hirschel}}}\ (\bibinfo  {publisher}
  {Vieweg+Teubner Verlag},\ \bibinfo {year} {1996})\ pp.\ \bibinfo {pages}
  {547--566}\BibitemShut {NoStop}%
\bibitem [{\citenamefont {Rabault}\ and\ \citenamefont
  {Kuhnle}(2019{\natexlab{a}})}]{rabaultAccelerating}%
  \BibitemOpen
  \bibfield  {author} {\bibinfo {author} {\bibfnamefont {J.}~\bibnamefont
  {Rabault}}\ and\ \bibinfo {author} {\bibfnamefont {A.}~\bibnamefont
  {Kuhnle}},\ }\bibfield  {title} {\enquote {\bibinfo {title} {Accelerating
  deep reinforcement learning strategies of flow control through a
  multi-environment approach},}\ }\href {\doibase 10.1063/1.5116415} {\bibfield
   {journal} {\bibinfo  {journal} {Physics of Fluids}\ }\textbf {\bibinfo
  {volume} {31}},\ \bibinfo {pages} {094105} (\bibinfo {year}
  {2019}{\natexlab{a}})}\BibitemShut {NoStop}%
\bibitem [{\citenamefont {Schaarschmidt}\ \emph {et~al.}(2018)\citenamefont
  {Schaarschmidt}, \citenamefont {Kuhnle}, \citenamefont {Ellis}, \citenamefont
  {Fricke}, \citenamefont {Gessert},\ and\ \citenamefont
  {Yoneki}}]{schaarschmidt2017tensorforce}%
  \BibitemOpen
  \bibfield  {author} {\bibinfo {author} {\bibfnamefont {M.}~\bibnamefont
  {Schaarschmidt}}, \bibinfo {author} {\bibfnamefont {A.}~\bibnamefont
  {Kuhnle}}, \bibinfo {author} {\bibfnamefont {B.}~\bibnamefont {Ellis}},
  \bibinfo {author} {\bibfnamefont {K.}~\bibnamefont {Fricke}}, \bibinfo
  {author} {\bibfnamefont {F.}~\bibnamefont {Gessert}}, \ and\ \bibinfo
  {author} {\bibfnamefont {E.}~\bibnamefont {Yoneki}},\ }\href
  {http://arxiv.org/abs/1808.07903} {\enquote {\bibinfo {title} {{LIFT}:
  Reinforcement learning in computer systems by learning from
  demonstrations},}\ } (\bibinfo {year} {2018}),\ \Eprint
  {http://arxiv.org/abs/1808.07903} {arXiv:1808.07903} \BibitemShut {NoStop}%
\bibitem [{\citenamefont {Abadi}\ \emph {et~al.}(2016)\citenamefont {Abadi},
  \citenamefont {Barham}, \citenamefont {Chen}, \citenamefont {Chen},
  \citenamefont {Davis}, \citenamefont {Dean}, \citenamefont {Devin},
  \citenamefont {Ghemawat}, \citenamefont {Irving}, \citenamefont {Isard},
  \citenamefont {Kudlur}, \citenamefont {Levenberg}, \citenamefont {Monga},
  \citenamefont {Moore}, \citenamefont {Murray}, \citenamefont {Steiner},
  \citenamefont {Tucker}, \citenamefont {Vasudevan}, \citenamefont {Warden},
  \citenamefont {Wicke}, \citenamefont {Yu},\ and\ \citenamefont
  {Zheng}}]{abadi2016tensorflow}%
  \BibitemOpen
  \bibfield  {author} {\bibinfo {author} {\bibfnamefont {M.}~\bibnamefont
  {Abadi}}, \bibinfo {author} {\bibfnamefont {P.}~\bibnamefont {Barham}},
  \bibinfo {author} {\bibfnamefont {J.}~\bibnamefont {Chen}}, \bibinfo {author}
  {\bibfnamefont {Z.}~\bibnamefont {Chen}}, \bibinfo {author} {\bibfnamefont
  {A.}~\bibnamefont {Davis}}, \bibinfo {author} {\bibfnamefont
  {J.}~\bibnamefont {Dean}}, \bibinfo {author} {\bibfnamefont {M.}~\bibnamefont
  {Devin}}, \bibinfo {author} {\bibfnamefont {S.}~\bibnamefont {Ghemawat}},
  \bibinfo {author} {\bibfnamefont {G.}~\bibnamefont {Irving}}, \bibinfo
  {author} {\bibfnamefont {M.}~\bibnamefont {Isard}}, \bibinfo {author}
  {\bibfnamefont {M.}~\bibnamefont {Kudlur}}, \bibinfo {author} {\bibfnamefont
  {J.}~\bibnamefont {Levenberg}}, \bibinfo {author} {\bibfnamefont
  {R.}~\bibnamefont {Monga}}, \bibinfo {author} {\bibfnamefont
  {S.}~\bibnamefont {Moore}}, \bibinfo {author} {\bibfnamefont {D.~G.}\
  \bibnamefont {Murray}}, \bibinfo {author} {\bibfnamefont {B.}~\bibnamefont
  {Steiner}}, \bibinfo {author} {\bibfnamefont {P.}~\bibnamefont {Tucker}},
  \bibinfo {author} {\bibfnamefont {V.}~\bibnamefont {Vasudevan}}, \bibinfo
  {author} {\bibfnamefont {P.}~\bibnamefont {Warden}}, \bibinfo {author}
  {\bibfnamefont {M.}~\bibnamefont {Wicke}}, \bibinfo {author} {\bibfnamefont
  {Y.}~\bibnamefont {Yu}}, \ and\ \bibinfo {author} {\bibfnamefont
  {X.}~\bibnamefont {Zheng}},\ }\bibfield  {title} {\enquote {\bibinfo {title}
  {{TensorFlow}: A system for {Large-Scale} machine learning},}\ }in\ \href
  {https://www.usenix.org/conference/osdi16/technical-sessions/presentation/abadi}
  {\emph {\bibinfo {booktitle} {12th USENIX Symposium on Operating Systems
  Design and Implementation (OSDI 16)}}}\ (\bibinfo  {publisher} {USENIX
  Association},\ \bibinfo {address} {Savannah, GA},\ \bibinfo {year} {2016})\
  pp.\ \bibinfo {pages} {265--283}\BibitemShut {NoStop}%
\bibitem [{\citenamefont {Brockman}\ \emph {et~al.}(2016)\citenamefont
  {Brockman}, \citenamefont {Cheung}, \citenamefont {Pettersson}, \citenamefont
  {Schneider}, \citenamefont {Schulman}, \citenamefont {Tang},\ and\
  \citenamefont {Zaremba}}]{brockman2016openai}%
  \BibitemOpen
  \bibfield  {author} {\bibinfo {author} {\bibfnamefont {G.}~\bibnamefont
  {Brockman}}, \bibinfo {author} {\bibfnamefont {V.}~\bibnamefont {Cheung}},
  \bibinfo {author} {\bibfnamefont {L.}~\bibnamefont {Pettersson}}, \bibinfo
  {author} {\bibfnamefont {J.}~\bibnamefont {Schneider}}, \bibinfo {author}
  {\bibfnamefont {J.}~\bibnamefont {Schulman}}, \bibinfo {author}
  {\bibfnamefont {J.}~\bibnamefont {Tang}}, \ and\ \bibinfo {author}
  {\bibfnamefont {W.}~\bibnamefont {Zaremba}},\ }\href@noop {} {\enquote
  {\bibinfo {title} {Openai gym},}\ } (\bibinfo {year} {2016}),\ \Eprint
  {http://arxiv.org/abs/1606.01540} {arXiv:1606.01540 [cs.LG]} \BibitemShut
  {NoStop}%
\bibitem [{\citenamefont {Johnson}, \citenamefont {Thompson},\ and\
  \citenamefont {Hourigan}(2004)}]{JOHNSON2004229}%
  \BibitemOpen
  \bibfield  {author} {\bibinfo {author} {\bibfnamefont {S.~A.}\ \bibnamefont
  {Johnson}}, \bibinfo {author} {\bibfnamefont {M.~C.}\ \bibnamefont
  {Thompson}}, \ and\ \bibinfo {author} {\bibfnamefont {K.}~\bibnamefont
  {Hourigan}},\ }\bibfield  {title} {\enquote {\bibinfo {title} {Predicted low
  frequency structures in the wake of elliptical cylinders},}\ }\href {\doibase
  https://doi.org/10.1016/j.euromechflu.2003.05.006} {\bibfield  {journal}
  {\bibinfo  {journal} {European Journal of Mechanics - B/Fluids}\ }\textbf
  {\bibinfo {volume} {23}},\ \bibinfo {pages} {229--239} (\bibinfo {year}
  {2004})},\ \bibinfo {note} {bluff Body Wakes and Vortex-Induced
  Vibrations}\BibitemShut {NoStop}%
\bibitem [{\citenamefont {Sahin}\ and\ \citenamefont
  {Owens}(2004)}]{sahin2004numerical}%
  \BibitemOpen
  \bibfield  {author} {\bibinfo {author} {\bibfnamefont {M.}~\bibnamefont
  {Sahin}}\ and\ \bibinfo {author} {\bibfnamefont {R.~G.}\ \bibnamefont
  {Owens}},\ }\bibfield  {title} {\enquote {\bibinfo {title} {A numerical
  investigation of wall effects up to high blockage ratios on two-dimensional
  flow past a confined circular cylinder},}\ }\href {\doibase
  10.1063/1.1668285} {\bibfield  {journal} {\bibinfo  {journal} {Physics of
  Fluids}\ }\textbf {\bibinfo {volume} {16}},\ \bibinfo {pages} {1305--1320}
  (\bibinfo {year} {2004})}\BibitemShut {NoStop}%
\bibitem [{\citenamefont {Pastoor}\ \emph {et~al.}(2008)\citenamefont
  {Pastoor}, \citenamefont {Henning}, \citenamefont {Noack}, \citenamefont
  {King},\ and\ \citenamefont {Tadmor}}]{pastoor2008feedback}%
  \BibitemOpen
  \bibfield  {author} {\bibinfo {author} {\bibfnamefont {M.}~\bibnamefont
  {Pastoor}}, \bibinfo {author} {\bibfnamefont {L.}~\bibnamefont {Henning}},
  \bibinfo {author} {\bibfnamefont {B.~R.}\ \bibnamefont {Noack}}, \bibinfo
  {author} {\bibfnamefont {R.}~\bibnamefont {King}}, \ and\ \bibinfo {author}
  {\bibfnamefont {G.}~\bibnamefont {Tadmor}},\ }\bibfield  {title} {\enquote
  {\bibinfo {title} {Feedback shear layer control for bluff body drag
  reduction},}\ }\href {\doibase 10.1017/S0022112008002073} {\bibfield
  {journal} {\bibinfo  {journal} {Journal of Fluid Mechanics}\ }\textbf
  {\bibinfo {volume} {608}},\ \bibinfo {pages} {161--196} (\bibinfo {year}
  {2008})}\BibitemShut {NoStop}%
\bibitem [{\citenamefont {Protas}\ and\ \citenamefont
  {Wesfreid}(2002)}]{protas2002drag}%
  \BibitemOpen
  \bibfield  {author} {\bibinfo {author} {\bibfnamefont {B.}~\bibnamefont
  {Protas}}\ and\ \bibinfo {author} {\bibfnamefont {J.~E.}\ \bibnamefont
  {Wesfreid}},\ }\bibfield  {title} {\enquote {\bibinfo {title} {Drag force in
  the open-loop control of the cylinder wake in the laminar regime},}\ }\href
  {\doibase 10.1063/1.1432695} {\bibfield  {journal} {\bibinfo  {journal}
  {Physics of Fluids}\ }\textbf {\bibinfo {volume} {14}},\ \bibinfo {pages}
  {810--826} (\bibinfo {year} {2002})}\BibitemShut {NoStop}%
\bibitem [{\citenamefont {Bergmann}, \citenamefont {Cordier},\ and\
  \citenamefont {Brancher}(2005)}]{bergmann2005optimal}%
  \BibitemOpen
  \bibfield  {author} {\bibinfo {author} {\bibfnamefont {M.}~\bibnamefont
  {Bergmann}}, \bibinfo {author} {\bibfnamefont {L.}~\bibnamefont {Cordier}}, \
  and\ \bibinfo {author} {\bibfnamefont {J.-P.}\ \bibnamefont {Brancher}},\
  }\bibfield  {title} {\enquote {\bibinfo {title} {Optimal rotary control of
  the cylinder wake using proper orthogonal decomposition reduced-order
  model},}\ }\href {\doibase 10.1063/1.2033624} {\bibfield  {journal} {\bibinfo
   {journal} {Physics of Fluids}\ }\textbf {\bibinfo {volume} {17}},\ \bibinfo
  {pages} {097101} (\bibinfo {year} {2005})}\BibitemShut {NoStop}%
\bibitem [{\citenamefont {Rabault}\ and\ \citenamefont
  {Kuhnle}(2019{\natexlab{b}})}]{rabault2024cylinder2dflowcontroldrlparallel}%
  \BibitemOpen
  \bibfield  {author} {\bibinfo {author} {\bibfnamefont {J.}~\bibnamefont
  {Rabault}}\ and\ \bibinfo {author} {\bibfnamefont {A.}~\bibnamefont
  {Kuhnle}},\ }\href@noop {} {\enquote {\bibinfo {title}
  {Cylinder2dflowcontroldrlparallel},}\ }\bibinfo {howpublished} {GitHub.
  \url{https://github.com/jerabaul29/Cylinder2DFlowControlDRLParallel}}
  (\bibinfo {year} {2019}{\natexlab{b}})\BibitemShut {NoStop}%
\bibitem [{\citenamefont {Li}(2021)}]{Nek5000}%
  \BibitemOpen
  \bibfield  {author} {\bibinfo {author} {\bibfnamefont {J.}~\bibnamefont
  {Li}},\ }\href@noop {} {\enquote {\bibinfo {title} {Rlcontrolnek5000},}\
  }\bibinfo {howpublished} {GitHub.
  \url{https://github.com/npuljc/RL_control_Nek5000}} (\bibinfo {year}
  {2021})\BibitemShut {NoStop}%
\bibitem [{\citenamefont {Qiulei~Wang}(2022)}]{DRLinFluids}%
  \BibitemOpen
  \bibfield  {author} {\bibinfo {author} {\bibfnamefont {W.~C.}\ \bibnamefont
  {Qiulei~Wang}, \bibfnamefont {Lei~Yan}},\ }\href@noop {} {\enquote {\bibinfo
  {title} {Drlinfluids},}\ }\bibinfo {howpublished} {GitHub.
  \url{https://github.com/venturi123/DRLinFluids}} (\bibinfo {year}
  {2022})\BibitemShut {NoStop}%
\end{thebibliography}%
\end{document}